\documentclass[a4paper,11pt]{article}
\pdfoutput=1 

\usepackage{jheppub} 

\usepackage[T1]{fontenc} 

\usepackage{graphicx}
\usepackage{epsfig}
\usepackage{rotating}
\usepackage{amssymb}
\usepackage{dsfont}
\usepackage{psfrag}
\usepackage{amsmath,euscript,array,mathrsfs}
\usepackage{bbold}
\usepackage{epsf}
\usepackage{multirow}

\usepackage{grffile}
\usepackage{subcaption}

\def\figdir{Figures/}

\newcommand{\beq}{\begin{equation}}
\newcommand{\eeq}{\end{equation}}
\newcommand{\beqs}{\begin{eqnarray}}
\newcommand{\eeqs}{\end{eqnarray}}
\newcommand{\lsim}{\mathrel{\raisebox{-
.6ex}{$\stackrel{\textstyle<}{\sim}$}}}
\newcommand{\gsim}{\mathrel{\raisebox{-
.6ex}{$\stackrel{\textstyle>}{\sim}$}}}
\newcommand{\Tr}{{\rm Tr}}

\def\hbar{\hspace{0pt}\raisebox{1pt}{$-$} \hspace{-7pt} h}

\def\r{\rho}

\newcommand{\be}{\begin{equation}}
\newcommand{\ee}{\end{equation}}

\newcommand{\bea}{\begin{eqnarray}}
\newcommand{\eea}{\end{eqnarray}}
\newcommand{\nn}{\nonumber}

\def\lbldef#1#2{\expandafter\gdef\csname #1\endcsname {#2}}

\def\href#1#2{#2}

\newcommand{\ber}{\begin{eqnarray}}
\newcommand{\eer}{\end{eqnarray}}

\newcommand{\beqar}{\begin{eqnarray}}

\newcommand{\eeqar}{\end{eqnarray}}

\newcommand{\dsl}
  {\kern.06em\hbox{\raise.15ex\hbox{$/$}\kern-.56em\hbox{$\partial$}}}

\newcommand{\eeqarr}{\end{eqnarray}}
\newcommand{\ZZ}{{\rm \kern 0.275em Z \kern -0.92em Z}\;}

\def\CC{{\mathchoice
{\rm C\mkern-8mu\vrule height1.45ex depth-.05ex
width.05em\mkern9mu\kern-.05em}
{\rm C\mkern-8mu\vrule height1.45ex depth-.05ex
width.05em\mkern9mu\kern-.05em}
{\rm C\mkern-8mu\vrule height1ex depth-.07ex
width.035em\mkern9mu\kern-.035em}
{\rm C\mkern-8mu\vrule height.65ex depth-.1ex
width.025em\mkern8mu\kern-.025em}}}

\def\RR{{\rm I\kern-1.6pt {\rm R}}}

\def\ZZ{{\rm Z}\kern-3.8pt {\rm Z} \kern2pt}
\def\IB{\relax{\rm I\kern-.18em B}}
\def\ID{\relax{\rm I\kern-.18em D}}
\def\II{\relax{\rm I\kern-.18em I}}
\def\IP{\relax{\rm I\kern-.18em P}}

\newcommand{\bear}{\begin{eqnarray}}
\newcommand{\eear}{\end{eqnarray}}

\def\r{\rho}                                     

\def\6{\partial}

\def\bea{\begin{eqnarray}}
\def\eea{\end{eqnarray}}

\def\beqx{\begin{displaymath}}
\def\eeqx{\end{displaymath}}

\newcommand{\bmat}{\left(\begin{array}}
\newcommand{\emat}{\end{array}\right)}

\newcommand\Sec[1]{Section~\ref{Sec:#1}}
\newcommand\App[1]{Appendix~\ref{Sec:#1}}
\newcommand\Tab[1]{Table~\ref{tab:#1}}
\newcommand\Fig[1]{Fig.~\ref{fig:#1}}
\newcommand\Eq[1]{Eq.~(\ref{eq:#1})}

\def\r{\rho}

\def\bo{{\raise-.3ex\hbox{\large$\Box$}}}               
\def\face{{\raise.2ex\hbox{$\displaystyle \bigodot$}\mskip-2.2mu \llap {$\ddot
        \smile$}}}                                   
\def\>{\rangle}                                      
\def\<{\langle}                                      

\def\leftrightarrowfill{$\mathsurround=0pt \mathord\leftarrow \mkern-6mu
        \cleaders\hbox{$\mkern-2mu \mathord- \mkern-2mu$}\hfill
        \mkern-6mu \mathord\rightarrow$}        
\def\dvec#1{\vbox{\ialign{##\crcr
        \leftrightarrowfill\crcr\noalign{\kern-1pt\nointerlineskip}
        $\hfil\displaystyle{#1}\hfil$\crcr}}}           
\def\Tr{{\rm Tr \,}}                                    

\def\-{\hphantom{-}}

\title{$Sp(4)$ gauge theories on the lattice: $N_f=2$  dynamical fundamental fermions}

\author[a]{Ed Bennett,}
\author[b]{Deog Ki Hong,}
\author[b,c]{Jong-Wan Lee,}
\author[d,e]{C.-J. David Lin,}
\author[f]{Biagio Lucini,}
\author[g]{Maurizio Piai,}
\author[h]{and Davide Vadacchino.}

\affiliation[a]{Swansea Academy of Advanced Computing, Swansea University,
Bay Campus, SA1 8EN, Swansea, Wales, UK}
\affiliation[b]{Department of Physics, Pusan National University, Busan 46241, Korea}
\affiliation[c]{Extreme Physics Institute, Pusan National University, Busan 46241, Korea}
\affiliation[d]{Institute of Physics, National Chiao-Tung University, 1001 Ta-Hsueh 
Road, Hsinchu 30010, Taiwan}
\affiliation[e]{Centre for High Energy Physics, Chung-Yuan Christian University,
Chung-Li 32023, Taiwan}
\affiliation[f]{Department of Mathematics, College of Science, Swansea University,
Bay Campus, SA1 8EN, Swansea, Wales, UK}
\affiliation[g]{Department of Physics, College of Science, Swansea University,
Singleton Park, SA2 8PP, Swansea, Wales, UK}
\affiliation[h]{INFN, Sezione di Pisa, Largo Pontecorvo 3, 56127 Pisa, Italy}

\emailAdd{e.j.bennett@swansea.ac.uk}
\emailAdd{dkhong@pusan.ac.kr}
\emailAdd{jwlee823@pusan.ac.kr}
\emailAdd{dlin@mail.nctu.edu.tw}
\emailAdd{b.lucini@swansea.ac.uk}
\emailAdd{m.piai@swansea.ac.uk}
\emailAdd{davide.vadacchino@pi.infn.it}

\abstract{
We  perform lattice studies of the gauge theory with $Sp(4)$ gauge group
and two flavours of (Dirac) fundamental matter.
The global $SU(4)$ symmetry is spontaneously broken by the fermion condensate.
The dynamical Wilson fermions in the  lattice action introduce a mass that  breaks
the global symmetry also explicitly.
The resulting pseudo-Nambu-Goldstone bosons describe the $SU(4)/Sp(4)$ coset,
and are relevant, in the context of physics beyond the Standard Model,
for  composite Higgs models.
We discuss scale setting, continuum extrapolation and finite volume effects in the lattice theory.
We study mesonic composite states, which span representations of the
unbroken $Sp(4)$ global symmetry,
and we measure masses and decay constants of the  (flavoured) spin-0 and spin-1 states accessible to
the numerical treatment, as a function of the fermion mass.
With help from
the effective field theory treatment of such mesons, 
we perform a first extrapolation towards the massless limit.
We assess our results by critically comparing to the literature on other models and to the quenched results,
and we conclude by outlining future avenues for further exploration.
The  results of our spectroscopic analysis
provide new input data for future phenomenological studies in the contexts of composite Higgs models,
and of dark matter models with a strongly coupled dynamical origin.
}

\preprint{PNUTP-19/A01}

\date{\today}

\begin{document}
\maketitle
\flushbottom



\section{Introduction}
\label{Sec:intro}


 The Large Hadron Collider (LHC) recently discovered a new scalar
  particle~\cite{Aad:2012tfa,Chatrchyan:2012xdj},
which has experimental properties  compatible with those of the Higgs boson
and mass $\sim 126$ GeV.
This discovery stands out against the current absence of clear evidence 
of new physics beyond the Standard Model (SM),
both in direct and indirect experimental searches, up to and beyond the TeV scale---evidence of 
a little desert in high energy physics.
Composite Higgs Models (CHMs)  implement the symmetry-based mechanism first proposed 
in Refs.~\cite{Kaplan:1983fs,Georgi:1984af,Dugan:1984hq}.
They provide a compelling framework  that allows to soften
the level of fine-tuning required to accommodate the little desert within an Effective Field Theory (EFT) 
description of Electro-Weak Symmetry Breaking (EWSB).
 
In the SM, EWSB is induced 
at the scale $v_W\sim 246$ GeV by the dynamics of a complex doublet of weakly-coupled 
Higgs fields. Composite Higgs models reinterpret its components
as some of the interpolating fields appearing in the EFT
that describes the low energy, weakly coupled dynamics of 
a set of composite pseudo-Nambu-Goldstone bosons (pNGBs).
They emerge from a more fundamental---possibly strongly coupled and UV complete---theory, 
that dynamically drives the spontaneous breaking, at scale $f > v_W$, of an extended
 approximate continuous global symmetry.
 After coupling this EFT  to the SM gauge bosons and fermions,
what results is a  natural and stable dynamical 
origin for  the little hierarchy between the masses of the SM particles (the Higgs boson in particular) 
and the higher scale beyond which new phenomena arise.
The ratio $v_W/f < 1$ is determined by the interplay of strong-coupling dynamics 
and  weakly coupled, symmetry-breaking 
perturbations.
Electroweak symmetry breaking is triggered
via what, in the  jargon of the field, is often referred to as
 {\it vacuum misalignment}---a phrase that highlights the intrinsic differences with other 
 models of EWSB with a strongly coupled dynamical origin~\cite{Peskin:1980gc}.

A wide variety of implementations of these ideas has been proposed in recent years,
and their model-building features, phenomenological implications, 
and dynamical properties are the subject of a rich, diverse,
and rapidly evolving  literature
(see for instance Refs.~\cite{Agashe:2004rs, Contino:2006qr,
Barbieri:2007bh,Lodone:2008yy,Marzocca:2012zn,Grojean:2013qca,Ferretti:2013kya,
Cacciapaglia:2014uja,Arbey:2015exa,Vecchi:2015fma,Panico:2015jxa,Ferretti:2016upr,
Agugliaro:2016clv,Alanne:2017rrs,Feruglio:2016zvt,Fichet:2016xvs,
Galloway:2016fuo,Alanne:2017ymh,Csaki:2017cep,
Chala:2017sjk,Csaki:2017jby,Ayyar:2018zuk,Ayyar:2018ppa,Cai:2018tet,
Agugliaro:2018vsu,Ayyar:2018glg,Cacciapaglia:2018avr,Witzel:2019jbe,
Cacciapaglia:2019bqz,Ayyar:2019exp,Cossu:2019hse,Cacciapaglia:2019ixa,BuarqueFranzosi:2019eee}). 
Particular attention has been devoted to models and dynamical theories 
characterised by the $SU(4)/Sp(4)$ coset 
(see for instance Refs.~\cite{Katz:2005au,Gripaios:2009pe,Barnard:2013zea,
Lewis:2011zb,Hietanen:2014xca,Arthur:2016dir,Arthur:2016ozw,
Pica:2016zst,Detmold:2014kba,Lee:2017uvl,Cacciapaglia:2015eqa,Bizot:2016zyu,Hong:2017smd,Golterman:2017vdj,Drach:2017btk,
Sannino:2017utc,Alanne:2018wtp,Bizot:2018tds,BuarqueFranzosi:2018eaj,Gertov:2019yqo}). 
The low-energy EFT Lagrangian 
contains, in this case,  five scalar fields, that describe the
 long-distance dynamics of the five pNGBs. With an appropriate  
choice of embedding for the $SU(2)\times U(1)$ gauge group of the SM,
four such fields reconstruct the complex 
Higgs doublet familiar to the Reader from the Standard Model, with the fifth scalar
a new real, neutral singlet, extending the  SM Higgs sector.

The investigations summarised in these pages contribute to the study of 
 the strong dynamics  underlying the $SU(4)/Sp(4)$ theory,
under the working assumption that it originates from a $Sp(2N)$ gauge theory
with two fundamental (Dirac) fermions, which is
amenable to lattice numerical treatment.
We ultimately  aim at computing the many free parameters 
of the low-energy EFT, by starting from fundamental principles.
This paper summarises the findings of  the second stage of development
of the programme outlined in Ref.~\cite{Bennett:2017kga} 
(see also~\cite{Bennett:2017tum,Bennett:2017ttu,
Bennett:2017kbp,Lee:2018ztv}), by focusing attention on the case in which the matter field content consists of
two dynamical Dirac fermions transforming in the fundamental representation of the $Sp(4)$ gauge group. 
Analogous to the $SU(3)$ gauge theory with two flavours of Dirac fermions in the fundamental representation, this theory is
expected to be asymptotically free and deep inside a chirally broken phase~\cite{Sannino:2009aw,Ryttov:2017dhd}. 
For instance, the first two coefficients in the perturbative beta function of the gauge coupling are positive.

The main step forwards we make here is that 
we move beyond the  quenched approximation adopted in earlier explorative
work~\cite{Bennett:2017kga}, as we
implement in its stead fully dynamical (Wilson)  fermions on the lattice.
In the presence of two fundamental lattice parameters,
the scale-setting process has to be reconsidered---especially 
in the dynamical regime away from the chiral limit.
We  compute the spectrum of  the lightest mesons, 
which are organised in  irreducible representations of the unbroken global $Sp(4)$ group.
We mostly discuss scalar S,  pseudoscalar PS, 
vector V and axial-vector AV composite flavoured particles.
For completeness we also analyse the mass of states sourced by 
both the antisymmetric
tensor T and its axial counterpart AT, 
although only the latter sources genuinely independent states---the 
tensor T and  vector V operators source the same states.
We extract masses and (appropriately renormalised) decay constants from 2-point functions of
 the relevant  interpolating operators. We
assess the size of finite volume effects and  perform  continuum limit extrapolations.

An important limitation  of this work is that the physical masses of the pNGBs 
are large enough that the vector mesons are effectively stable. 
The corresponding ranges of pseudoscalar mass and decay constant with respect to vector mass in the continuum limit 
are $0.54 \lesssim m_{\rm PS} / m_{\rm V} \lesssim 0.72$ and $0.129 \lesssim f_{\rm PS} / m_{\rm V} \lesssim 0.136$, respectively. 
While we attempt an extrapolation  towards light masses
of relevance to phenomenologically viable CHMs,
the large-mass regime is interesting in itself, being  relevant 
 for models of dark matter with a  strongly coupled
dynamical origin, along the lines discussed in
 Refs.~\cite{Hochberg:2014dra,Hochberg:2014kqa,Berlin:2018tvf}.
A crucial piece of dynamical information in this context turns out to be the strength of the
coupling $g_{\rm VPP}$ between V and two PS mesons, which plays an important role in 
controlling the relic abundance.

The $g_{\rm VPP}$ coupling can in principle be extracted
by careful analysis of  4-pNGB amplitudes~\cite{Luscher:1991cf,Feng:2010es,Alexandrou:2017mpi} 
(see also Ref.~\cite{Gasbarro:2017fmi} for an application
in the context of new physics),  but the amount of data
generated for this paper is not sufficient to even approach this gargantuan task,
which we leave for future dedicated studies.
We instead perform a first, preliminary extrapolation of our results towards the massless limit,
with help from low-energy EFT instruments.
The EFT treatment we proposed in Ref.~\cite{Bennett:2017kga} 
is based on the ideas of hidden local symmetry (HLS), adapted from 
Refs.~\cite{Bando:1984ej,Casalbuoni:1985kq,Bando:1987br,Casalbuoni:1988xm,Harada:2003jx}
(see also Refs.~\cite{Georgi:1989xy,Appelquist:1999dq,Piai:2004yb,Franzosi:2016aoo}), 
and supplemented by
some additional, simplifying working assumptions.
This process  allows us not only to estimate
the masses and decay constants of the spin-1 states in the regime relevant to electroweak models, but also
to extract an estimate of  $g_{\rm VPP}$, hence providing a first,
possibly rough measurement of its size based on a numerical, dynamical calculation.
In particular, we obtain this coupling in the massless limit, $g_{\rm VPP}^\chi=6.0(4)(2)$,
which is not far from the experimental value of real world QCD.
We also discuss several non-trivial features of the spectra, and compare them to previously published
results obtained in other related gauge theories as well as the
results of quenched calculations, that are reported elsewhere~\cite{Sp4ASFQ}.
A result of particular interest concerns the ratio between vector mass and pseudoscalar decay constant, 
for which we find that $m_{\rm V}/\sqrt{2}f_{\rm PS}=5.47(11)$ for the lightest ensemble and $5.72(18)(13)$ in the massless limit. 

The paper is organised as follows.
In Section~\ref{Sec:latticemodel} we introduce the model by defining its lattice action.
We recall some useful notions about the Hybrid Monte Carlo (HMC)
algorithm we employ, and present our choices of lattice parameters.
In Section~\ref{Sec:systematics} we employ the gradient flow method to set the physical scales.
The mass dependence of the flow scale can be understood in  EFT terms~\cite{Bar:2013ora}.
We also discuss the size of finite-volume artefacts.
In Section~\ref{Sec:mesons} we present our numerical lattice results for the spectra of mesons and renormalised decay constants. 
We define the corresponding interpolating operators and analyse their 2-point correlation functions.
Details on the HMC algorithm, diquark operators and  the fits of 
 correlation functions are presented in 
Appendix~\ref{Sec:technique}.
We present our strategy and perform the continuum limit extrapolation by employing 
a mass-dependent prescription~\cite{Ayyar:2017qdf} introduced through the Wilson chiral perturbation 
theory (W$\chi$PT)~\cite{Sheikholeslami:1985ij,Rupak:2002sm} (see also Ref.~\cite{Sharpe:1998xm},
and the literature on improvement~\cite{Symanzik:1983dc, Luscher:1996sc}),
and report the results in Section~\ref{Sec:continuum}.
Given the extensive amount of information we are communicating,  
we find it useful to conclude this part of the paper with a short summary
of the lattice numerical results in Section~\ref{Sec:summary1}.

From Section~\ref{Sec:result_con} onwards, we restrict the discussion to the study of the 
continuum limit extrapolations
  obtained by using only a set of eleven ensembles selected in Sec.~\ref{Sec:continuum}.
We deploy our EFT tools and perform  extrapolations towards the massless limit
to determine the Low Energy Constants (LECs).
We also critically discuss implications, applications, and limitations of the resulting numerical fits.
Details of the numerical results  are presented in  the histograms of Appendix~\ref{Sec:LECs_hist}.
Section~\ref{Sec:discussion} is devoted to comparing our results
 to the analogous observables in other theories, by borrowing
published data available in the literature, as well as to the results obtained 
within the quenched approximation~\cite{Sp4ASFQ}.
As we shall see, besides providing an important sanity check, the latter also
 allows us to assess the impact of quenching on 2-point correlators,
 information that might be of general 
value, as it provides guidance towards future studies of $Sp(2N)$ theories with $N>2$.
We provide a summary of the most important results in the continuum in Section~\ref{Sec:summary2},
based upon the detailed information provided in Sections~\ref{Sec:result_con} and~\ref{Sec:discussion}.
We conclude the paper with a short list of open 
avenues for future exploration in Section~\ref{Sec:conclusion}.

\section{Lattice model}
\label{Sec:latticemodel}

A distinctive  feature of $Sp(2N)$ gauge theories with $N_f$ massless 
Dirac fermions in the fundamental representation 
is the enhancement of the global symmetry to $SU(2N_f)$, which originates 
from the pseudo-real character of the representation. 

The lattice formulation in terms of Wilson fermions introduces an 
operator that breaks  the global symmetry, and introduces a
 (degenerate) mass term for the fermions.
The global symmetry is expected  to be spontaneously broken by the formation
 of a non-zero fermion bilinear condensate.
With $N_f=2$, both explicit and spontaneous breaking follow the (aligned) $SU(4)\rightarrow Sp(4)$ pattern. 
The resulting low-energy dynamics is governed by five pNGBs.
They describe the $SU(4)/Sp(4)$ coset, and  have degenerate masses.

\subsection{Lattice action}

The four-dimensional Euclidean-space lattice action
contains the gauge-field term ${S}_g$, together with the fermion matter-field term ${S}_f$:
\beq
S=S_g+S_f.
\label{eq:lattice_action}
\eeq
We use the standard Wilson plaquette action for the discretised gauge fields, 
with the gauge links $U_\mu$  group elements of $Sp(4)$ in the fundamental representation:
\beq
S_g\equiv\beta \sum_x \sum_{\mu <\nu} 
\left(
1-\frac{1}{4}\textrm{Re}\,{\Tr} \,\mathcal{P}_{\mu\nu}
\right)\,.
\label{eq:gauge_action}
\eeq
The plaquette $\mathcal{P}_{\mu\nu}$ is defined by
\beq
\mathcal{P}_{\mu\nu}(x)\equiv U_\mu (x)U_\nu(x+\hat{\mu})U^\dagger_\mu (x+\hat{\nu})U^\dagger_\nu(x)\,.
\label{eq:plaquette}
\eeq
The trace $\Tr$ is over the fundamental of $Sp(4)$, and the lattice coupling is given by $\beta=8/g^2$. 
We define the fermion sector by using the (unimproved) Wilson action 
for two mass-degenerate Dirac fermions $Q$
in the fundamental representation
\bea
S_f&\equiv&a^3\sum_x\bar{Q}(x)\left(4+a m_0\right)Q(x)\nn\,+\, \\
&& -\frac{1}{2}a^3\sum_{x,\mu} \bar{Q}(x)\left((1-\gamma_\mu)U_\mu(x)\psi(x+\hat{\mu})
+(1+\gamma_\mu)U_\mu^\dagger(x-\hat{\mu})Q(x-\hat{\mu})\right)\,,
\label{eq:fermion_action}
\eea
where $a$ is the lattice spacing and $a m_0$ is the bare mass in lattice units.

\subsection{Numerical Monte Carlo treatment}

\begin{table}
\begin{center}
\begin{tabular}{|c|c|c|c|c|c|c|c|}
\hline\hline
Ensemble & $\beta$ & $am_0$  & $N_t\times N_s^3$  & $N_{\rm configs}$ & $\delta_{\rm traj}$ & $\langle P \rangle$ & $w_0/a$ \\
\hline
DB1M1 & 6.9 & -0.85 & $32\times 16^3$ & 
100 & 24 &  0.54675(5) & 0.8149(7) \\
DB1M2 & 6.9 & -0.87 & $32\times 16^3$ & 
100 & 24 & 0.55052(6) & 0.8654(9) \\
DB1M3 & 6.9 & -0.89 & $32\times 16^3$ & 
100 & 20 & 0.55478(6) & 0.9342(11) \\
DB1M4 & 6.9 & -0.9 & $32\times 16^3$ & 
100 & 20 & 0.55696(6) & 0.9784(18) \\
DB1M5 & 6.9 & -0.91 & $32\times 16^3$ & 
100 & 20 & 0.55951(5) & 1.0413(19) \\
DB1M6 & 6.9 & -0.92 & $32\times 24^3$ & 
80 & 28 & 0.56204(3) & 1.1196(14) \\
DB1M7 & 6.9 &-0.924 & $32\times 24^3$ & 
62 & 12 & 0.56328(4) & 1.1618(13) \\
\hline
DB2M1 & 7.05 &-0.835 & $36\times 20^3$ & 
100 & 20 & 0.575267(29)  & 1.2939(19) \\
DB2M2 & 7.05 &-0.85 & $36\times 24^3$ & 
100 & 24 & 0.577371(23) & 1.4148(21) \\
DB2M3 & 7.05 &-0.857 & $36\times 32^3$ & 
102 & 20 & 0.578324(13) & 1.4836(15) \\
\hline
DB3M1 & 7.2 &-0.7 & $36\times 16^3$ & 
100 & 20 & 0.58333(4) & 1.2965(25) \\
DB3M2 & 7.2 &-0.73 & $36\times 16^3$ & 
100 & 20 & 0.58548(4) & 1.3884(36) \\
DB3M3 & 7.2 &-0.76 & $36\times 16^3$ & 
100 & 20 & 0.58767(4) & 1.5155(28) \\
DB3M4 & 7.2 &-0.77 & $36\times 24^3$ & 
100 & 20 & 0.588461(19) & 1.5625(21) \\
DB3M5 & 7.2 &-0.78 & $36\times 24^3$ & 
96 & 12 & 0.589257(20) & 1.6370(29) \\
DB3M6 & 7.2 &-0.79 & $36\times 24^3$ & 
100 & 20 & 0.590084(18)  & 1.7182(32) \\
DB3M7 & 7.2 &-0.794 & $36\times 28^3$ & 
195 & 12 & 0.590429(9) & 1.7640(18) \\
DB3M8 & 7.2 &-0.799 & $40\times 32^3$ & 
150 & 12 & 0.590869(9) & 1.8109(23) \\
\hline
DB4M1 & 7.4 &-0.72 & $48\times 32^3$ & 
150 & 12 & 0.604999(7)  & 2.1448(25) \\
DB4M2 & 7.4 &-0.73 & $48\times 32^3$ & 
150 & 12 & 0.605519(7)  & 2.2390(34) \\
\hline
DB5M1 & 7.5 &-0.69 & $48\times 24^3$ & 
100 & 12 & 0.611900(13)  & 2.3463(84) \\
\hline\hline
\end{tabular}
\end{center}
\caption{%
\label{tab:ensembles}%
List of ensembles generated for this study. 
The lattice parameters $\beta$ and $a m_0$ are, respectively, the bare coupling and bare fermion mass.
The lattice sizes are denoted by  $N_t\times N_s^3$, separately highlighting the time-like and space-like dimensions. 
The number of configurations used to estimate the average plaquette $\langle P \rangle$ 
and the gradient flow scale $w_0/a$ is denoted by $N_{\rm configs}$. 
The separation of trajectories between adjacent configurations is denoted by $\delta_{\rm traj}$. 
In the results for $\langle P \rangle$  and $w_0/a$, in parenthesis we indicate the statistical error.
}
\end{table}

We use the lattice action in \Eq{lattice_action} to study  the $Sp(4)$ theory with $N_f=2$ Dirac fundamental
fermions, with the gauge configurations generated by the standard Hybrid Monte Carlo (HMC) algorithm.
In Ref.~\cite{Bennett:2017kga} we extensively discussed the relevant numerical techniques adopted, 
including  the need to project back onto the symplectic group after each HMC update
 of the configurations,
and the associated modifications to the HiRep code~\cite{DelDebbio:2008zf}. 
During this study, we have further improved the code to implement arbitrary values of $N\geq 2$ 
and to reduce the storage size of an individual 
gauge configuration by a factor of two---details are presented in~\App{HMC}. 

Pioneering lattice studies of $Sp(2N)$ Yang-Mills showed that 
a bulk phase transition is absent in the $Sp(4)$ theory, implying that 
one can in principle take the continuum limit by choosing any values of $\beta$ \cite{Holland:2003kg}. 
By contrast, in the case of dynamical simulations with two Wilson-Dirac fermions,
 the preliminary study of 
the average plaquette value 
$\langle P \rangle \equiv (24 N_t N_s^3)^{-1} {\rm Re} \sum_x \sum_{\mu <\nu} {\rm Tr} \,
\mathcal{P}_{\mu\nu} (x)$ detected evidence of a first-order bulk phase
 transition~\cite{Bennett:2017kga}---$N_t$ and $N_s$ are 
the temporal and spatial extents of the lattice, respectively.
Hybrid Monte Carlos trajectories of $\langle P \rangle$ 
started from cold (unit) and hot (random) configurations 
at small lattice volume 
show signs of hysteresis. 
Careful study of the volume dependence of the plaquette susceptibilities
indicates that the continuum extrapolation can be carried out safely
when $\beta\gsim6.8$.
In this regime, the desired continuum limit can be reached safely,
and subsequently the  fermion mass can be lowered smoothly, avoiding
the  unphysical Aoki phase near the massless limit~\cite{DelDebbio:2008wb}. 
For all ensembles considered in this work,
we operate far enough from the massless limit and
 no sign of the Aoki phase is visible. 

The parameters characterising the ensembles generated by the dynamical simulations 
are summarised in \Tab{ensembles}. 
In the table we present the values of lattice coupling $\beta$ and bare fermion mass $am_0$.
The former is chosen to be in the range $6.9\leq\beta\leq 7.5$.
The choices of the latter will be discussed later in the paper, where we will see that some of the 
larger choices  of $a m_0$ will not be used in the analysis.

The four-dimensional Euclidean lattice has size $N_t\times N_s^3$, and we impose
 periodic boundary conditions in all directions for the gauge fields.
The physical volume $V=T\times L^3$ is obtained by setting $T=N_t a$ and $L=N_s a$. 
For the Dirac fields we implement periodic boundary conditions in the spatial directions, but
 anti-periodic boundary conditions in the temporal one. 

We anticipate that all lattice volumes satisfy the condition $m_{\rm PS} L \gsim 7.5$, 
where $a m_{\rm PS}$ denotes the mass of the lightest pseudo-scalar meson (expressed  in lattice units). 
The latter is extracted from the two-point correlation functions, as will be discussed in details in \Sec{mesons}
(see also Table~\ref{tab:meson_spec_spin0}). 
As we shall see later in the paper, 
this choice guarantees that the volumes are large enough that the finite-size effects are under control.
In the table we also present the results for the average plaquette $\langle P \rangle$ and
 the gradient flow scale $w_0$ defined in \Sec{GF}, 
which are measured from $N_{\rm configs}$ configurations separated by $\delta_{\rm traj}$ trajectories. 
Throughout this work we estimate the statistical uncertainties by using a standard
 bootstrapping method for resampling \cite{Efron:1982}.

\section{Scale setting, topology and finite volume effects}
\label{Sec:systematics}

Lattice calculations yield dimensionless numbers.
The inverse of the lattice spacing $a$ 
acts as a hard momentum cut-off $\Lambda_{\rm cut}$,
and all lattice measurements in lattice units can be written 
as a function of $a$. For example, a dimensionless mass $m^{\rm lat}$
can be expressed as $m^{\rm lat} =m a$, with $m$ having canonical units.
But this is not sufficient  to take the non-trivial limit $a\rightarrow 0$,
as the lattice spacing does not have a precise counterpart in the continuum theory.
A physical quantity that can be measured both in the continuum 
as well as in the discretised theory must be used 
to set a common reference scale, and yield a {\it scale setting} procedure
within the continuum extrapolation.

We adopt L\"{u}scher's Gradient Flow (GF) technique~\cite{Luscher:2010iy}.
Besides achieving good accuracy with modest numerical effort,
this procedure has  two other major advantages. 
The reference scale is defined on fully theoretical grounds, 
which is convenient for theories that have not been tested experimentally. 
Furthermore, it preserves the topological charge $Q$, 
while strongly suppressing  ultraviolet (UV) fluctuations. 
In this section, after carrying out the scale-setting programme,
 we also discuss the topology of the ensembles.
We conclude by studying finite volume effects 
and arguing that they are smaller than the statistical uncertainties. 

\subsection{Gradient flow and scale setting}
\label{Sec:GF}

We denote by $A_\mu(x)$  the four-dimensional non-abelian gauge field 
evaluated at the space-time coordinates $x$. 
The gradient flow is defined in the continuum theory
by a diffusion equation (in Euclidean five-dimensional space)
for a new gauge field $B_{\mu}(t,x)$ at the fictitious flow time $t$. The equation reads: 
\beq
\frac{\textrm{d}B_\mu (t,x)}{\textrm{d}t} = D_\nu G_{\nu\mu}(t,x),~\textrm{with}~B_\mu(0,x)=A_\mu(x), 
\label{Eq:gf}
\eeq
where $D_\nu$ is the covariant derivative in terms of $B_\nu$, while 
$G_{\mu\nu}=[D_\mu,D_\nu]$ is the field-strength tensor. 
Repeated indices are summed over.
Along the flow time the gauge fields evolve into renormalised gauge fields, smoothed 
over a radius of $\sqrt{8t}$, the characteristic scale of the diffusion process. 
As shown in Ref.~\cite{Luscher:2011bx}, at $t>0$ the correlation functions of the renormalised fields are finite 
to all orders in perturbation theory. 
In particular, the following gauge-invariant observable does not require any additional renormalisation 
other than that at zero flow time ($t=0$):
\beq
E(t,x)\equiv-\frac{1}{2}\Tr\,G_{\mu\nu}(t,x)G_{\mu\nu}(t,x)\,.
\label{eq:action_density}
\eeq
The expectation value of $E(t)$ is proportional to the inverse of the flow time squared. 

We consider two different proposals for defining the gradient flow scale, 
and denote them by $t_0$~\cite{Luscher:2010iy} and $w_0$~\cite{Borsanyi:2012zs}. 
We first define the dimensionless observables 
at positive flow time $t$ as
\beqs
\mathcal{E}(t) \equiv  t^2 \langle E(t) \rangle, 
\eeqs
and
\beqs
\mathcal{W}(t) \equiv \frac{\textrm{d}}{\textrm{d}\,\textrm{ln} t} t^2\langle E(t) \rangle. 
\label{eq:GF_schemes}
\eeqs
Then the scales are set by imposing the conditions
\beqs
\mathcal{E}|_{t=t_0}\equiv \mathcal{E}_0,
\eeqs
and
\beqs
\mathcal{W}|_{t=w_0^2} \equiv \mathcal{W}_0. 
\eeqs
Here $\mathcal{E}_0$ and $\mathcal{W}_0$ are common, dimensionless reference values.
In numerical studies,  we measure the dimensionless quantities  $t_0/a^2$ and $w_0/a$, 
which  determine the relative size of the lattice spacing between ensembles 
obtained by using different (bare) lattice parameters. 
  In this project, consistently with our previous work~\cite{Bennett:2017kga},
   we employ  the Wilson-flow method~\cite{Luscher:2010iy} 
   to proceed with the lattice implementation of Eq.~(\ref{Eq:gf}).  

\begin{figure}
\begin{center}
\includegraphics[width=.89\textwidth]{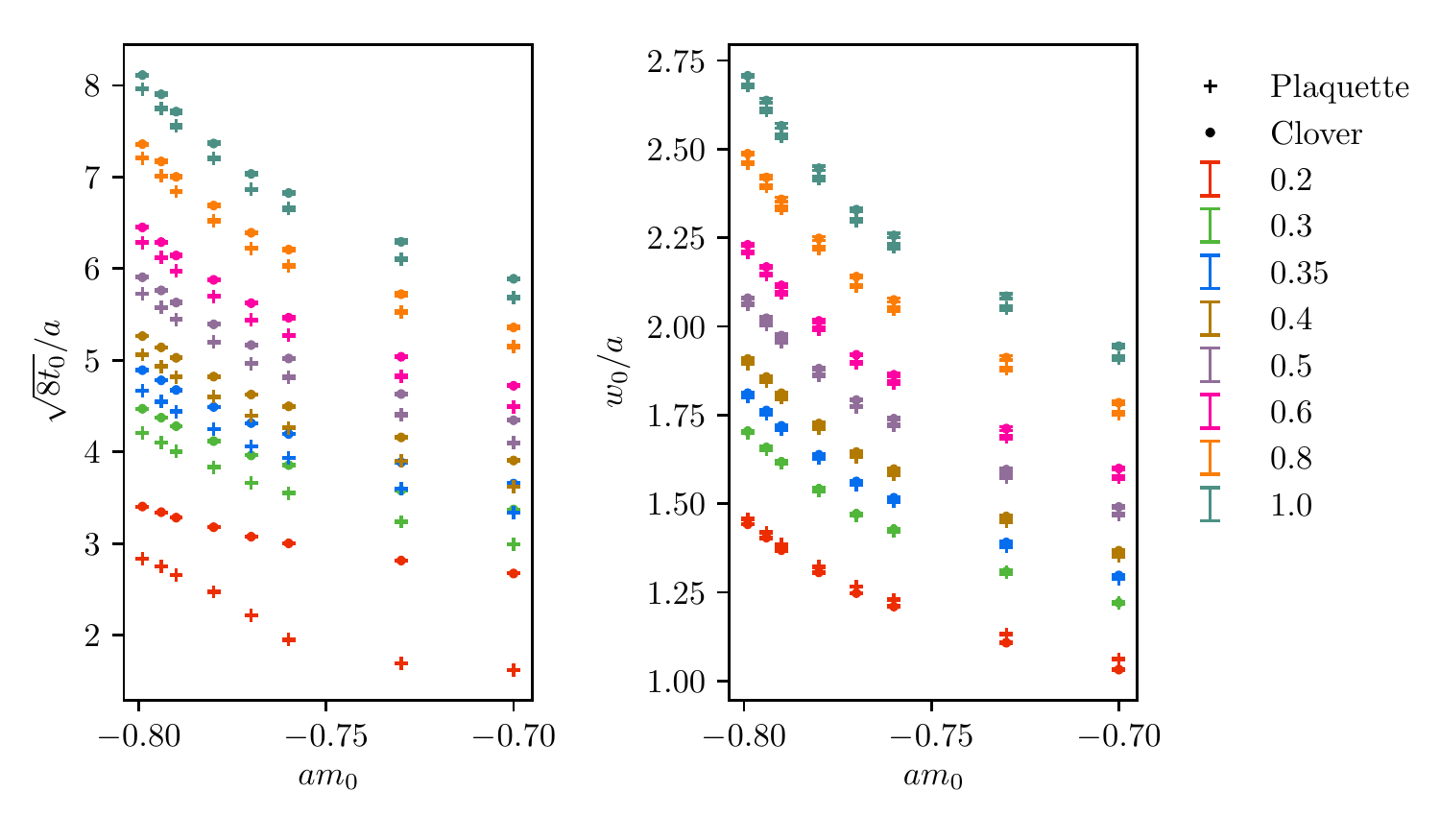}
\caption{%
\label{fig:GF_scales}%
Gradient flow scales $t_0$ (left panel) and $w_0$ (right panel) as a
 function of the bare quark mass $a m_0$,
 for $\beta=7.2$. 
Different symbols denote the different definitions of an action density (plaquette or clover). 
Different colours denote the reference values chosen for $\mathcal{E}_0$ and $\mathcal{W}_0$. 
The choices of mass and coupling identify the ensembles from Table~\ref{tab:ensembles}.
}
\end{center}
\end{figure}

In our previous publication~\cite{Bennett:2017kga}, we performed detailed numerical studies of
 the GF scheme for the quenched theory, as well as full dynamical calculations for $\beta=6.9$.
We found that $w_0$ shows smaller cut-off-dependent effects, compared to $t_0$. 
In particular, no significant deviation was found between the values of $w_0$ obtained by using 
the action density at non-zero flow time $E(t)$ constructed from the average plaquette and from the
 symmetric four-plaquette clover, as
 defined  in~\cite{Luscher:2010iy}. 

In this study, we consider a finer lattice with $\beta=7.2$. 
The results are presented in \Fig{GF_scales}. 
We find that while the values of $t_0$ show significant discrepancies, 
the measured values of $w_0$ from the two definitions 
of $E(t)$ are in good agreement over the wide range of $\mathcal{W}_0$ and $m_0$ we considered, 
in particular  for $\mathcal{W}_0=0.3\sim0.4$. 
The agreement in the flow scales has improved with  respect
to the results from  coarser lattices in~\cite{Bennett:2017kga}.
In \Tab{ensembles}, and in subsequent calculations,
we elect to use the gradient flow scale $w_0$, which we compute
 with the reference value of $\mathcal{W}_0=0.35$,
on  the four-plaquette clover action density---for which smaller lattice artefacts are observed.
For convenience, we introduce the following notation: 
$\hat{m} \equiv m^{\rm lat} w_0^{\rm lat} = m w_0$ denotes the dimensionless quantity corresponding to a mass. 
We use $\hat{a}\equiv a/w_0$ when we discuss lattice-spacing artefacts  in \Sec{spectroscopy}. 

\subsection{Chiral perturbation theory for gradient flow observables}
\label{Sec:w0}

Figure~\ref{fig:GF_scales} shows that the scales $\sqrt{8t_0}/a$ and 
$w_0/a$ depend on the fermion mass $a m_0$.
The title of this subsection
is borrowed from Ref.~\cite{Bar:2013ora}, to reflect the fact that we employ the EFT treatment 
suggested in this reference 
and we apply it to our numerical results.
The EFT treatment assumes that the square root of the flow scale $t_0$ is 
much smaller than the Compton wavelength of the pseudoscalar meson. 

Following~\cite{Bar:2013ora}, we use the leading order (LO) relation in the chiral expansion
$m_{\rm PS}^2=2Bm_f$ (where $m_f$ is the fermion mass), 
to write the next-to-leading-order (NLO) result for the GF scale $w_0^{\rm NLO}$ as
\beqs
\label{eq:w0_NLO}
w_0^{\rm NLO}(m_{PS}^2)=w_0^\chi 
\left(1+k_1\frac{m_{PS}^2}{(4\pi f_{\rm PS})^2} \right), 
\eeqs
where $w_0^\chi$ is the GF scale and $f_{\rm PS}$ is the pseudoscalar decay constant,
both defined in the chiral limit. It is convenient to 
rescale this expression by writing
\beqs
w_0^{\rm NLO}(m_{PS}^2)/a=(1+\tilde{k}_1 \hat{m}_{PS}^2)w_0^\chi/a,
\label{eq:fit_w0}
\eeqs
where $\tilde{k}_{1}$, $w_0^\chi/a$ and $\hat{m}_{\rm PS}=w_0 \,m_{\rm PS}$ are 
dimensionless parameters\footnote{The difference between $w_0 \,m_{\rm PS}$ and
$w_0^\chi \, m_{\rm PS}$ is a sub-leading effect, which would 
appear at next-to-next-to-leading order (NNLO).}. 
We also find it convenient to report here on the extraction of  the
constants $\tilde{k}_1$ and $w_0^{\chi}$ from the dynamical ensembles in \Tab{ensembles}. 
This requires that we anticipate the use of numerical data for the measurement of $m_{\rm PS}/a$ that
will  be discussed extensively in \Sec{mesons} and \App{meff}, and will be reported in \Tab{meson_spec_spin0}.

\begin{figure}
\begin{center}
\includegraphics[width=.49\textwidth]{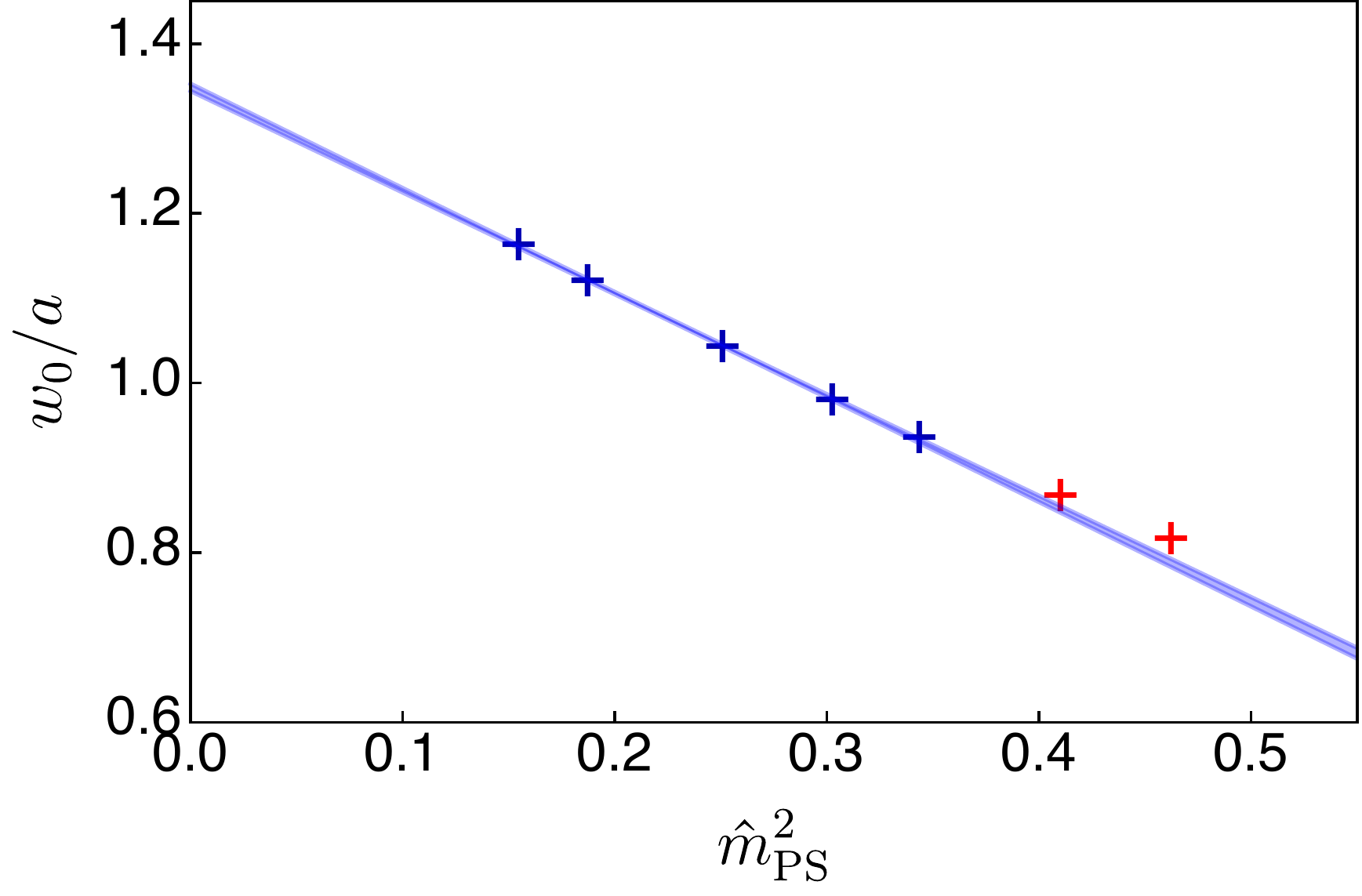}
\includegraphics[width=.49\textwidth]{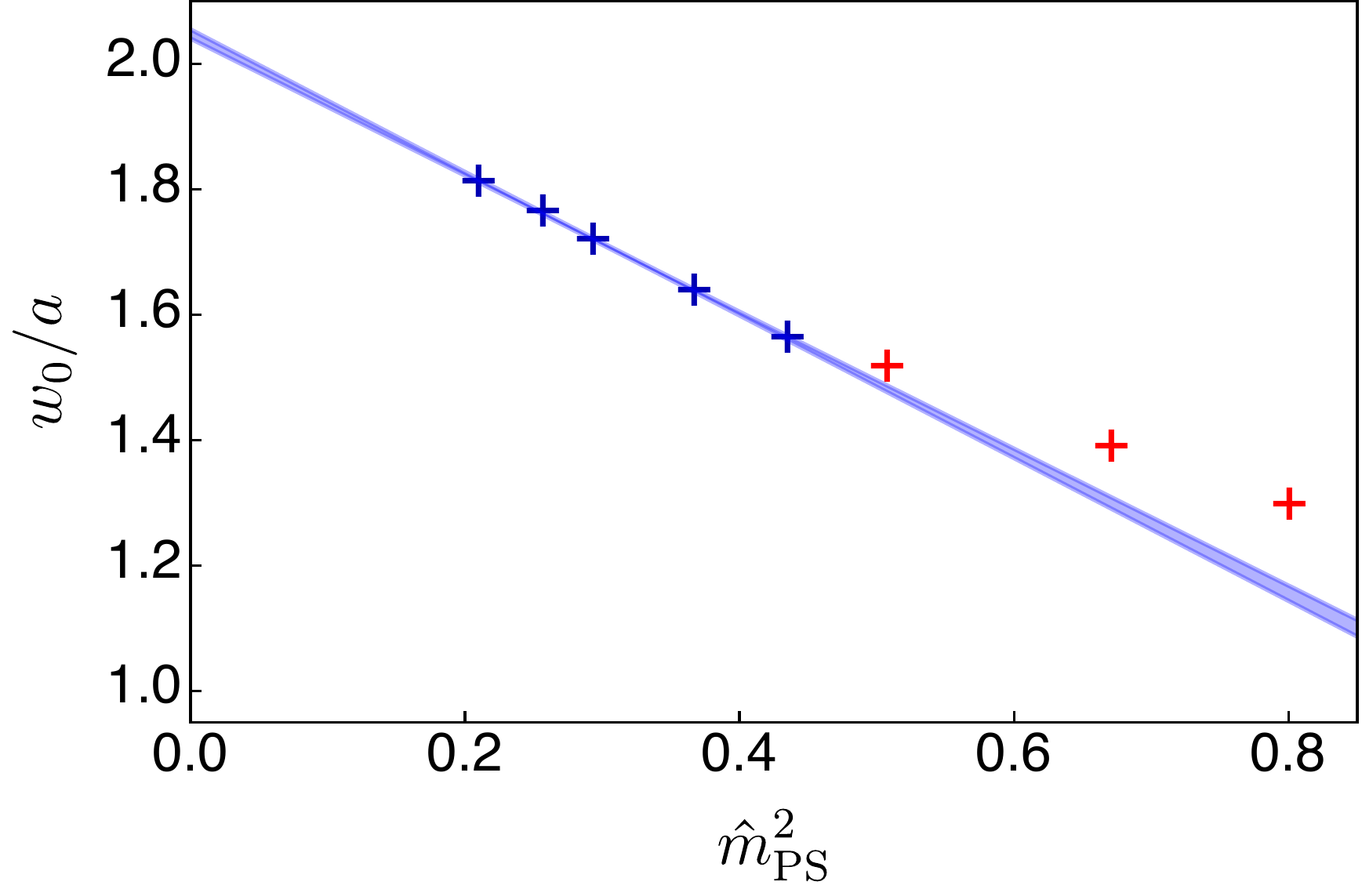}
\caption{%
\label{fig:GF_scale_fit}%
Gradient flow scale dependence on the pseudoscalar meson mass for $\beta=6.9$ (left panel) 
and $\beta=7.2$ (right panel). Numerical data from  \Tab{ensembles} and from \Tab{meson_spec_spin0}.
The blue  bands as based upon Eq.~(\ref{eq:fit_w0}), with the fit parameters in \Tab{GF_LECs}.
The band width represents the  ($1\sigma$) statistical uncertainties. The fits are restricted to the five
ensembles with lightest PS mass (shown in blue), while larger $\hat{m}_{\rm PS}^2$
(shown in red) are not included.
}
\end{center}
\end{figure}

\begin{table}
\begin{center}
\begin{tabular}{|c|c|c|c|}
\hline\hline
$\beta$ &
$w_0^\chi/a$  & $\tilde{k}_1$ & $\chi^2/N_{\rm d.o.f.}$ \\
\hline
$6.9$ & $1.347(4)$ & $-0.896(12)$ &  $0.7$ \\
\hline
$7.2$ & $2.047(8)$ & $-0.545(10)$ &  $0.5$ \\
\hline\hline
\end{tabular}
\end{center}
\caption{%
\label{tab:GF_LECs}%
Results of the NLO fits for $w_0/a$ from \Tab{ensembles} and $\hat{m}_{\rm PS}$ from the combination
with \Tab{meson_spec_spin0}. The fit uses the five ensembles with smallest mass
to extract the parameters $\tilde{k}_1$ and $w_0^{\chi}/a$ in Eq.~(\ref{eq:fit_w0}).
}
\end{table}

Figure~\ref{fig:GF_scale_fit} shows data from \Tab{ensembles} combined with 
\Tab{meson_spec_spin0}, together with the result of the two separate fits 
(for $\beta=6.9$ and $7.2$) to
Eq.~(\ref{eq:fit_w0}) of the five ensembles with smallest $\hat{m}_{\rm PS}$.
The  resulting values of the fit parameters
are reported in \Tab{GF_LECs}. 
The values of $\chi^2/N_{\rm d.o.f.}$ at the minima indicate that chiral perturbation theory for $w_0$ 
well describes the data. 
Deviations from  linear mass dependence of $w_0/a$ appear at around $\hat{m}_{\rm PS}^2\sim 0.4$. 
We anticipate that this scale is in broad agreement with the upper bound inferred by studying the
 pseudoscalar decay constant, and will be  discussed in Section~\ref{Sec:mesons}.

\begin{figure}
\begin{center}
\includegraphics[width=.65\textwidth]{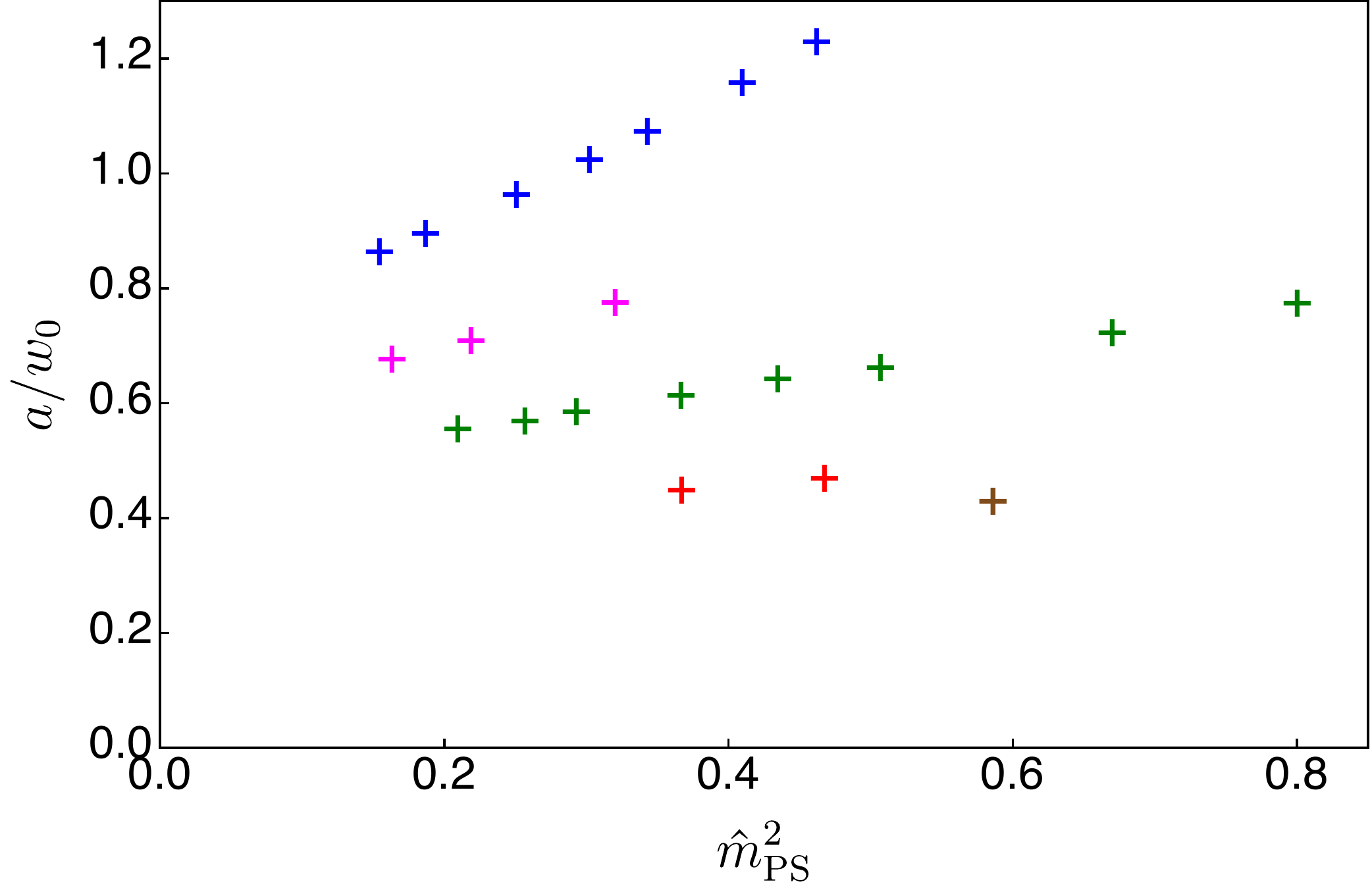}
\caption{%
\label{fig:latt_spacing}%
Inverse of the gradient flow scales $w_0$ with respect to the pseudoscalar meson mass squared,
for data taken from  \Tab{ensembles} and from \Tab{meson_spec_spin0}.
Blue, purple, green, red and brown colours (top to bottom, approximately linear series) 
have $\beta=6.9$, $7.05$, $7.2$, $7.4$ and $7.5$, respectively. 
}
\end{center}
\end{figure}

We observe that the value of $\tilde{k}_1$ is smaller for the finer lattice. 
We have too few ensembles for other lattice couplings to extract the values of $\tilde{k}_1$, 
yet the generic trend is consistent with  expectations, as visible in \Fig{latt_spacing}, 
with the mass-dependence becoming milder for larger choices of $\beta$.\footnote{
Numerical studies of $SU(2)$ lattice gauge theory with two fundamental Dirac fermions show that
the resulting values of low energy constant $\hat{k}_1$
 in the chiral expansion of $w_0$ obtained from fine lattices 
are not affected by large discretisation effects~\cite{Arthur:2016dir}.} 
Later in the paper, we will perform simultaneous continuum and massless extrapolations
via a global fit of all measurements of physical quantities---masses and decay 
constants of mesons---expressed in units of $w_0$.

\subsection{Topology}
\label{Sec:topology}

By analogy with the continuum definition ($ \frac{1}{32\pi^2} \int \mathrm{d}^4x \,
\varepsilon^{\mu\nu\rho\sigma}\,
\mathrm{Tr}\, \left\{ F_{\mu\nu}(x) 
{F}^{\rho\sigma}(x)\right\}$), the lattice topological charge of a gauge configuration is defined 
by summing over lattice sites $i$ as
\begin{equation}
    Q \equiv \frac{1}{32\pi^2} \sum_i \varepsilon^{\mu\nu\rho\sigma} \,\mathrm{Tr} \,\left\{ U_{\mu\nu}(i) U_{\rho\sigma}(i)\right\}\;.
\end{equation}

The HMC algorithm yields gauge configurations in which ultraviolet fluctuations 
have typical sizes that are  orders of magnitude larger than the desired signal.
The resulting large cancellations prevent a reliable extraction of $Q$. 
A smoothing procedure must be applied, that preserves the topological 
charge of a configuration while removing UV fluctuations, 
in order  to measure $Q$. The gradient flow described in the earlier 
subsections can be used for this purpose. 
For each ensemble, we measure $Q$ from uncorrelated, thermalised configurations,
 after flowing to a flow time $t / a^2 = {N_t^2}/32$ (equivalent to $\sqrt{8t} = T/2$).

With our choice of boundary conditions,
at finite lattice volume, $Q$ is quantised, and assumes only integer values.
Infinitesimal changes in the field configuration cannot alter the topological charge. 
The change at each Monte Carlo time-step 
required to yield a good acceptance rate in the Metropolis step of the HMC becomes smaller
for fine lattice spacings and small fermion masses.
For this reason, at large volumes and small masses, there is a risk that the topological charge 
freezes at a single value,
 for the finite HMC trajectories that one implements in practice.
 Since values of observables may depend on which topological sector is 
being observed, this effect may introduce a systematic error. 
To ensure that our measurements are not
 (heavily) affected by this type of systematic effect, we plot 
 and inspect histories of $Q$ for each 
ensemble used.

\begin{table}
\begin{center}
\begin{tabular}{|c|c|c|c|}
    \hline
    \hline
    Ensemble & $~~Q_0~~$ & $~~\sigma~~$ & $~~\tau_{\exp}~~$ \\
    \hline
    DB1M1 & $0.02(95)$ & $8.97(96)$ & $\ll 1$ \\
    DB1M2 & $0.20(85)$ & $8.32(85)$ & $0.34(23)$ \\
    DB1M3 & $-0.76(77)$ & $8.23(77)$ & $\ll 1$ \\
    DB1M4 & $-0.44(87)$ & $8.48(87)$ & $\ll 1$ \\
    DB1M5 & $-0.26(65)$ & $6.92(67)$ & $0.542(65)$ \\
    DB1M6 & $1.1(1.7)$ & $13.8(1.8)$ & $0.36(15)$ \\
    \hline
    DB2M1 & $1.28(72)$ & $8.06(73)$ & $0.82(17)$ \\
    DB2M2 & $-0.9(1.2)$ & $8.9(1.2)$ & $0.90(29)$ \\
    DB2M3 & $-0.7(1.4)$ & $12.1(1.4)$ & $0.38(20)$ \\
    \hline
    DB3M1 & $-0.27(71)$ & $6.98(74)$ & $1.39(19)$ \\
    DB3M2 & $-0.50(41)$ & $4.61(41)$ & $0.90(17)$ \\
    DB3M3 & $-0.34(48)$ & $4.35(48)$ & $1.18(25)$ \\
    DB3M4 & $0.12(89)$ & $7.41(93)$ & $0.80(17)$ \\
    DB3M5 & $-0.48(86)$ & $7.96(86)$ & $1.98(65)$ \\
    DB3M6 & $-0.45(63)$ & $6.24(63)$ & $1.21(19)$ \\
    DB3M7 & $-0.40(60)$ & $4.76(60)$ & $1.160(81)$ \\
    DB3M8 & $-0.06(99)$ & $8.2(1.0)$ & $2.33(32)$ \\
    \hline
    DB4M1 & $0.77(60)$ & $6.27(60)$ & $5.39(28)$ \\
    DB4M2 & $0.51(93)$ & $9.05(94)$ & $10.45(18)$ \\
    \hline
    DB5M1 & $0.11(24)$ & $2.53(24)$ & $7.33(36)$\\
    \hline
    \hline
\end{tabular}
\caption{%
\label{tab:topology}%
Fit results of the topological charge to a Gaussian function for all ensembles. 
The quantities $Q_{0}$ and $\sigma$ are defined in \Eq{nQ}. 
They are the mean and the standard deviation for the distribution of the topological charge, respectively.
In the last column, we present the results of the exponential autocorrelation time $\tau_{\exp}$ 
defined in \Eq{autocorr}. 
}
\end{center}
\end{table}

\begin{figure}
  \center
\captionsetup[subfigure]{aboveskip=-10pt}
  \begin{subfigure}{\textwidth}
    \center
    \includegraphics[width=0.9\textwidth]{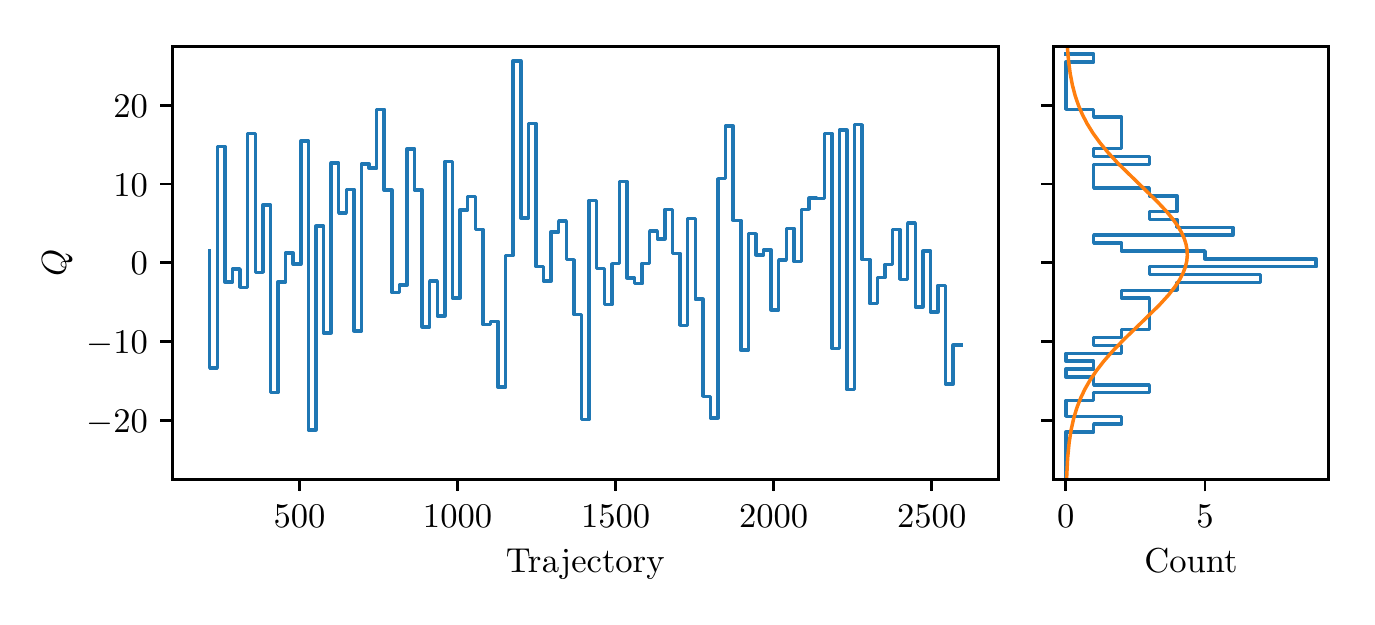}
    \caption{$\beta=6.9,m=-0.85,V=32\times16^3$}
  \end{subfigure}
  \begin{subfigure}{\textwidth}
    \center
    \includegraphics[width=0.9\textwidth]{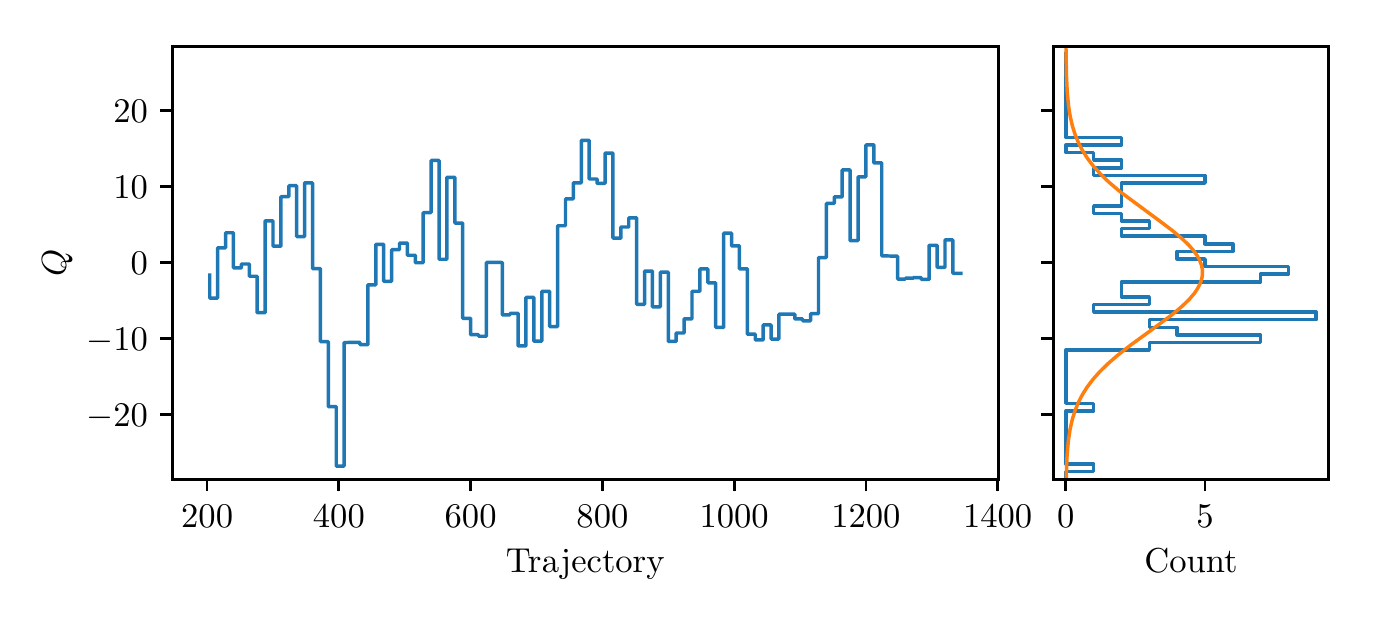}
    \caption{$\beta=7.2,m=-0.78,V=36\times24^3$}
  \end{subfigure}
  \begin{subfigure}{\textwidth}
    \center
    \includegraphics[width=0.9\textwidth]{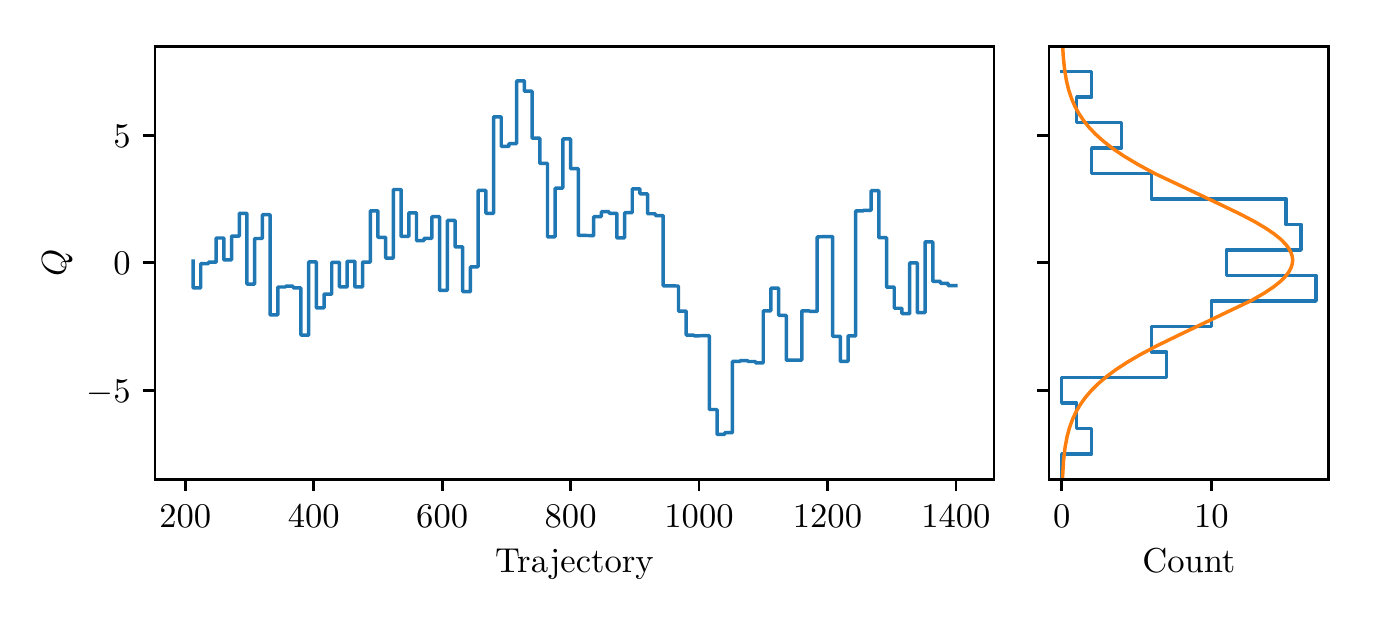}
    \caption{$\beta=7.5,m=-0.69,V=48\times24^3$}
  \end{subfigure}
  \caption{
Topological charge histories (left), and histograms (right), for the ensembles
 DB1M1, DB3M5, and DB5M1, respectively. Fitted parameters are
(a) $Q_0=0.02(95)$, $\sigma=8.97(96)$,
$\tau_{\exp}\ll 1$;
(b) $Q_0=-0.48(86)$, $\sigma=7.96(86)$,
$\tau_{\exp}=1.9(7)$;
(c) $Q_0=0.11(24)$, $\sigma=2.53(24)$,
$\tau_{\exp}=7.9(6)$.}
  \label{fig:topcharge}
\end{figure}

We expect the measurements of the charge $Q$ to obey a Gaussian 
distribution about zero in the limit of infinite simulation time. 
With a finite number of trajectoris, we use the fitting form
\begin{equation}
    n(Q) = A_n \exp \left( -\frac{(Q - Q_0)^2}{2\sigma^2} \right)
\label{eq:nQ}
\end{equation}
in order to fit the histogram data. The two fit parameters $Q_0$ and $\sigma$ are the mean and the standard deviation of the Gaussian distribution. 
The resulting values are presented in \Tab{topology}, 
while a sample of topological charge histories is shown in Fig.~\ref{fig:topcharge}.
 
We also calculate the autocorrelation function $C_Q(\tau)$ of the topological charge history as
\begin{equation}
    C_Q(\tau) = \sum_{\mathfrak{t}=1}^{N-\tau} \frac{(Q_\mathfrak{t} - \langle Q\rangle)(Q_{\mathfrak{t}+\tau}-\langle Q\rangle)}{\mathrm{var}(Q)}\;,
\end{equation}
where $\mathfrak{t}$ indexes the configurations analysed from 1 to $N$, and $\langle Q\rangle$ and $\mathrm{var}(Q)$ 
are the mean and variance of $Q$ for these configurations without any assumptions on the distribution. 
The exponential autocorrelation time $\tau_{\exp}$ is then calculated from this by fitting 
\begin{equation}
    C_Q(\tau) = \exp \left(-\frac{\tau}{\tau_{\exp}}\right)\;.
\label{eq:autocorr}
\end{equation}
We report the fit results for $\tau_{\exp}$ in the last column of \Tab{topology}. 
In the ideally decorrelated case, $C_Q(\tau)$ is consistent with zero for all $\tau>0$, and so the fit fails to converge; this is checked explicitly, and we report the autocorrelation time as $\tau_{\exp}\ll1$ in these cases.

All ensembles used to generate the results presented in this paper were found to be free of significant topological freezing, with $Q_0$ always within 1$\sigma$ from zero, and the autocorrelation time typically $\tau_{\exp} < 2$ and always $\tau_{\exp} \lesssim 10$.

\subsection{Finite size effects}
\label{Sec:FV}

The finiteness of the lattice volume represents an inherent source of systematic uncertainties 
in  lattice calculations. 
In a confining gauge theory, finite-volume (FV) contaminations of  the masses and decay constants
 are expected 
to be exponentially suppressed as a function of  the length of the spatial lattice $L$,
at least 
when the volume is much larger than the inverse of the Compton wavelength of the lightest state,
i.e. as long as $m_{\rm PS} L \gg 1$.
The size of FV effects can be estimated in a systematic way within chiral perturbation theory,
as long as the volume is larger than the hadronic scale. For $f_{\rm PS} L \gtrsim 1$, 
the dominant contribution is expected to arise from one-loop tadpole
 integrals of the pseudoscalar mesons~\cite{Gasser:1986vb, Golterman:2009kw}. 
The leading-order FV corrections to the meson masses can be written as
\beq
m_{\rm M}(L)=m_{\rm M}^{\textrm{inf}}\left(1+
A_M \frac{e^{-m_{\rm PS}^{\rm inf} L}}{(m_{\rm PS}^{\rm inf}L)^{3/2}}\right)\,, 
\label{eq:FV_correction}
\eeq
where $m_{\rm M (PS)}^{\rm inf}$ is the meson (pseudoscalar) mass in the infinite volume limit. 
In principle, the coefficient $A_M$ can be determined within chiral perturbation theory 
without introducing any new parameters in the infinite-volume chiral Lagrangian. 
Nevertheless, since our data is far from the massless limit, we treat $A_{M}$ as  free. 

\begin{table}
\begin{center}
\begin{tabular}{|c|c|c|c|c|c|}
\hline\hline
Ensemble & $~~~a m_0~~~$  & $N_t\times N_s^3$  
& $~~~~~a m_{\rm PS}~~~~~$ & $~~~~~a m_{\rm V}~~~~~$ & $~~~~m_{\rm PS}^{\rm inf} \, L~~~~$ \\
\hline
${\rm DB3M4}^*$ & & $36\times 16^3$ & 0.4267(16) & 0.521(4) & 6.75 \\
${\rm DB3M4}^{**}$ & -0.77 & $36\times 20^3$ & 0.4224(12) & 0.5153(28)& 8.43 \\
DB3M4 & & $36\times 24^3$ &  0.4222(8) & 0.5112(16) & 10.12 \\
\hline
${\rm DB3M6}^*$ &  & $36\times 16^3$ & 0.3309(16) & 0.445(4) & 5.02\\
${\rm DB3M6}^{**}$ & -0.79 & $36\times 20^3$ & 0.3183(10) & 0.4290(28) & 6.28 \\
DB3M6 & & $36\times 24^3$ & 0.3153(9) & 0.4264(19) & 7.53\\
\hline\hline
\end{tabular}
\caption{%
\label{tab:FV_ensembles}%
Ensembles and numerical results used to estimate the size of finite volume effects. 
The number of configurations and the separation of trajectories between adjacent configurations 
are given by $N_{\rm configs}=200$ and $\delta_{\rm traj}=12$, respectively, for all these ensembles. 
The pseudoscalar masses in the infinite volume limit $m_{\rm PS}^{\rm inf}$ are estimated 
from the exponential fits as discussed in the text.
}
\end{center}
\end{table}

\begin{figure}
\begin{center}
\includegraphics[width=.49\textwidth]{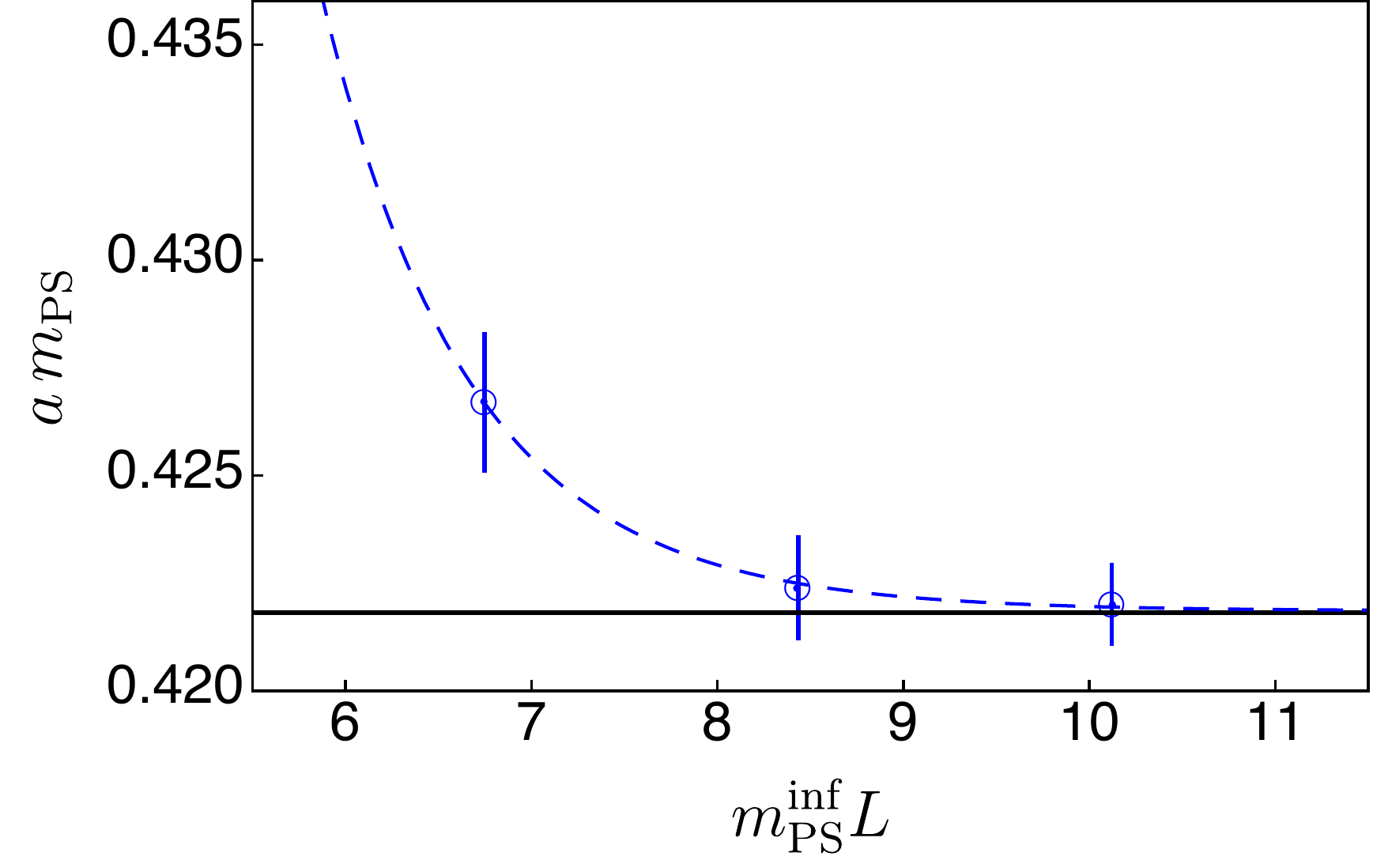}
\includegraphics[width=.49\textwidth]{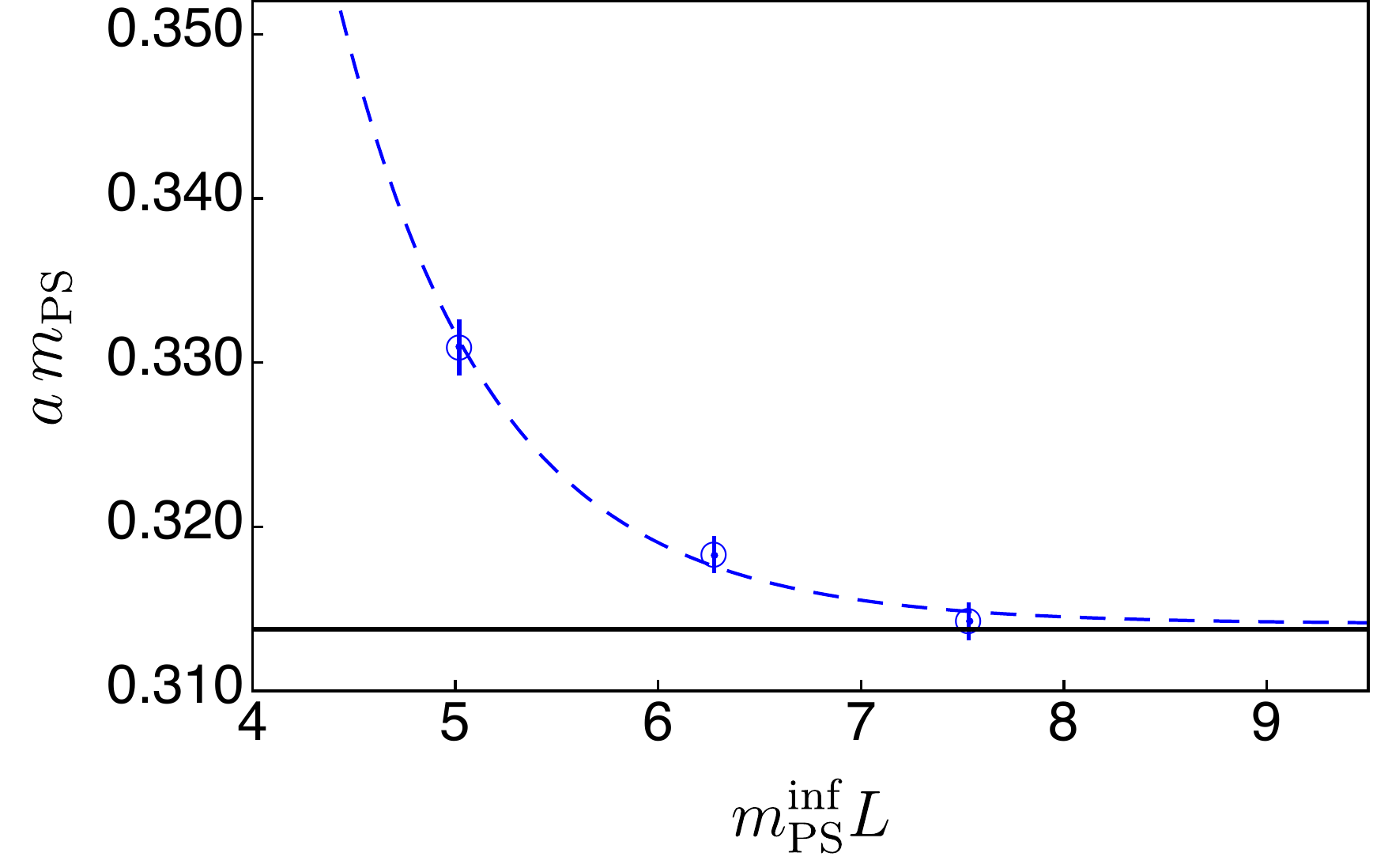}
\caption{%
\label{fig:mps_FV}%
Volume dependence of the pseudoscalar masses, as in Table~\ref{tab:FV_ensembles}.
Lattice parameters are $(\beta,\,m_0)=(7.2,\,-0.77)$ in the left panel 
and $(7.2,\,-0.79)$ in the right panel. 
The (blue) dashed and (black) solid lines denote the fit results and 
the extrapolated values in the infinite volume limit, respectively. 
}
\end{center}
\end{figure}

To quantify the size of FV effects, we calculate masses of pseudoscalar PS and vector V 
mesons $a m_{\rm PS (V)}$ 
on three spatial lattice volumes of $N_s^3=16^3,\,20^3,\,24^3$ 
at two sets of lattice parameters, $(\beta,\, a \,m_0)=(7.2,\,-0.77)$ and $(\beta,\, a \,m_0)=(7.2,\, -0.79)$. 
Numerical results are reported in \Tab{FV_ensembles}. 
At the smallest volume $m_{\rm PS}^{\rm inf}\, L\sim 5$, the masses deviate 
from those at  the largest volume by $4\sim 5\%$, 
more  than expected  from statistically uncertainties. 
At $m_{\rm PS}^{\rm inf}\, L\sim 6.3$ the deviations decrease to the level of $1\sim2\%$,
compatible with the statistical uncertainties for $a m_{\rm V}$, but not for $a m_{\rm PS}$. 
We performed a fit of the pseudoscalar masses to \Eq{FV_correction}. 
As shown in \Fig{mps_FV}, the data are well described by the exponential fit function: 
the blue dashed lines correspond to the best fits, 
while the black solid lines indicate the masses in the infinite volume limit. 
From this analysis we conclude that 
 the size of FV effects is no larger than $\sim 0.3\%$---smaller than the
 typical size of statistical uncertainties in the spectroscopic measurements 
reported in the following sections---if we require that $m_{\rm PS}\, L \gtrsim 7.5$. 
In the rest of this work we restrict attention to ensembles that satisfy this requirement---see 
\Tab{meson_spec_spin0}.  

The gradient flow scale also receives a correction from the finite size of the lattice volume. 
The flow along the fictitious time $t$ can be understood as a smearing procedure with scale $\sqrt{8t}$,
hence FV effects are controlled by the dimensionless ratio $c_\tau=\sqrt{8t}/L$. 
At the reference value of $\mathcal{E}_0=0.35$ we find that 
$c_\tau$ does not exceed $0.2$ in most ensembles.
Using the results in Ref.~\cite{Fodor:2012td} 
we estimate the size of FV corrections to \Eq{GF_schemes} to be at most 
at a few per-mille level. 
The only exception are the ensembles with
 $\beta=7.2$ on a $36\times 16^3$ lattice and the one with $\beta=7.5$. 
These ensembles do not play an important role in the analyses that follow,  because of the large
physical masses associated with them.

\section{Meson spectroscopy and decay constants}
\label{Sec:mesons}

In this section we summarise the main results of our lattice study, by focusing
on the properties (masses and decay constants) of the mesons in the dynamical theory.
We start in Section~\ref{Sec:corr} by defining the operators we are interested in, 
the correlation functions we measure,
and the renormalisation procedure we apply. We then present all the main spectroscopy results
 in Section~\ref{Sec:spectroscopy}, and discuss the continuum limit extrapolation in Section~\ref{Sec:continuum}.
Useful supplementary details are relegated to Appendix~\ref{Sec:diquark} and~\ref{Sec:meff}.

\subsection{Two-point correlation functions}
\label{Sec:corr}

Following established procedure, 
we extract the masses and the decay constants of flavoured mesons by studying 
the behaviour of the relevant  two-point correlation functions at large Euclidean time $t$. 
The interpolating operators which carry the same quantum numbers with 
the desired meson states take the generic form
\beq
\mathcal{O}_M(x)\equiv\overline{Q^i}(x) \Gamma_M Q^j(x),
\label{eq:meson_ops}
\eeq
where $i,j=1,2$ are the flavour indices, 
and $\Gamma_M$ refer to the Dirac structures  summarised in \Tab{mesons}. 
Summations over spinor and colour indices are understood. 

The lightest spin-0 and spin-1 mesons, denoted by 
PS, V and AV for pseudoscalar, vector and axial-vector mesons, respectively, 
appear in the low-energy EFT described in~\cite{Bennett:2017kga}. 
In Section~\ref{Sec:result_con} we will use their masses and decay constants to test the EFT
and to extrapolate towards the chiral limit.
We also consider additional interpolating operators with the Dirac 
structures  $\mathbb{1},\, \gamma^0\gamma^\mu,\, \gamma^5\gamma^0\gamma^\mu$, 
referred to as scalar S, (antisymmetric) tensor T,
and axial tensor AT,
though from the related correlation functions we only
extract the meson masses. 
We restrict our attention to the flavoured mesons (by choosing $i\neq j$ in flavour 
space)--- the analogous mesons in QCD are $\pi$, $\rho$, $a_1$, $a_0$, and $b_1$. 
As the global symmetry is broken, the states created by the interpolating operators denoted by
 V and T mix, 
with the low-lying state corresponding to the $\rho$ meson 
in QCD (see \cite{Glozman:2007ek} and references therein,
as well as Fig.~1 of both Refs.~\cite{Glozman:2015qva} and ~\cite{Glozman:2018jkb}).

\begin{table}
\begin{center}
\begin{tabular}{|c|c|c|c|c|}
\hline\hline
{\rm ~~~Label~}($M$) & {\rm ~Interpolating operator ($\mathcal{O}_M$)~} & {\rm ~~~Meson~~~} & {\rm ~~~$J^{P}$~~~}
& $Sp(4)$ \cr
\hline
PS & $\overline{Q^i}\gamma_5 Q^j$ & $\pi$ & $0^{-}$ & $5$ \cr
S & $\overline{Q^i} Q^j$ & $a_0$ & $0^{+}$ & $5$ \cr
V & $\overline{Q^i}\gamma_\mu Q^j$ & $\rho$ & $1^{-}$ & $10$ \cr
T & $\overline{Q^i}\gamma_0\gamma_\mu Q^j$ & $\rho$ & $1^{-}$ & $10 (+5)$  \cr
AV & $\overline{Q^i}\gamma_5\gamma_\mu Q^j$ & $a_1$ & $1^{+}$ & $5$ \cr
AT & $\overline{Q^i}\gamma_5\gamma_0\gamma_\mu Q^j$ & $b_1$ & $1^{+}$ & $10 (+5)$ \cr
\hline\hline
\end{tabular}
\end{center}
\caption{
Interpolating operators $\mathcal{O}_M$ sourcing the lightest mesons
 in the six channels considered in the main text. 
To avoid mixing with the flavour singlets, we restrict to  $i\neq j$ the flavour indices of the Dirac fermions, 
while colour and spinor indices are summed and omitted. 
For completeness, we also show the $J^P$ quantum numbers and the 
corresponding particle in the QCD classification of mesons. 
Notice that two of the operators source the same particles ($\rho$ meson)
because of the breaking of chiral symmetry. In the last column 
we report the irreducible representation  of  the unbroken global 
$Sp(4)$ spanned by the meson (see also~\cite{Lewis:2011zb}).
In brackets are irreducible representations of $Sp(4)$ that are sourced by operators with the
same Lorentz structure, but that we do not discuss in this context.
}
\label{tab:mesons}
\end{table}

For all  the meson interpolating operators $\mathcal{O}_M$ listed  in \Tab{mesons}, we define the 
zero-momentum Euclidean two-point correlation functions at positive Euclidean time $t$ as
\beqs
C_{\mathcal{O}_M}(t)\equiv\sum_{\vec{x}} 
\langle 0 | \mathcal{O}_M(\vec{x},t) \mathcal{O}_M^\dagger(\vec{0},0) | 0\rangle.
\label{eq:meson_corr}
\eeqs
In the numerical study, the resulting mesonic two-point correlation functions  are studied by replacing the 
point-like sources in Eq.~(\ref{eq:meson_ops}) with
 $Z_2\times Z_2$ single time slice stochastic sources~\cite{Boyle:2008rh},
 with number of hits $3$.

Because of the pseudoreal nature of the representation of the symplectic gauge group,
diquark operators are indistinguishable from the mesonic operators. We report in \Tab{mesons}
also the multiplicity of each state (the size of the irreducible representation of $Sp(4)$).
For instance, five pNGBs form a multiplet of the unbroken $Sp(4)$, 
in the $SU(4)\rightarrow Sp(4)$ enhanced symmetry pattern
of the  gauge theory considered here.
Compared with what happens with gauge group $SU(N)$, these five pNGBs include the three 
associated with the 
breaking $SU(2)_L\times SU(2)_R\rightarrow SU(2)_V$, together with
two diquarks\footnote{
The full expressions of spin-$0$ and spin-$1$ meson operators in the bases of both four-component Dirac 
and two-component Weyl spinors will be presented in a separate publication~\cite{Sp4ASFQ}. See also
the analysis in Ref.~\cite{DeGrand:2015lna}}.
In \App{diquark} we explicitly show the equivalence of meson 
and diquark correlators by using the lattice action in \Eq{fermion_action}.

At large Euclidean time $t$ 
the correlation functions in \Eq{meson_corr} are dominated by the lowest excitation
 at zero spatial momentum 
so that the mass $m_M$ appears in the asymptotic expression:
\beqs
C_{\mathcal{O}_M}(t)\xrightarrow{t\rightarrow \infty}\langle 0 |\mathcal{O}_M | M 
\rangle \langle 0 |\mathcal{O}_M | M \rangle^* 
\frac{1}{2m_M} \left[e^{-m_M t}+e^{-m_M(T-t)}\right]\,,
\label{eq:corr_ground}
\eeqs
where $T$ is the temporal  extent of the lattice. 
The decay constants $f_M$ are determined from the matrix elements, which are parameterised as
\beqs
\langle 0 | \overline{Q_1} \gamma_5 \gamma_\mu Q_2 | {\rm PS} \rangle &=& f_{\rm PS} p_\mu\,,\nn \\
\langle 0 | \overline{Q_1} \gamma_\mu Q_2 | {\rm V} \rangle &=& f_{\rm V} m_{\rm V} \epsilon_\mu\,,\nn \\
\langle 0 | \overline{Q_1} \gamma_5 \gamma_\mu Q_2 | {\rm AV} \rangle 
&=& f_{\rm AV} m_{\rm AV} \epsilon_\mu\,.
\label{eq:matrix_element}
\eeqs
The polarisation vector $\epsilon_\mu$ is transverse to the momentum $p_\mu$ 
and normalised by $\epsilon_\mu^* \epsilon^\mu=1$. 
The meson states $|M\rangle$ are  conventionally defined by the self-adjoint isospin fields, 
as in $M=M^A T^A$, where $T^A$ are the generators of the group. 
We adopt conventions such that in QCD the analogous experimental value 
of the pion (pseudoscalar) decay constant is $f_\pi \simeq 93\,{\rm MeV}$. 
In \Eq{matrix_element}, the pseudoscalar decay constant $f_{\rm PS}$ is defined via 
the local axial current.
To calculate the decay constant $f_{PS}$, 
we introduce an additional two-point correlation function 
\beqs
C_{\Pi}(t)&=&\sum_{\vec{x}} 
\langle 0 | [\overline{Q_1} \gamma_5 \gamma_\mu Q_2(\vec{x},t)]\,[\overline{Q_1} \gamma_5 Q_2(\vec{0},0)] | 0\rangle \nn \\
&\xrightarrow{t\rightarrow\infty}&
\frac{f_{\rm PS} \langle 0 | \mathcal{O}_{\rm PS} | {\rm PS} \rangle^*}{2}\left[
e^{-m_{\rm PS} t}-e^{-m_{\rm PS} (T-t)}\right],
\label{eq:axial_corr}
\eeqs
where $\langle 0 | \mathcal{O}_{\rm PS} | {\rm PS} \rangle^*$ 
can be obtained from $C_{\mathcal{O}_{\rm PS}}(t)$ in \Eq{corr_ground}. 
In practice, we calculate $m_{\rm PS}$ and $f_{\rm PS}$ by performing 
a simultaneous fit to the numerical data for $C_{\mathcal{O}_{\rm PS}}(t)$ 
and $C_{\Pi}(t)$. 
The details of the fit of the meson correlators, including the effective masses and best-fit ranges, 
are provided in Appendix~\ref{Sec:meff}. 

The matrix elements in \Eq{matrix_element}, calculated from the lattice at finite lattice spacing $a$,
 must be converted to those renormalised in the continuum. 
For Wilson fermions the decay constants in the continuum are determined from lattice ones via
\beqs
f_{\rm PS} = Z_{\rm A} f^{\rm bare}_{\rm PS},
~~~f_{\rm V} = Z_{\rm V} f_{\rm V}^{\rm bare},
~~~{\rm and}~~~ f_{\rm AV} = Z_{\rm A} f_{\rm AV}^{\rm bare},
\label{eq:f_con}
\eeqs
where $Z_{\rm V}$ and $Z_{\rm A}$ are the renormalisation factors for vector and axial-vector currents 
which are expected to approach unity in the continuum. 
Since the pseudoscalar decay constant $f_{\rm PS}$ is defined using the axial current as in Eq.~(\ref{eq:matrix_element}), 
it receives renormalisation with the factor of $Z_{\rm A}$. 
The renormalisation factors are determined by the  one-loop renormalisation 
procedure in lattice perturbation theory for Wilson fermions, 
and the expressions for the matching factors are 
the following~\cite{Martinelli:1982mw}:
\beqs
Z_A&=&1+C(F)\left(\Delta_{\Sigma_1}+\Delta_{\gamma_5\gamma_\mu}\right)\frac{\tilde{g}^2}{16\pi^2}, \nn \\
Z_V&=&1+C(F)\left(\Delta_{\Sigma_1}+\Delta_{\gamma_\mu}\right)\frac{\tilde{g}^2}{16\pi^2}.
\label{eq:zfactor}
\eeqs
The eigenvalue of the quadratic Casimir operator with fundamental fermions is $C(F)=5/4$ for the $Sp(4)$ gauge theory. 
The one-loop factor $\Delta_{\Sigma_1}$ arises
 from the wave-function renormalisation of the external fermion lines, 
while the other $\Delta$'s arise from the one-loop computations of the vertex functions. 
The numerical values obtained by one-loop integrals within 
the continuum $\overline{\rm MS}$ (modified minimal subtraction) renormalisation scheme are as follows: 
$\Delta_{\Sigma_1}=-12.82$, $\Delta_{\gamma_\mu}=-7.75$ and $\Delta_{\gamma_5\gamma_\mu}=-3.0$. 
The coupling used in \Eq{zfactor} is defined via the mean field approach to the link variable,
which effectively removes the tadpole diagrams, as
$\tilde{g}^2=g^2/\langle P \rangle$ \cite{Lepage:1992xa}, 
where $\langle P \rangle$ is the average plaquette value and $g$ the bare gauge coupling. \footnote{
This tadpole-improved coupling is a convenient choice for evaluating one-loop matching coefficients in \Eq{zfactor} 
in our exploratory work. 
Drawing experience from QCD calculations, its value is normally very 
close to that of reasonable choices of renormalised couplings which are 
determined by more complicated procedures.
}

\subsection{Masses and decay constants}
\label{Sec:spectroscopy}

\begin{table}
\begin{center}
\begin{tabular}{|c|c|c|c|c|c|}
\hline\hline
Ensemble & $am_{\rm PS}$ & $af_{\rm PS}$ & $am_{\rm S}$ & $m_{\rm PS}\,L$ & $f_{\rm PS}\,L$ \\
\hline
DB1M1 & 0.8344(11) & 0.1431(7) & 1.52(4) & 13.351(17) & 2.290(10)\\
DB1M2 & 0.7403(12) & 0.1299(11) & 1.44(4) & 11.845(19) & 2.079(17)\\
DB1M3 & 0.6276(14) & 0.1147(8) & 1.15(5) & 10.042(23) & 1.836(13) \\
DB1M4 & 0.5625(21) & 0.1052(11) & 1.290(20) & 9.00(3) & 1.683(18) \\
DB1M5 & 0.4813(10) & 0.0943(6) & 1.04(5) & 7.701(16) & 1.509(10) \\
DB1M6 & 0.3867(11) & 0.0823(6) & 1.032(25) & 9.28(26) & 1.977(13) \\
DB1M7 & 0.3388(12) & 0.0765(6) & 0.92(5) & 8.13(3) & 1.835(14)\\
\hline
DB2M1 & 0.4376(14) & 0.0822(9) & 0.88(3) & 8.752(28) & 1.645(17) \\
DB2M2 & 0.3311(11) & 0.0670(5) & 0.830(16) & 7.946(26) & 1.609(13) \\
DB2M3 & 0.2729(9) & 0.0612(4) & 0.777(13) & 8.732(27) & 1.958(12) \\
\hline
DB3M1 & 0.6902(11) & 0.0994(9) & 1.046(25) & 11.043(18) & 1.590(14) \\
DB3M2 & 0.5898(13) & 0.0905(8) & 0.994(16) & 9.437(21) & 1.449(13) \\
DB3M3 & 0.4700(13) & 0.0772(6) & 0.838(13) & 7.521(21) & 1.235(10) \\
DB3M4 & 0.4222(8) & 0.0726(3) & 0.792(11) & 10.133(18) & 1.743(8)\\
DB3M5 & 0.3702(9) & 0.0666(4) & 0.744(13) & 8.884(21) & 1.598(9) \\
DB3M6 & 0.3153(9) & 0.0604(4) & 0.646(18) & 7.568(22) & 1.448(9) \\
DB3M7 & 0.2874(7) & 0.05755(28) & 0.665(12) & 8.048(19) & 1.611(8) \\
DB3M8 & 0.2532(7) & 0.0536(3) & 0.598(17) & 8.102(24) & 1.714(10)\\
\hline
DB4M1 & 0.3190(5) & 0.05452(23) & 0.576(9) & 10.208(15) & 1.745(7) \\
DB4M2 & 0.2707(6) & 0.04999(27) & 0.548(8) & 8.663(20) & 1.600(9) \\
\hline
DB5M1 & 0.3264(9) & 0.0529(4) & 0.562(7) & 7.835(23) & 1.270(9)\\
\hline\hline
\end{tabular}
\caption{%
\label{tab:meson_spec_spin0}%
Masses and (renormalised) decay 
constants for flavoured mesons sourced by pseudoscalar (PS) and scalar (S) operators
 in units of the lattice spacing $a$.
The pseudoscalar decay constant  $f_{PS}$ is
 renormalised via the one-loop perturbative matching in \Eq{f_con}. 
 Statistical uncertainties are indicated in parenthesis.
}

\end{center}
\end{table}

\begin{table}
\begin{center}
\begin{tabular}{|c|c|c|c|c|c|c|}
\hline\hline
Ensemble & $am_{\rm V}$ & $af_{\rm V}$ & $am_{\rm AV}$ & $af_{\rm AV}$ & $am_{\rm T}$ & $am_{\rm AT}$  \\
\hline
DB1M1 & 0.9275(17) & 0.2326(15) & 1.561(29) & 0.228(14) & 0.9277(21) & 1.512(24)\\
DB1M2 & 0.8475(19) & 0.2224(14) & 1.445(26) & 0.215(11) & 0.8475(27) & 1.434(29)\\
DB1M3 & 0.7494(25) & 0.2056(18) & 1.315(24) & 0.211(10) & 0.754(3) & 1.276(25) \\
DB1M4 & 0.692(4) & 0.1917(25) & 1.27(3) & 0.217(14) & 0.688(5) & 1.205(29)\\
DB1M5 & 0.622(3) & 0.1777(21) & 1.16(4) & 0.191(21) & 0.619(4) & 1.169(21)\\
DB1M6 & 0.546(4) & 0.1629(20) & 1.059(14) & 0.193(5) & 0.547(5) & 0.97(3) \\
DB1M7 & 0.517(4) & 0.1564(23) & 0.97(4) & 0.163(14) & 0.527(6) & 0.96(4) \\
\hline
DB2M1 & 0.5517(24) & 0.1445(14) & 0.918(28) & 0.139(10) & 0.554(4) & 0.933(26)\\
DB2M2 & 0.470(3) & 0.1284(20) & 0.825(23) & 0.135(8) & 0.466(5) & 0.834(24) \\
DB2M3 & 0.4237(29) & 0.1202(14) & 0.75(3) & 0.120(11) & 0.424(5) & 0.736(27)\\
\hline
DB3M1 & 0.7490(17) & 0.1478(16) & 1.142(16) & 0.146(6) & 0.7484(23) & 1.154(9)\\
DB3M2 & 0.6591(25) & 0.1408(14) & 1.036(15) & 0.146(6) & 0.659(3) & 1.038(13)\\
DB3M3 & 0.5529(26) & 0.1269(15) & 0.879(18) & 0.127(7) & 0.555(3) & 0.905(14)\\
DB3M4 & 0.5112(16) & 0.1208(10) & 0.847(13) & 0.129(5) & 0.5138(16) & 0.788(16)\\
DB3M5 & 0.4664(25) & 0.1121(15) & 0.789(14) & 0.125(5) & 0.4750(28) & 0.785(21)\\
DB3M6 & 0.4264(19) & 0.1083(9) & 0.720(20) & 0.114(8) & 0.426(3) & 0.696(24)\\
DB3M7 & 0.4019(23) & 0.1040(11) & 0.698(10) & 0.116(3) & 0.4013(27) & 0.715(13)\\
DB3M8 & 0.3772(24) & 0.0990(13) & 0.650(15) & 0.107(5) & 0.386(3) & 0.643(17)\\
\hline
DB4M1 & 0.3974(11) & 0.0905(6) & 0.623(11) & 0.092(5) & 0.3976(15) & 0.635(9)\\
DB4M2 & 0.3548(17) & 0.0844(9) & 0.560(8) & 0.0853(27) & 0.3528(21) & 0.555(11)\\
\hline
DB5M1 & 0.3941(20) & 0.0832(12) & 0.594(8) & 0.0851(26) & 0.391(3) & 0.596(11) \\
\hline\hline
\end{tabular}
\caption{%
\label{tab:meson_spec_spin1}%
Masses and (renormalised) decay constants for flavoured mesons sourced by 
vector (V), axial-vector (AV), and tensor (T) and axial-tensor (AT)
operators in lattice units. 
The V and AV  decay constants are renormalised via the one-loop perturbative matching in \Eq{f_con}. 
 Statistical uncertainties are indicated in parenthesis.
}
\end{center}
\end{table}

\begin{figure}
\begin{center}
\begin{picture}(340,360)
\put(0,185){\includegraphics[width=.715\textwidth]{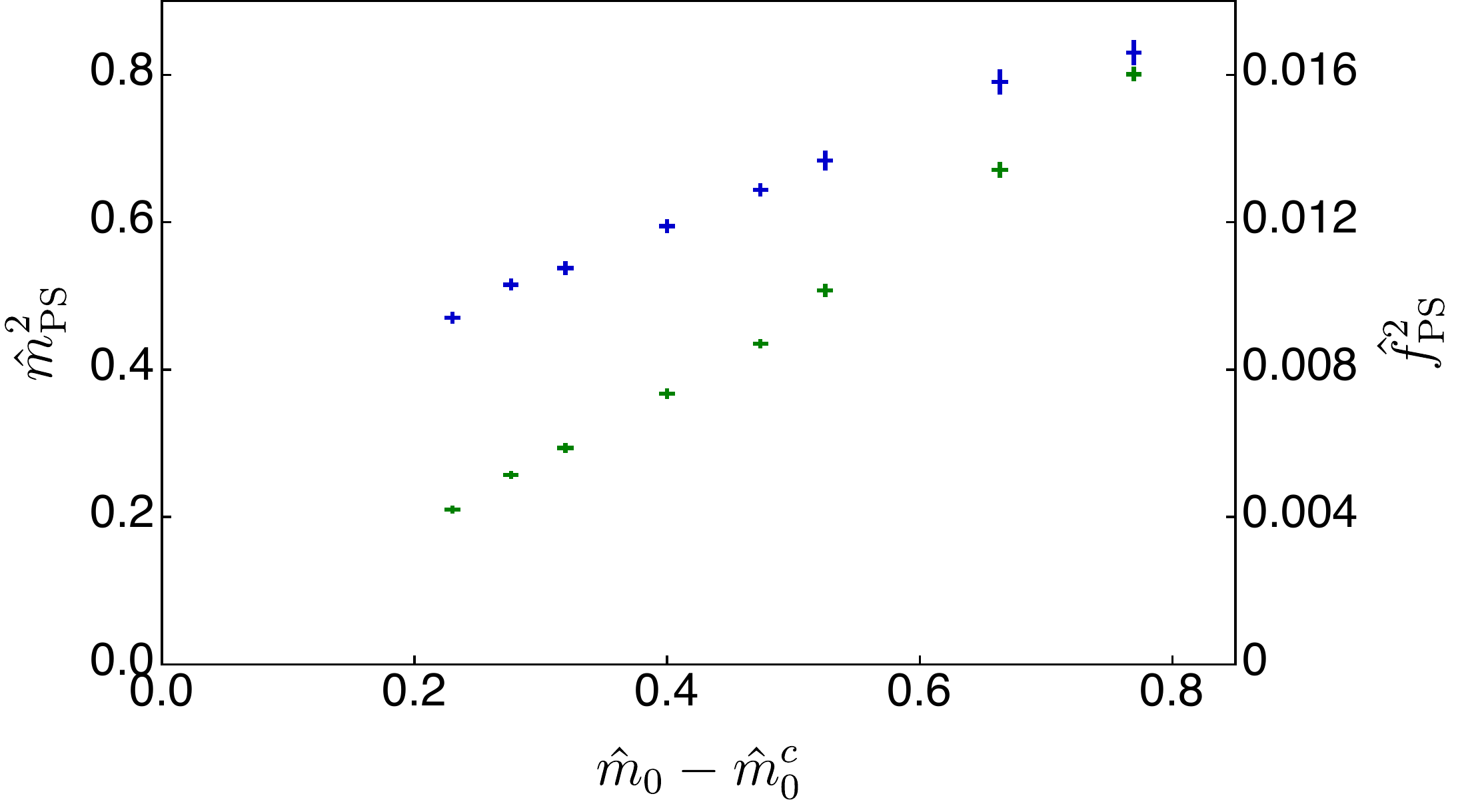}}
\put(-17,0){\includegraphics[width=.65\textwidth]{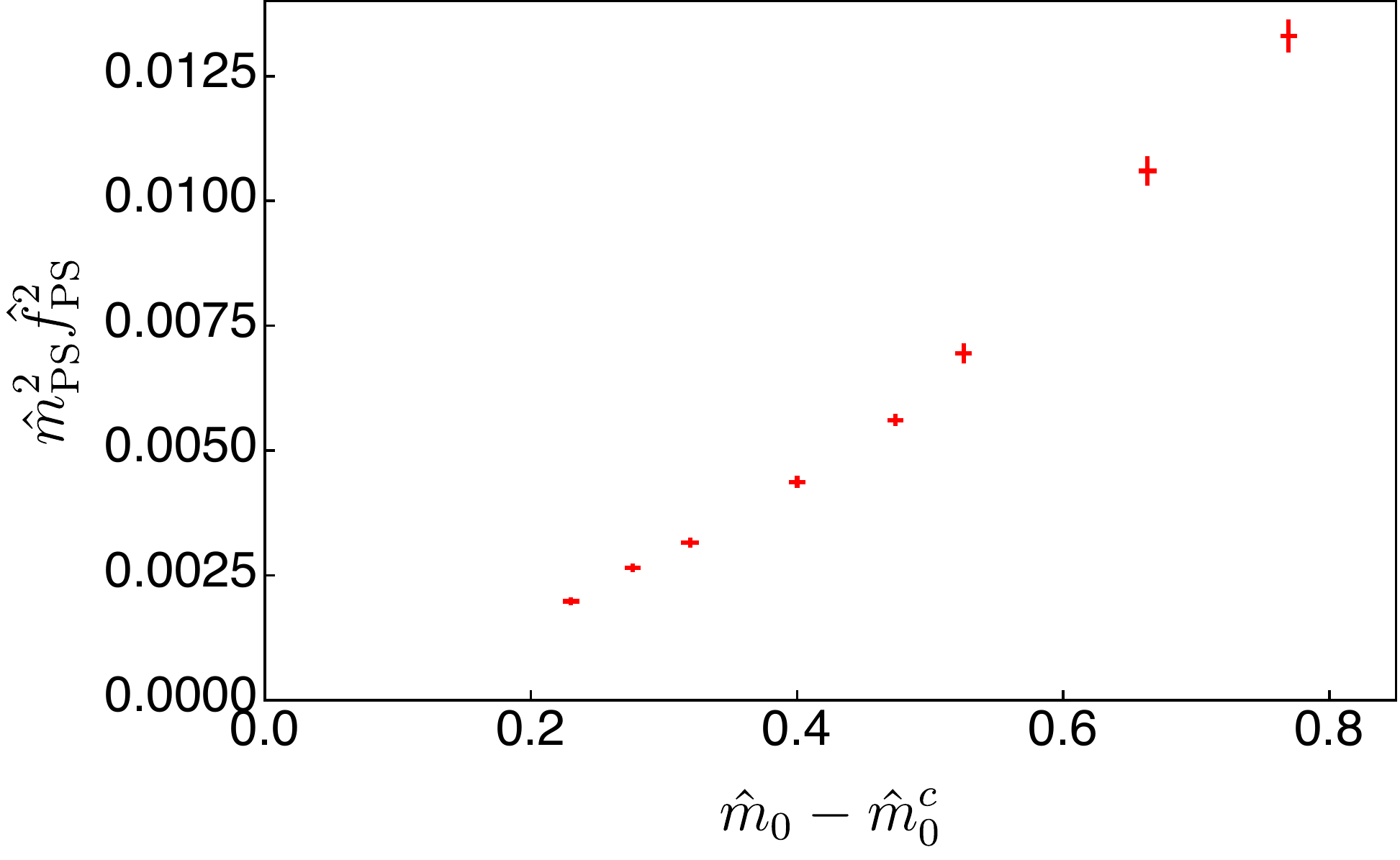}}
\end{picture}
\caption{%
\label{fig:LO_relation}%
Top panel: pseudoscalar masses squared $\hat{m}_{\rm PS}$ 
and decay constants squared  $\hat{f}^2_{\rm PS}$ 
as a function of the subtracted bare fermion mass. 
Green and blue symbols stand for  $\hat{m}_{\rm PS}^2$ 
and $\hat{f}_{\rm PS}^2$, respectively. 
Bottom panel: left-hand side of the GMOR relation,
 as a function of the subtracted fermion mass.
All points computed with $\beta=7.2$.
The uncertainty on the horizontal axis descends form the determination of $\hat{m}_0^c$.
}
\end{center}
\end{figure}

\begin{figure}
\begin{center}
\includegraphics[width=.65\textwidth]{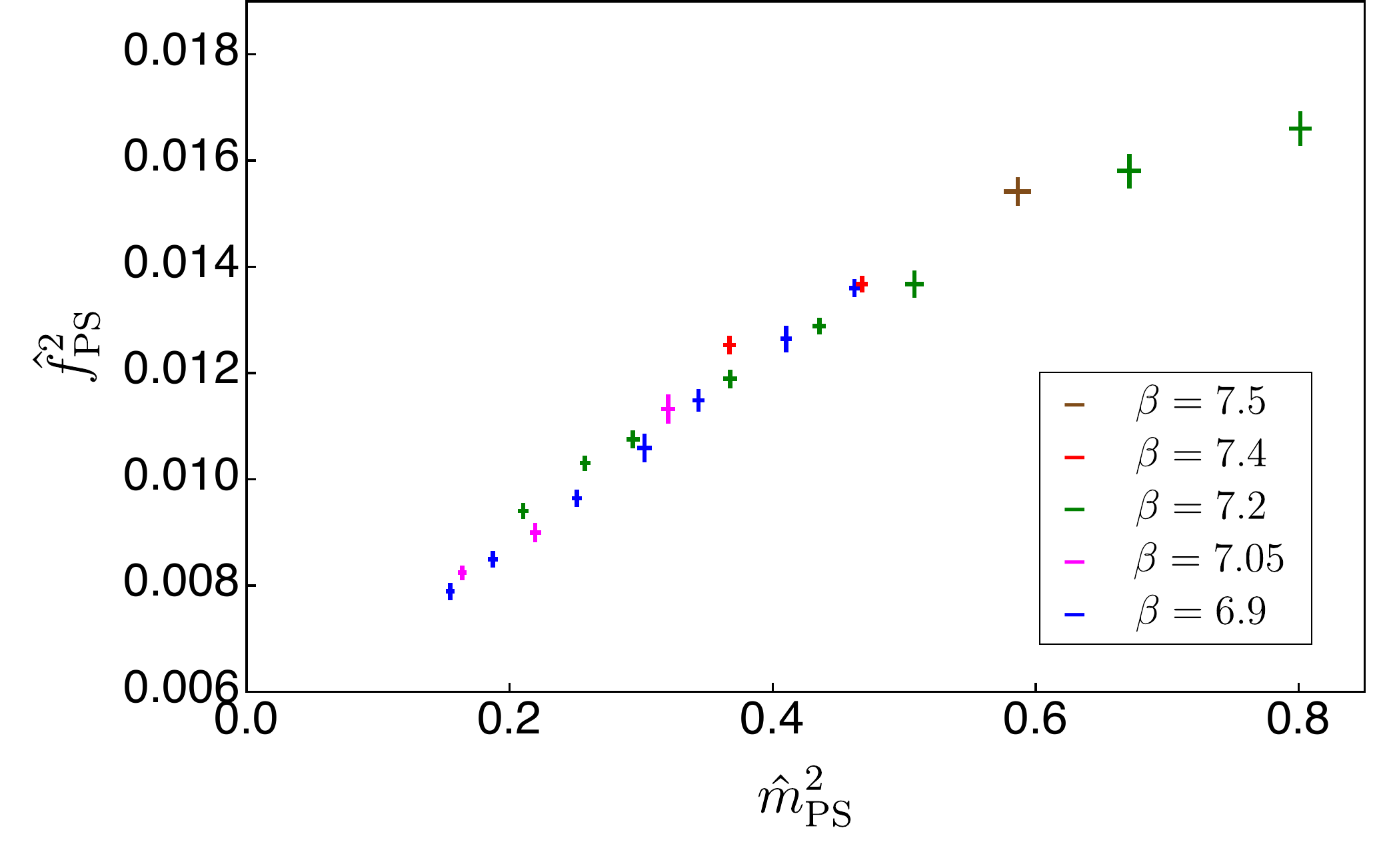}
\includegraphics[width=.65\textwidth]{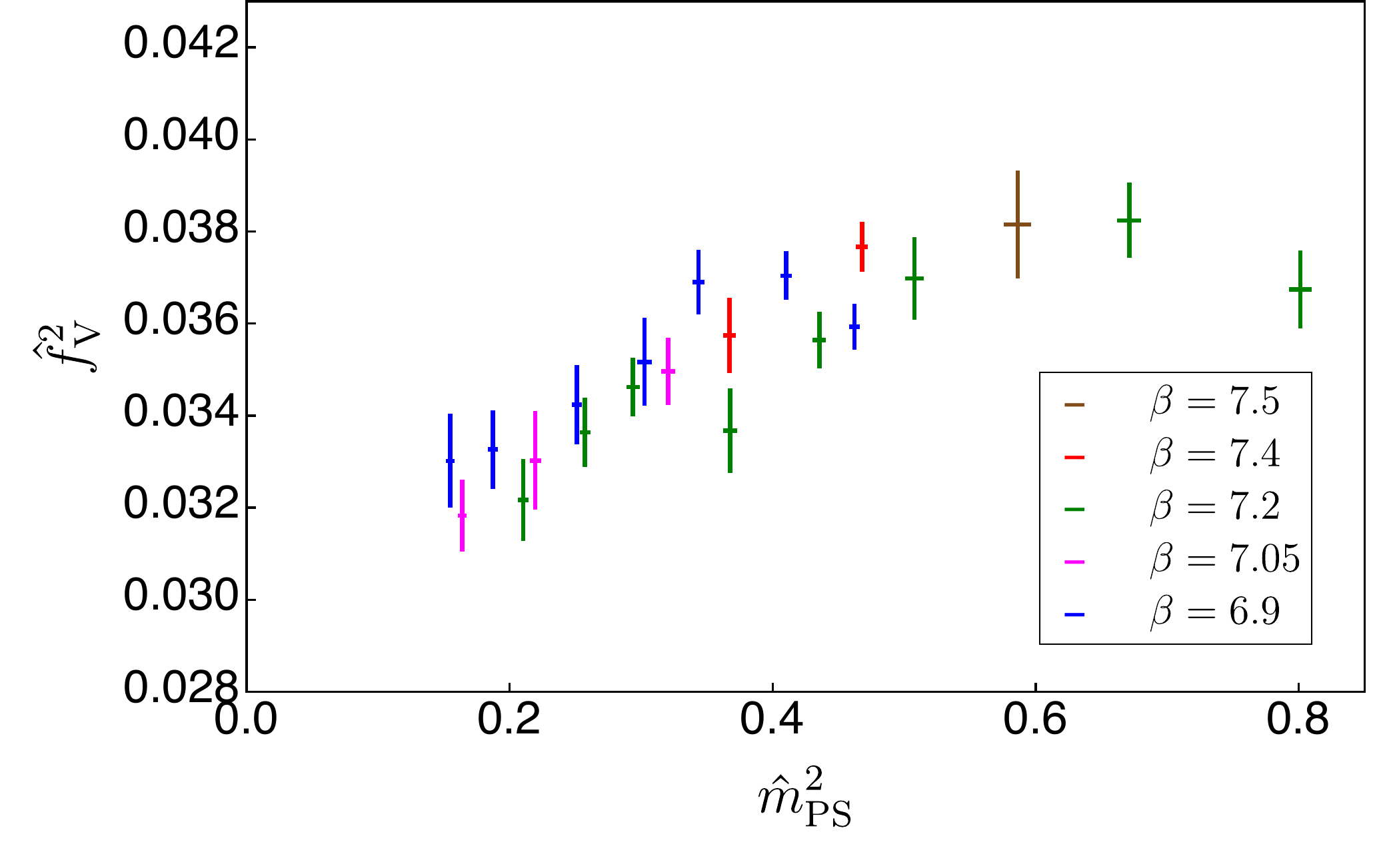}
\includegraphics[width=.65\textwidth]{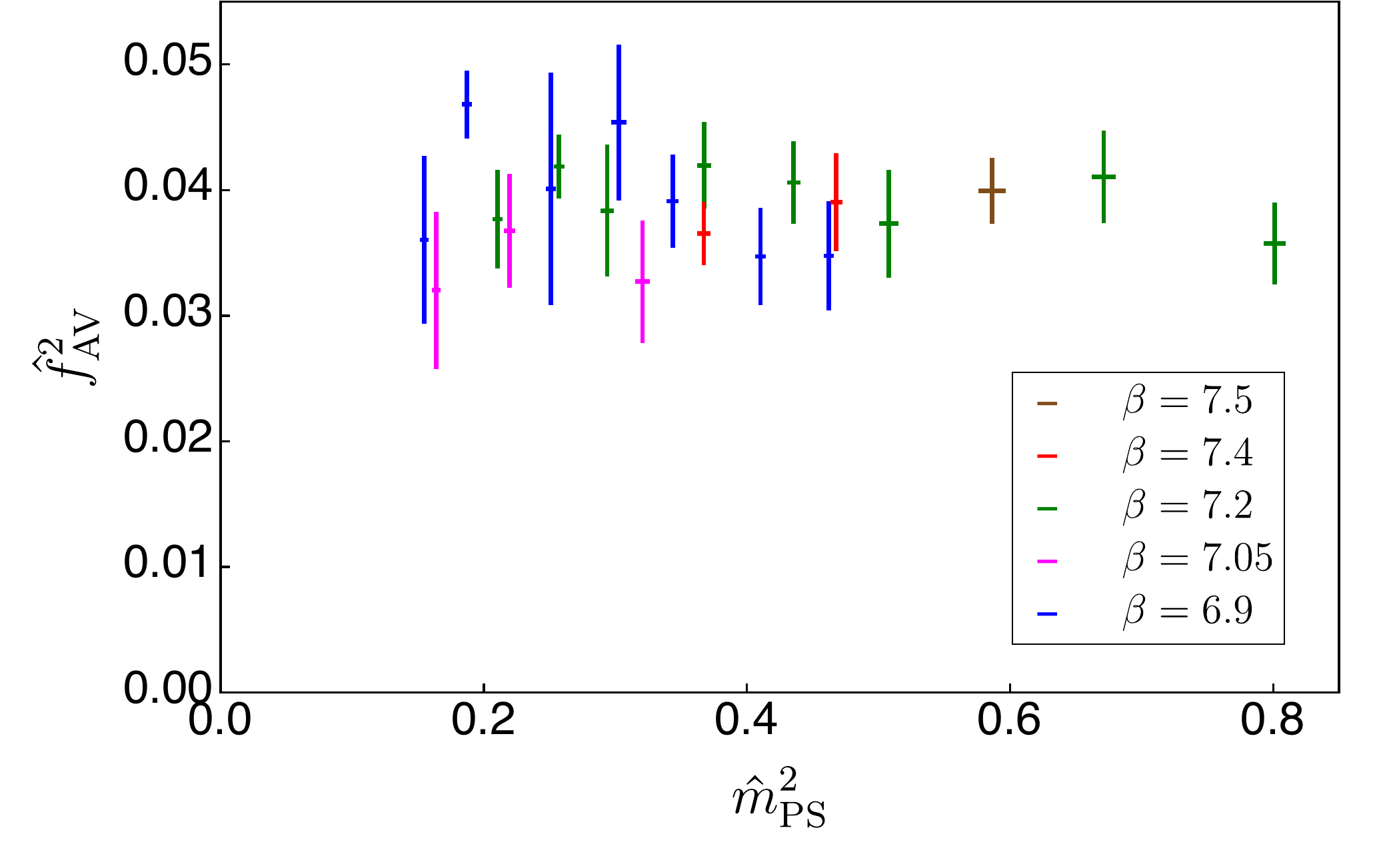}
\caption{%
\label{fig:f2}%
Decay constant squared for pseudoscalar (PS, top), vector (V, middle) and axial-vector (AV, bottom) mesons 
as a function of the  pseudoscalar mass squared $\hat{m}_{\rm PS}^2$.
Different colours refer to different lattice couplings as shown in the legends.
The error bars represent the size of statistical uncertainties. (See \App{meff} for the details.) 
}
\end{center}
\end{figure}

\begin{figure}
\begin{center}
\includegraphics[width=.65\textwidth]{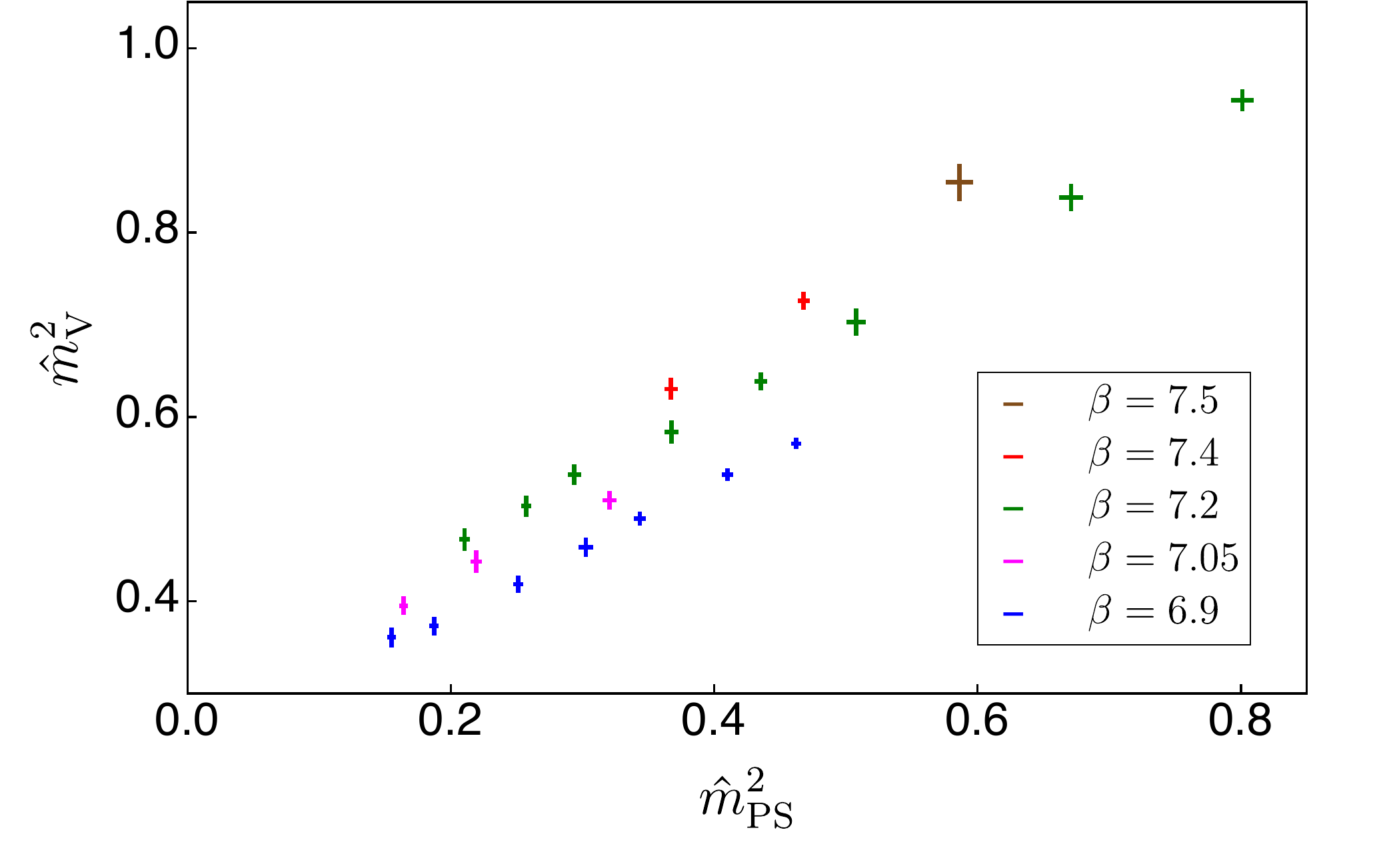}
\includegraphics[width=.65\textwidth]{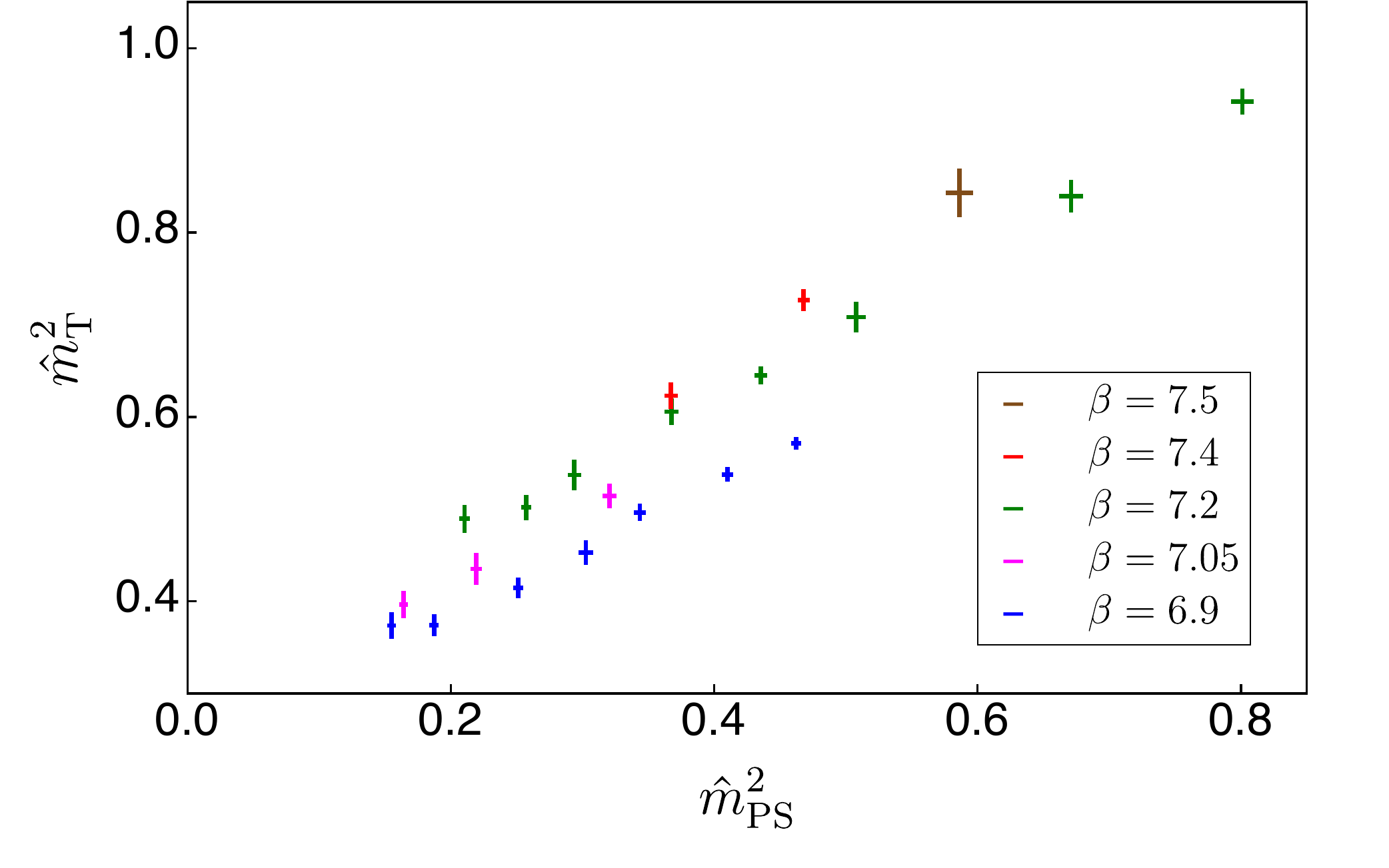}
\caption{%
\label{fig:m2v}%
Mass squared of mesons sourced by vector (V, top) and tensor (T, bottom) operators
as a function of the  pseudoscalar mass squared $\hat{m}_{\rm PS}^2$.
Different colours refer to different lattice couplings as shown in the legends.
The error bars represent the size of statistical uncertainties. (See \App{meff} for the details.) 
}
\end{center}
\end{figure}

\begin{figure}
\begin{center}
\includegraphics[width=.65\textwidth]{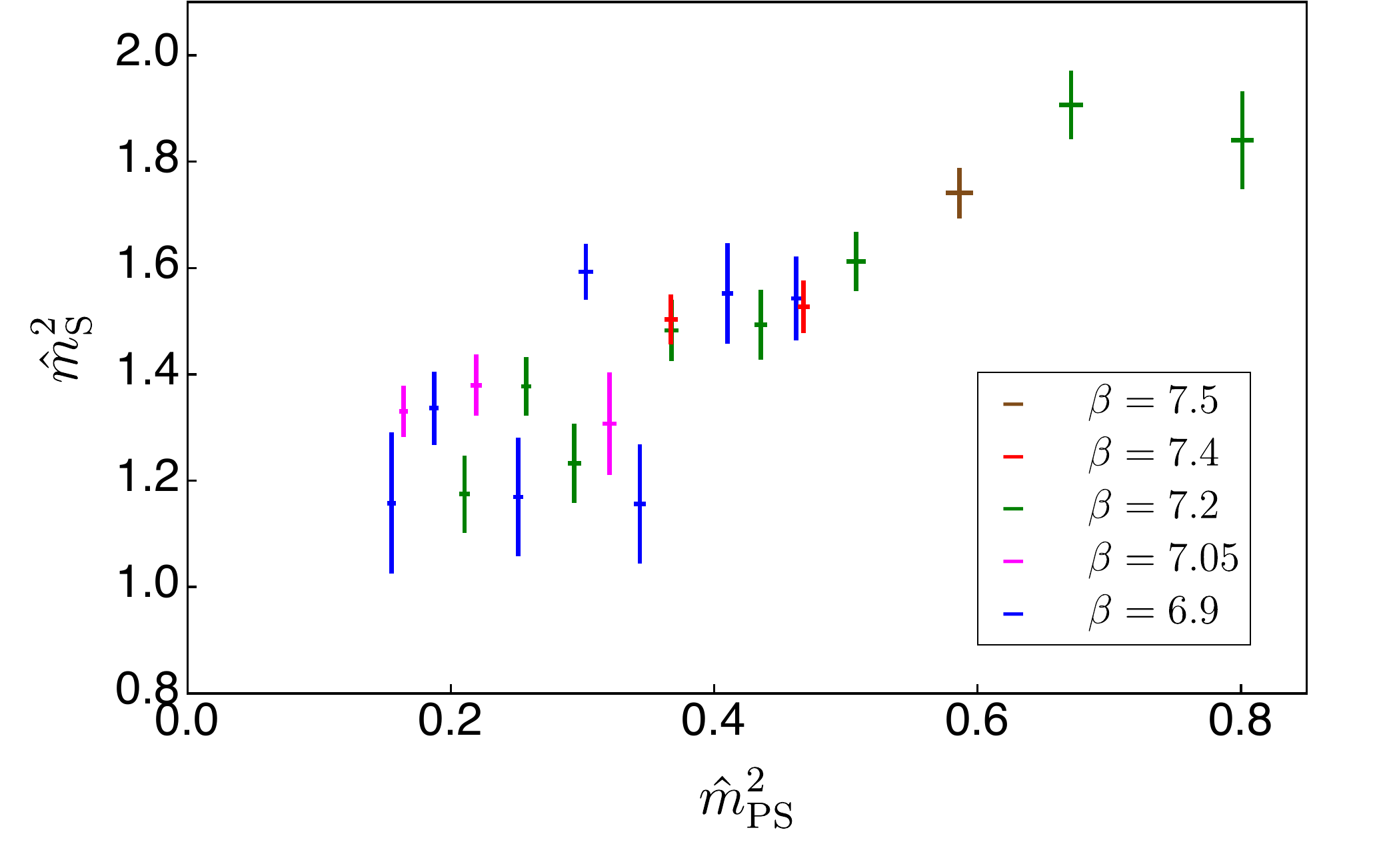}
\includegraphics[width=.65\textwidth]{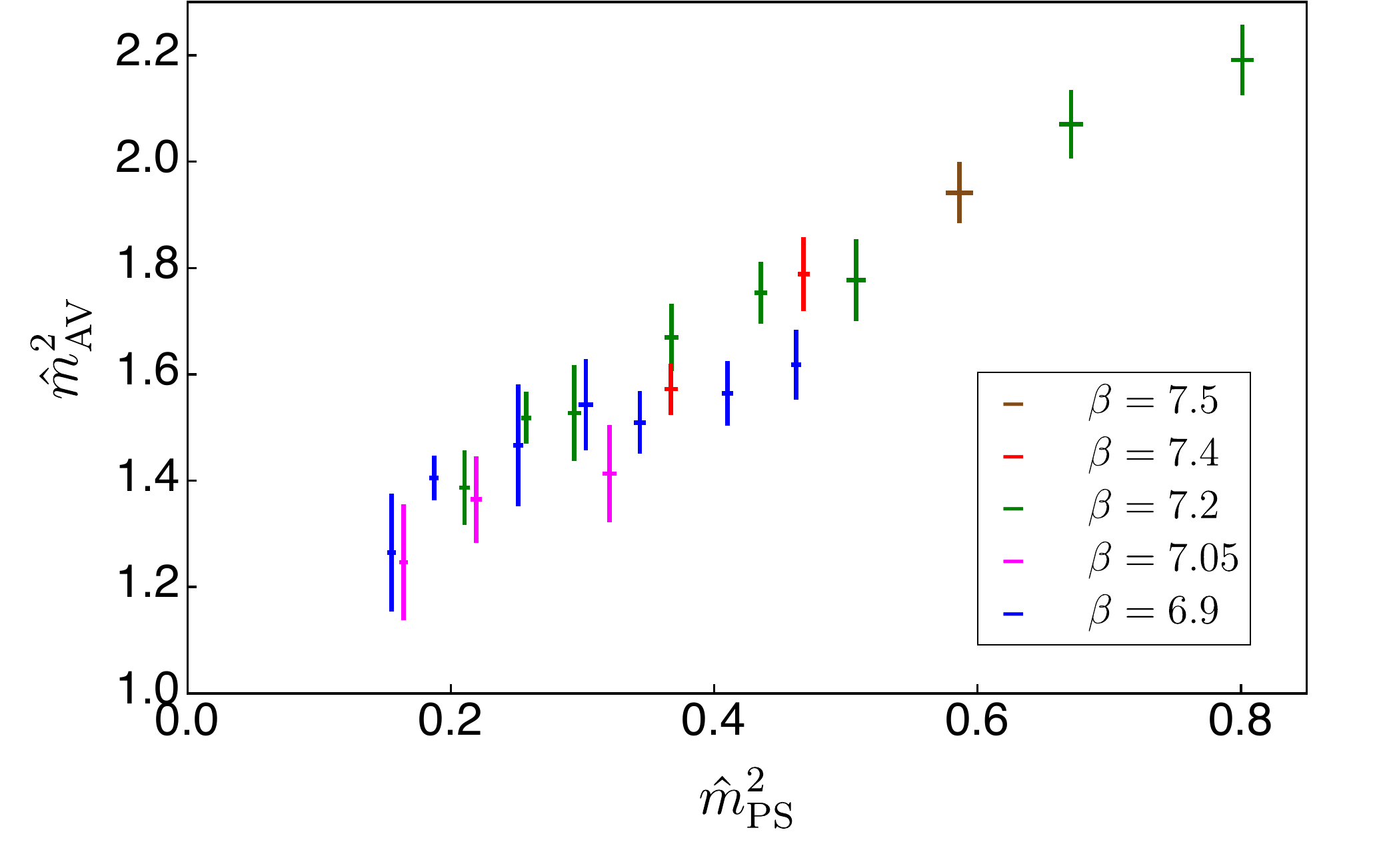}
\includegraphics[width=.65\textwidth]{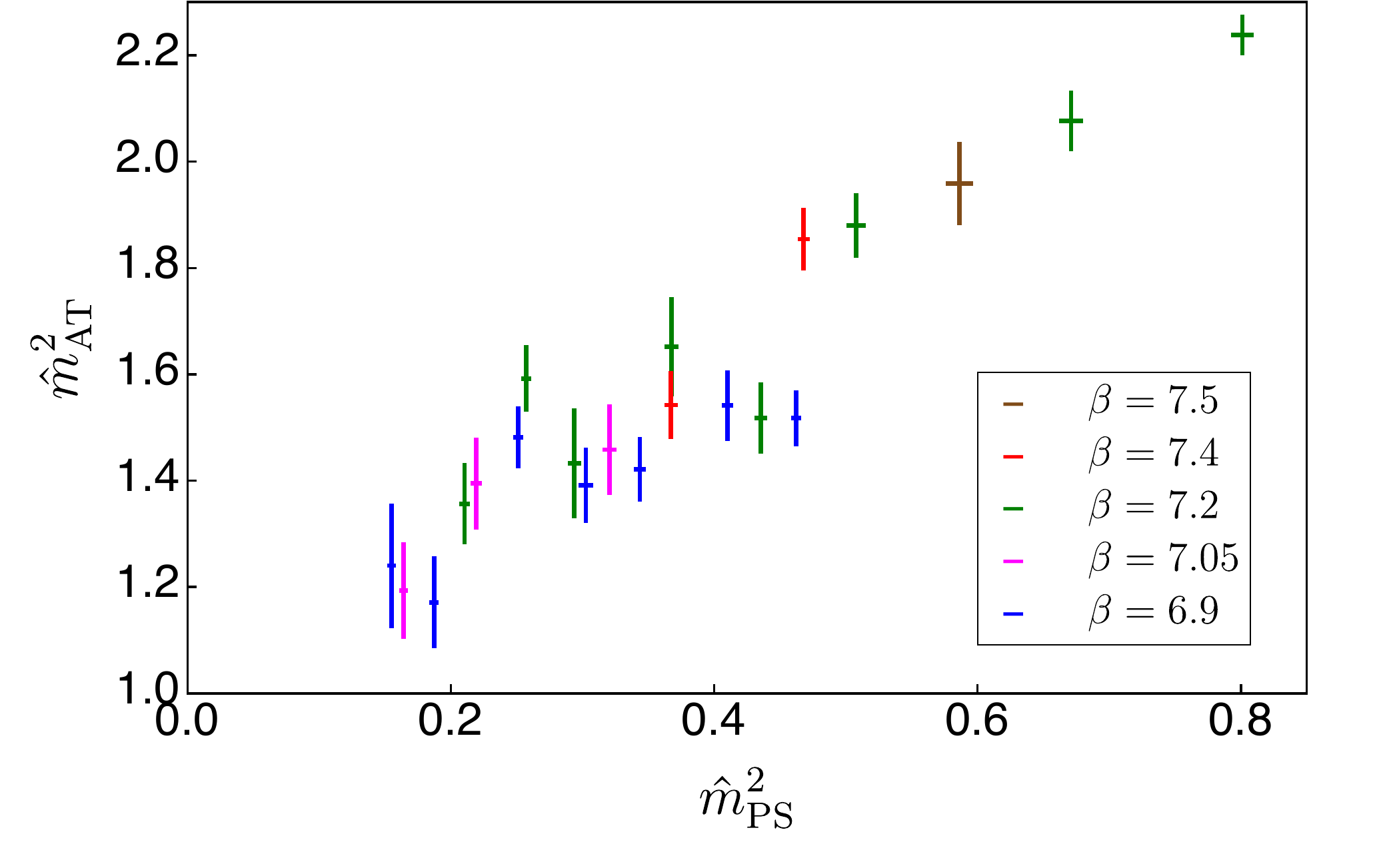}
\caption{%
\label{fig:m2}%
Mass squared of mesons  sourced by scalar (S, top), axial-vector (AV, middle) and axial-tensor (AT, bottom) operators
as a function of the  pseudoscalar mass squared $\hat{m}_{\rm PS}^2$.
Different colours refer to different lattice couplings as shown in the legends.
The error bars represent the size of statistical uncertainties. (See \App{meff} for the details.) 
}
\end{center}
\end{figure}

Using the techniques described in the previous subsection, 
we calculate meson masses and decay constants for the ensembles in \Tab{ensembles}. 
The resulting values in lattice units are summarised in \Tab{meson_spec_spin0} for mesons sourced by
PS and S operators, 
and in \Tab{meson_spec_spin1} for those sourced by V, AV, T, and AT operators. 
The decay constants in the tables are  renormalised  as in \Eq{f_con}. 
As expected, the pseudoscalar mesons are the lightest states for all ensembles. 
In \Tab{meson_spec_spin0} we also present the numerical values of $m_{\rm PS}L$ and $f_{\rm PS}L$. 
The lattice volumes in all ensembles are large enough that $m_{\rm PS} L\gtrsim 7.5$, 
and, as we discussed in \Sec{FV}, 
finite volume effects are  negligible. 
Furthermore, all the values of $f_{\rm PS}$ satisfy the condition $f_{\rm PS} L > 1$, 
 to ensure that the lattice volume is large enough to capture the chiral symmetry breaking scale. 
The ensembles to be used for the massless and continuum extrapolations 
are further restricted to $f_{\rm PS} L \gtrsim 1.5$.

Mesons with higher mass (and spin) can in principle decay 
into 2 and/or 3 pseudoscalars \cite{Drach:2017btk}, 
but for all the ensembles we considered they cannot decay due to large pseudoscalar masses. 
\Tab{meson_spec_spin1} shows that the masses of mesons sourced by V and T operators are 
in good agreement, within current statistical uncertainties. 
As is the case within QCD, low-lying states identified by a given set of quantum numbers $(I,J^{PC})$ 
can result from admixture of more than one possible operator 
when the global symmetry is strongly broken~\cite{Glozman:2007ek}. 
We also find that the masses of three heavier states, sourced by S, AV, and AT operators,
and that have different quantum numbers,  are close to one another. 
They are affected by  large statistical and systematical uncertainties when their masses are extracted 
from the two-point correlation functions (see \App{meff} for technical details).

In order to compare to one another  results obtained from ensembles at different bare parameters, 
we must relate them to the  corresponding renormalised quantities in the continuum limit. 
We do so by adopting the GF scheme as explained in \Sec{GF}: 
we define the meson masses and decay constants in units of $w_0$ using the notation 
\beq
\hat{m}_M\equiv m_M w_0 =m^{\rm lat}_M w_0^{\rm lat}~{\rm and}~\hat{f}_M 
\equiv f_M w_0 = f^{\rm lat}_M w_0^{\rm lat}.
\eeq
We can compare measurements at different lattice couplings $\beta$, and use
the EFT to remove residual  artefacts due to the discretisation. This procedure will be explained in detail in 
section~\ref{Sec:continuum}. 

On the lattice, the bare fermion mass $m_f$ is  a free parameter.
It can vary from being very small (yielding very light pseudoscalars) 
to assuming somewhat large values (yielding stable vector mesons). 
A detailed study of the renormalisation of the fermion mass $m_f$ 
requires a dedicated study that goes beyond the purposes of this work.
In its stead,
we replaced $m_f$ by the pseudoscalar mass squared $m_{\rm PS}^2$, which is a physical quantity. 
In the low-energy EFT, the two are related through the LO relation $m_{\rm PS}^2=2B m_f$, with $B$ one of
the coefficients in the chiral Lagrangian.

Before proceeding, we must check what is the  regime of parameters for which the LO mass relation 
between fermion mass and pseudoscalar mass 
is a good approximation to the data. 
To illustrate this point, in the top panel of \Fig{LO_relation} we present 
the pseudoscalar masses squared $\hat{m}_{\rm PS}^2$ 
and decay constants squared $\hat{f}_{\rm PS}^2$ 
against the bare mass---after subtracting the effects of 
lattice additive renormalisation $(\hat{m}_0-\hat{m}_0^c)$---for 
ensembles at $\beta=7.2$, with various bare masses.\footnote{
Notice that all  dimensional quantities are normalised by the flow scale $w_0$.
The transition from $m_{\rm PS}^2=2B m_f$ to $\hat{m}_{\rm PS}^2=2 \hat{B} \hat{m}_f$ is understood up to 
higher order corrections of $\mathcal{O}(\hat{m}_{\rm PS}^4)$.
}
The critical mass $\hat{m}_0^c$ is determined numerically, extrapolating 
 from the linear fit to the lightest five data points to the value for which 
 $\hat{m}_{\rm PS}^2=0$. 
As shown in the figure,  deviation from  linearity appear for $\hat{m}_{\rm PS}^2>0.4$. 
The decay constant squared also shows a linear behaviour, and its slope is not negligible. 
The bottom panel of \Fig{LO_relation} shows deviations from linearity of the dependence of the
combination $m_{\rm PS}^2 f_{\rm PS}^2$ on the fermion mass.
The Gell-Mann-Oakes-Renner (GMOR) relation~\cite{GellMann:1968rz}, 
$m_{\rm PS}^2 f_{\rm PS}^2=m_f\langle \bar{\psi}\psi \rangle$, 
would imply a mass dependence of the fermion condensate over the range of mass considered.
We alert the reader that a rigorous discussion of the GMOR relation would require first
to determine the values of $\hat{m}_0^c$ for each fixed choice of $\hat{a}$ (obtained by adjusting both
the bare mass and coupling, while keeping the lattice spacing in units of $w_0$ fixed), 
while in this simplified discussion we kept the lattice coupling $\beta=7.2$ fixed.
This is adequate for the purposes of this subsection, but we refer the reader to Section~\ref{Sec:global_fit}
for an assessment of the validity of the GMOR relation in the continuum limit.

In \Fig{f2} we show the numerical results of the decay constants squared of PS, V and AV mesons, 
while in Figs.~\ref{fig:m2v} and~\ref{fig:m2} we present the masses squared of  mesons sourced by
the operators V and T,  and by S, AV, and AT, respectively,
as functions of $\hat{m}_{\rm PS}^2$.
Our first observation is that discretisation effects
 in $\hat{f}_{\rm PS}^2$ and $\hat{m}_{\rm V}^2$ (or $\hat{m}_{\rm T}^2$) 
are significant, 
given the visible difference   between data collected at 
different lattice couplings (and denoted by different colours). 
For other quantities the deviations are no larger than the statistical uncertainties.
Also, the masses and the decay constants  decrease as we 
approach the massless limit, with the exception of $\hat{f}_{\rm AV}^2$. 
Overall, the masses and decay constants show linear dependence on $\hat{m}_{\rm PS}^2$
 in a wide range inside the small-mass region.

\subsection{Continuum extrapolation}
\label{Sec:continuum}

We are now in a position to perform the continuum extrapolation, 
and to eliminate discretisation artefacts in the meson masses and decay constants. 
In order to do so, we introduce the important tool of 
W$\chi$PT~\cite{Sheikholeslami:1985ij,Rupak:2002sm} (see also Ref.~\cite{Sharpe:1998xm},
and~\cite{Symanzik:1983dc, Luscher:1996sc}).
It extends the continuum effective field theory by a double expansion,
both in small fermion mass $m$,
as well as lattice spacing $a$, as both of them break chiral symmetry
and can be introduced in the EFT as spurions.
We denote the lattice spacing in units of $w_0$ by
\beq
\hat{a}\equiv a/w_0 = 1/w_0^{\rm lat}.
\eeq
This yields  the natural size of discretisation effects, 
consistently with the fact that we measure all other dimensional quantities in units of $w_0$. 

At NLO in  W$\chi$PT~\cite{Rupak:2002sm}, the tree-level expression 
for the pseudoscalar decay constant leads to
\beq
\hat{f}_{\rm PS}^{\rm NLO}=\hat{f}^\chi\left(1 + \hat{b}_f^\chi \hat{m}_{\rm PS}^2\right) + \hat{W}_f^\chi \hat{a}, 
\label{eq:f_chipt}
\eeq
where $\hat{f}^\chi=f^\chi w_0^\chi$ is the pseudoscalar decay constant in the massless and continuum limit. 
The fermion masses used in this study are comparatively large, and hence it is legitimate to 
omit from Eq.~(\ref{eq:f_chipt}) the chiral logs, which are important for small values of $\hat{m}_{\rm PS}$.
The coefficients $\hat{b}_f^\chi$ and $\hat{W}_f^\chi$ control the size of corrections due
 to finite mass and finite lattice spacing, respectively. 
In principle one should  measure all observables  in units of $w_0^\chi$, 
while  we instead use the mass-dependent $w_0$, as measured at the finite mass of the individual ensembles, 
hence avoiding the need to extrapolate to the massless limit~\cite{Ayyar:2017qdf}---we collected
enough data to attempt such extrapolation only for  two values of $\beta$. 
The replacement of $w_0^\chi$ by $w_0$ does not affect  the NLO EFT, 
the difference appearing  at higher orders in $m_{\rm PS}^2$. 
Compared to the continuum NLO expression in Ref. \cite{Bijnens:2009qm}, it results in a 
shift  $\hat{b}_f^\chi$ by $\tilde{k}_1$ in \Eq{fit_w0}, 
due to fitting the measurements of $f_{\rm PS} w_0$.

\begin{table}
\begin{center}
\begin{tabular}{|c|c|c|c|c|}
\hline\hline
 & $~~\hat{f}^{2,\,\chi}_M~~$ & $~~~L^0_{f,M}~~~$
& $~~~~~W^0_{f,M}~~~~~$ & $~~~~~\chi^2/N_{\rm d.o.f}~~~~~$ \cr \hline
PS & $0.00618(28)(33)$ & $3.01(21)(33)$ & $-0.00135(29)(19)$ & $1.6$\cr
V & $0.0296(15)(8)$ & $0.51(9)(6)$ & $0.0004(16)(8)$ & $1.0$\cr
AV & $0.032(7)(2)$ & $0.17(35)(14)$ & $0.012(8)(2)$ & $1.1$\cr
\hline\hline
 & $~~\hat{m}^{2,\,\chi}_M~~$ & $~~~L^0_{m,M}~~~$
& $~~~~~W^0_{m,M}~~~~~$ & $~~~~~\chi^2/N_{\rm d.o.f}~~~~~$ \cr \hline
V & $0.404(13)(9)$ & $2.18(10)(7)$ & $-0.220(15)(12)$ & $0.9$\cr
T & $0.418(18)(2)$ & $2.08(12)(17)$ & $-0.229(22)(30)$ & $0.8$\cr
AV & $1.07(13)(2)$ & $1.37(32)(7)$ & $0.04(13)(2)$ & $0.8$ \cr
AT & $1.08(13)(8)$ & $1.49(34)(16)$ & $-0.08(13)(13)$ & $2.4$ \cr
S & $1.16(12)(12)$ & $0.85(21)(20)$ & $-0.08(14)(16)$ & $1.8$ \cr
\hline\hline
\end{tabular}
\end{center}
\caption{
\label{tab:LECs}
Fit results of the continuum and massless extrapolations for masses squared and decay constants squared
of mesons in the dynamical simulations. 
The low-energy constants appearing in W$\chi$PT are defined
 in Eqs.~(\ref{eq:f2_chipt}) and (\ref{eq:m2_chipt}).
The fits of $\hat{f}_{\rm PS}^2$ measurements are restricted to 
include only the  eleven ensembles identified in the main text.
For the other quantities, additional ensembles satisfying
 $\hat{m}_{\rm PS}^2\lsim 0.6$ and $\hat{a}\lsim 1$ have been included.
In parenthesis, we report the statistical and systematic uncertainties, respectively.
}
\end{table}

\begin{figure}
\begin{center}
\includegraphics[width=.49\textwidth]{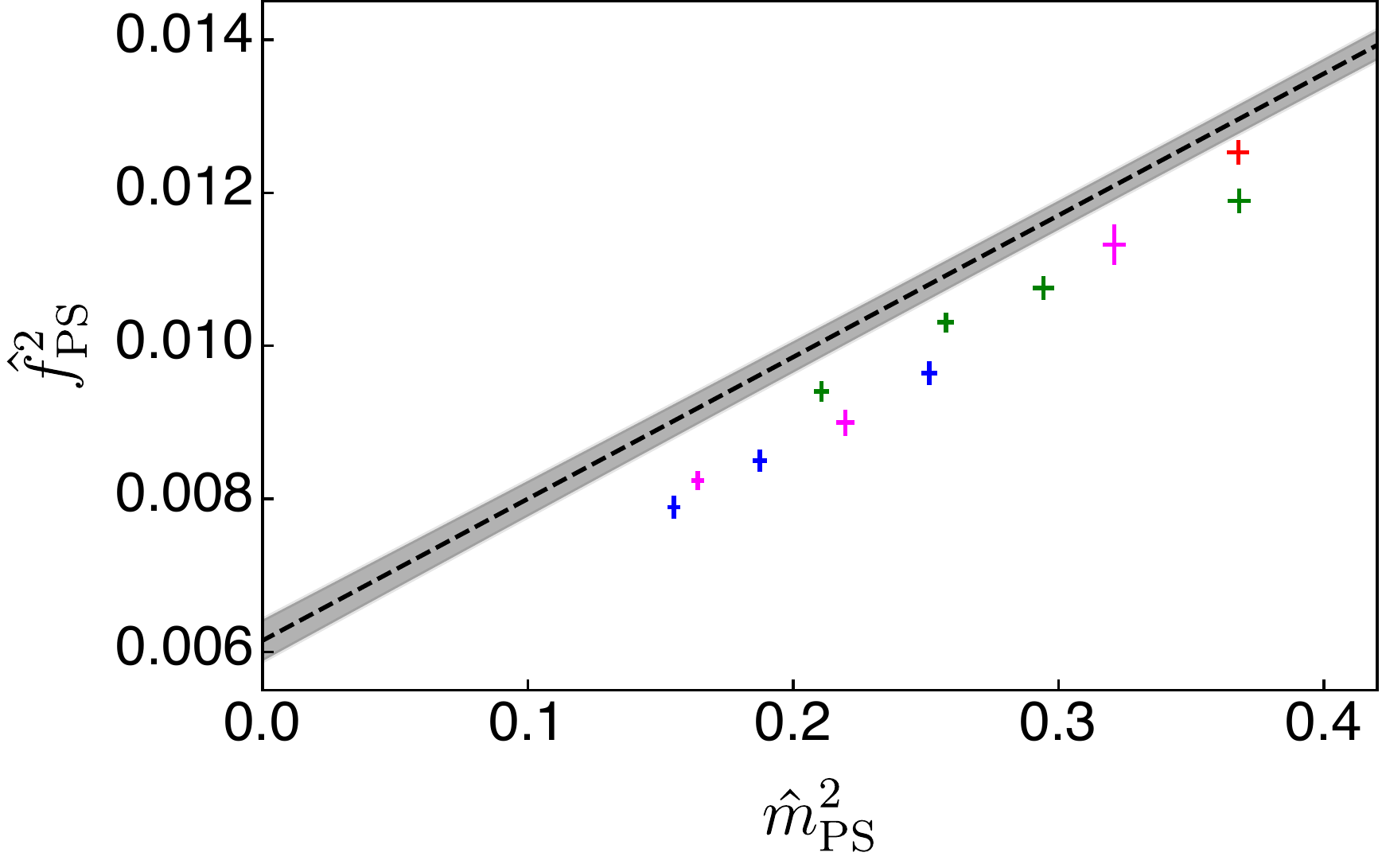}
\includegraphics[width=.49\textwidth]{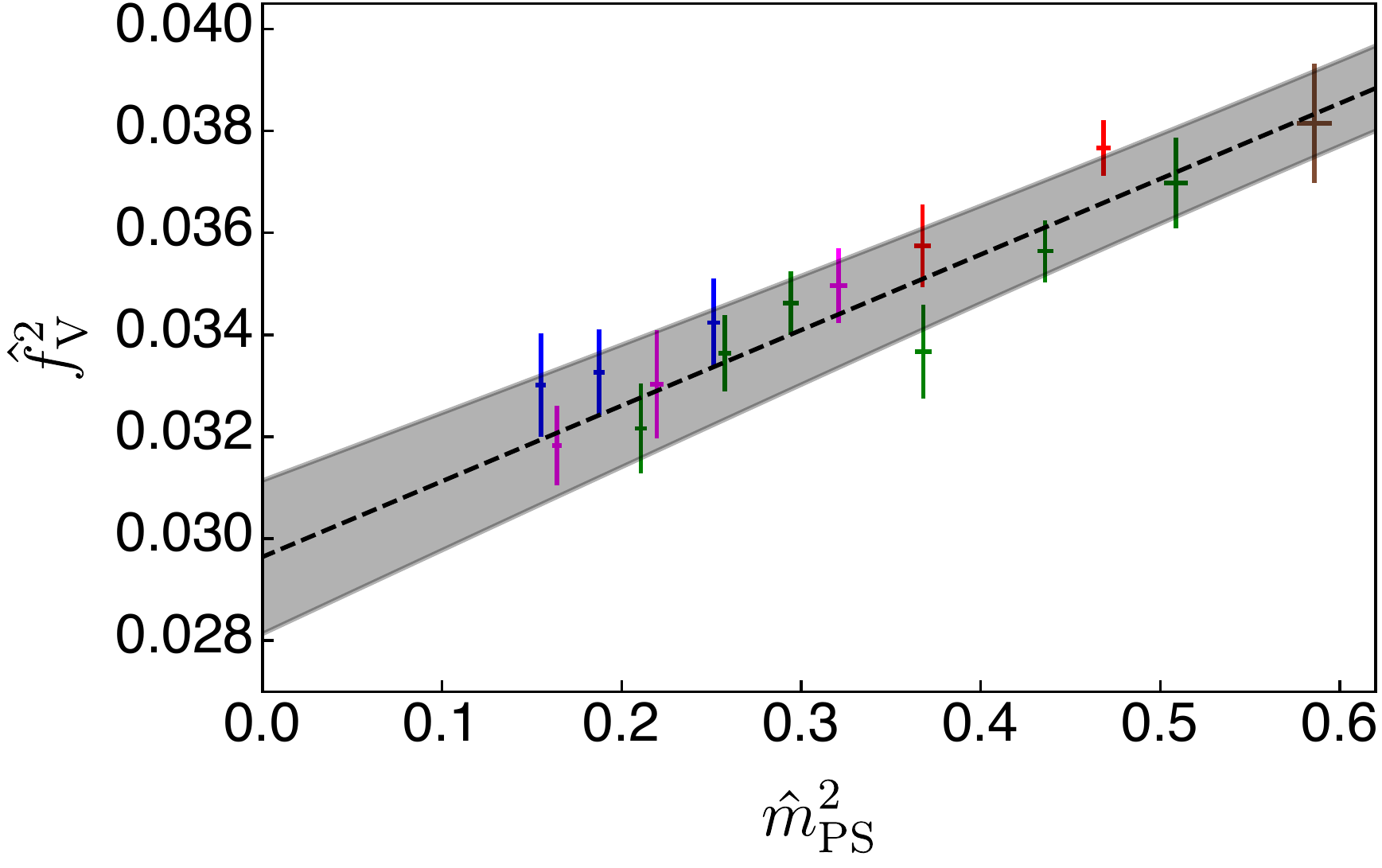}
\includegraphics[width=.49\textwidth]{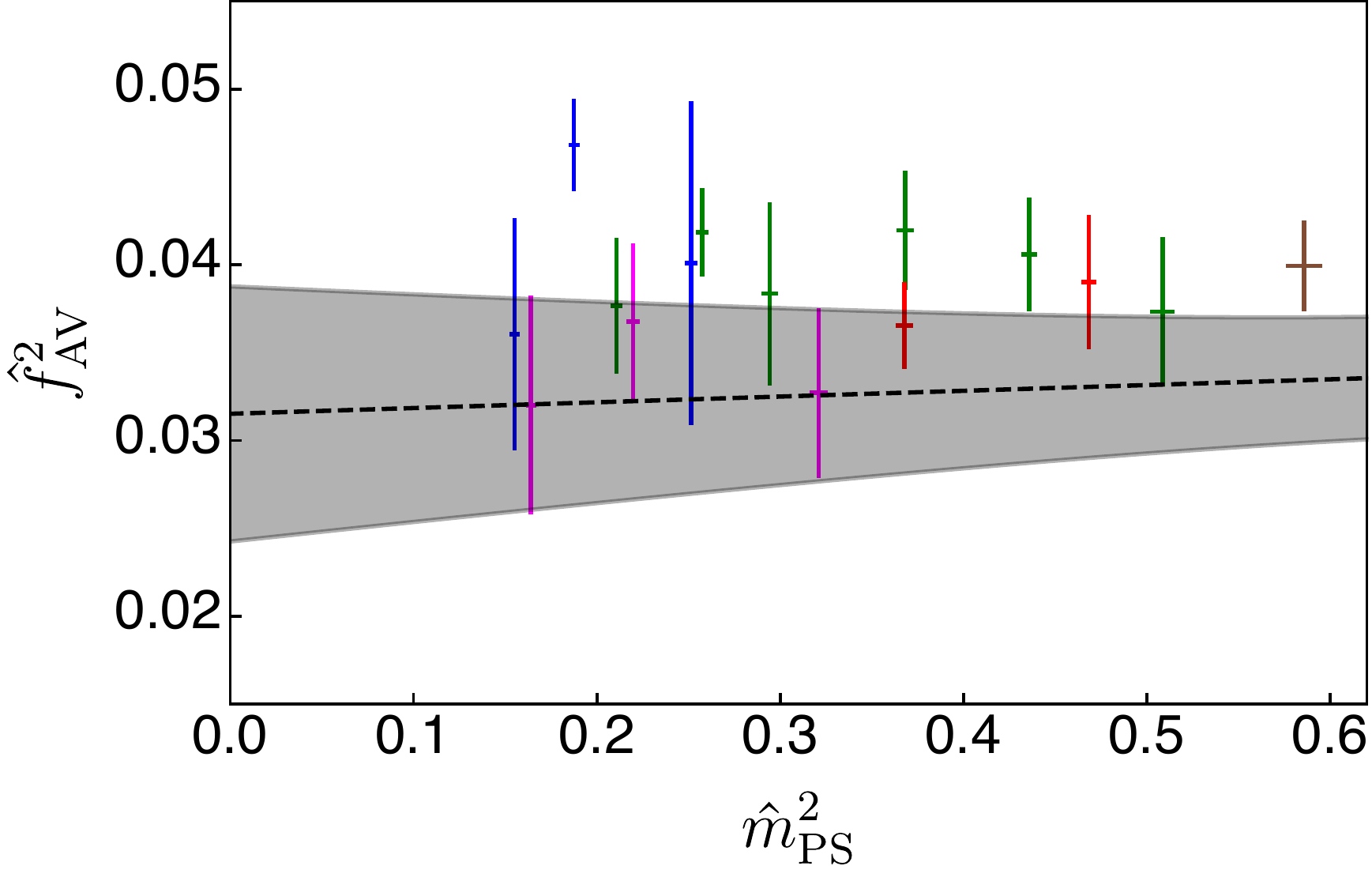}
\caption{%
\label{fig:f2_NLO_fit}%
Top to bottom, left to right: decay constants squared of the 
pseudoscalar (PS), vector (V), and axial-vector (AV) mesons, 
as a function of the pseudoscalar mass squared $\hat{m}_{\rm PS}^2$.
Various colours denote different lattice couplings: $\beta=6.9$ 
(blue), $7.05$ (purple), $7.2$ (green), $7.4$ (red), and $7.5$ (brown). 
The grey bands (the width of which indicates the statistical error)
show the results of the continuum and massless extrapolations
after subtracting the discretisation artefact, as
 discussed in the text.  
}
\end{center}
\end{figure}

\begin{figure}
\begin{center}
\includegraphics[width=.49\textwidth]{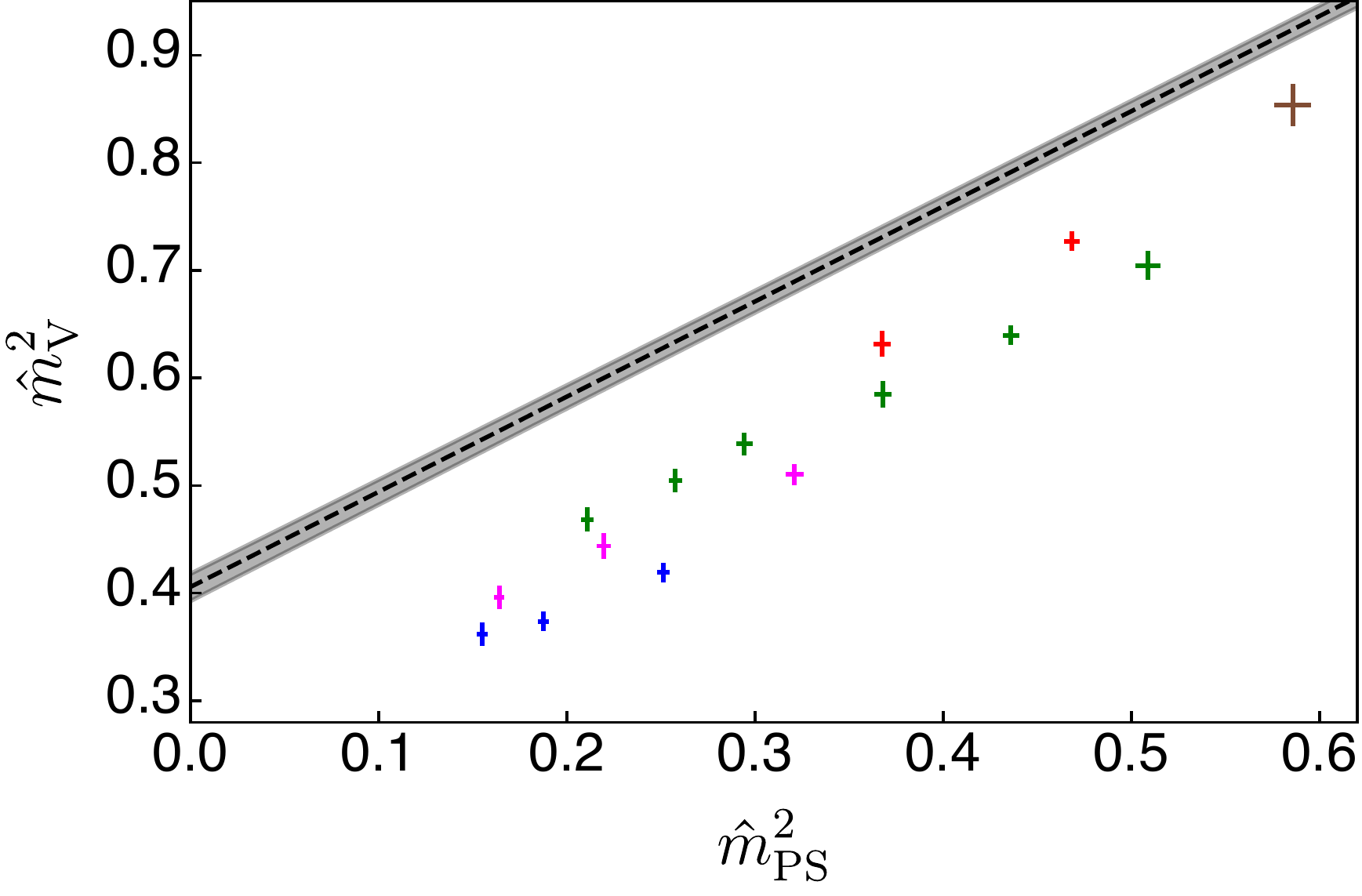}
\includegraphics[width=.49\textwidth]{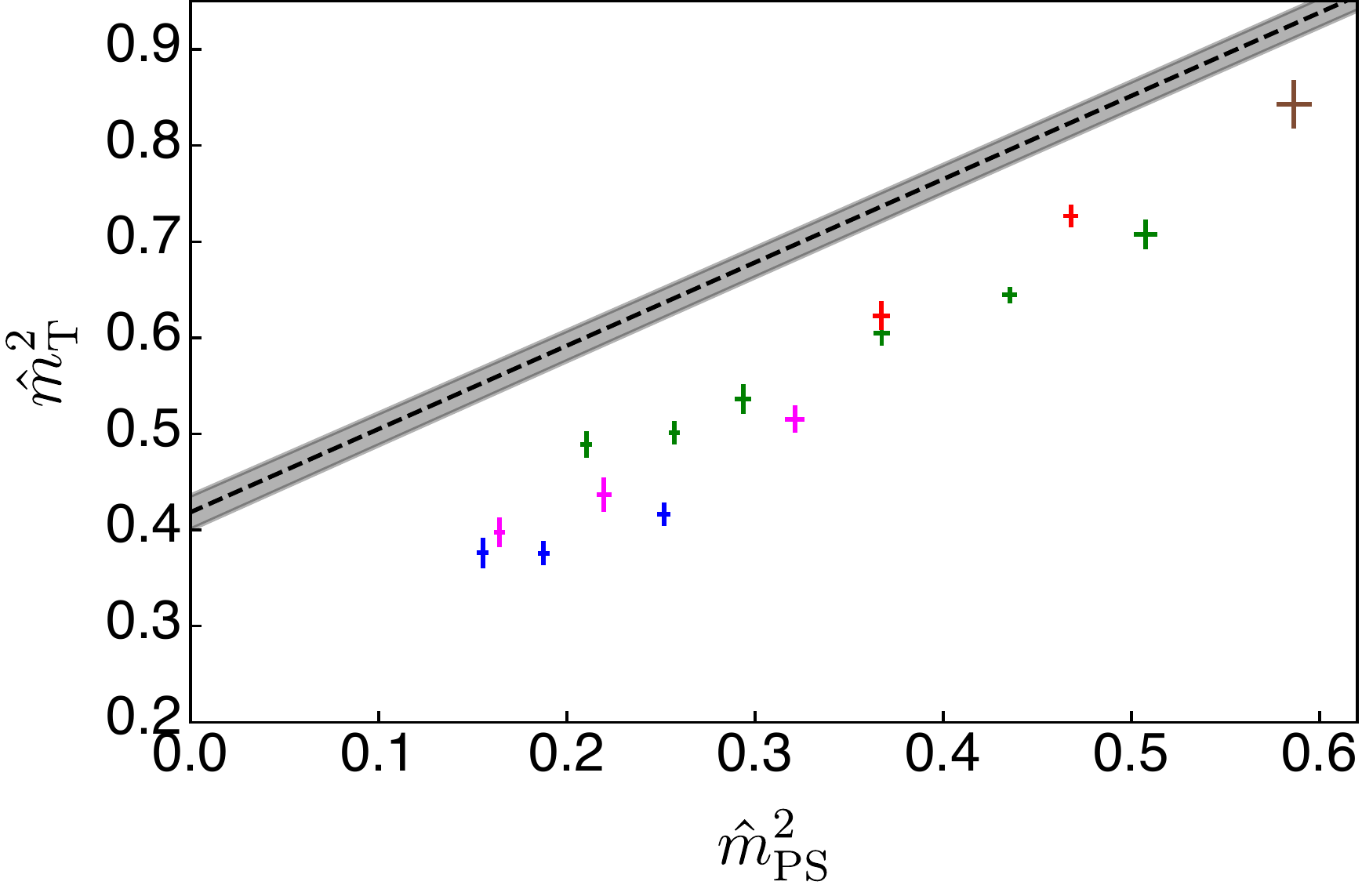}
\includegraphics[width=.49\textwidth]{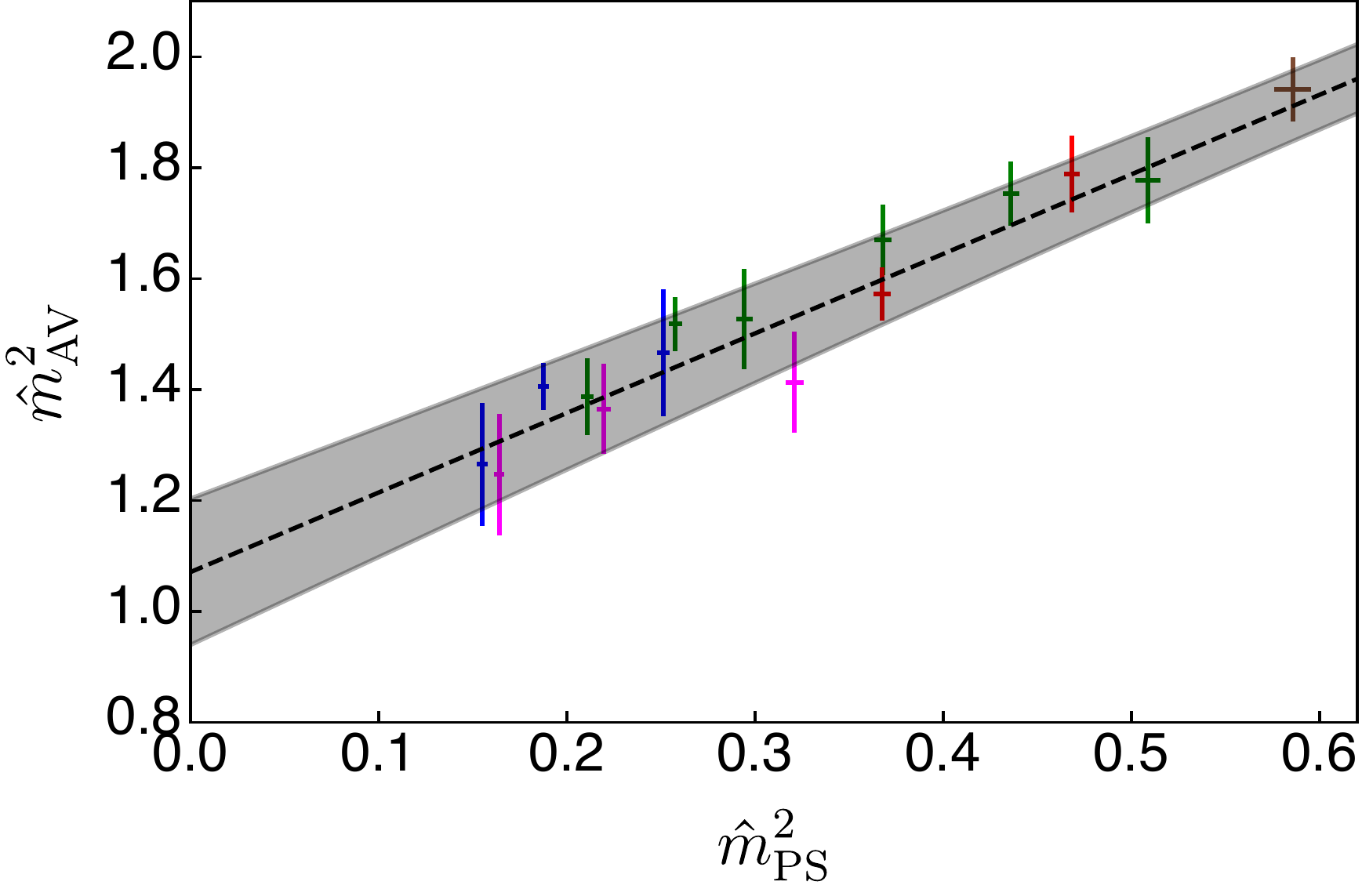}
\includegraphics[width=.49\textwidth]{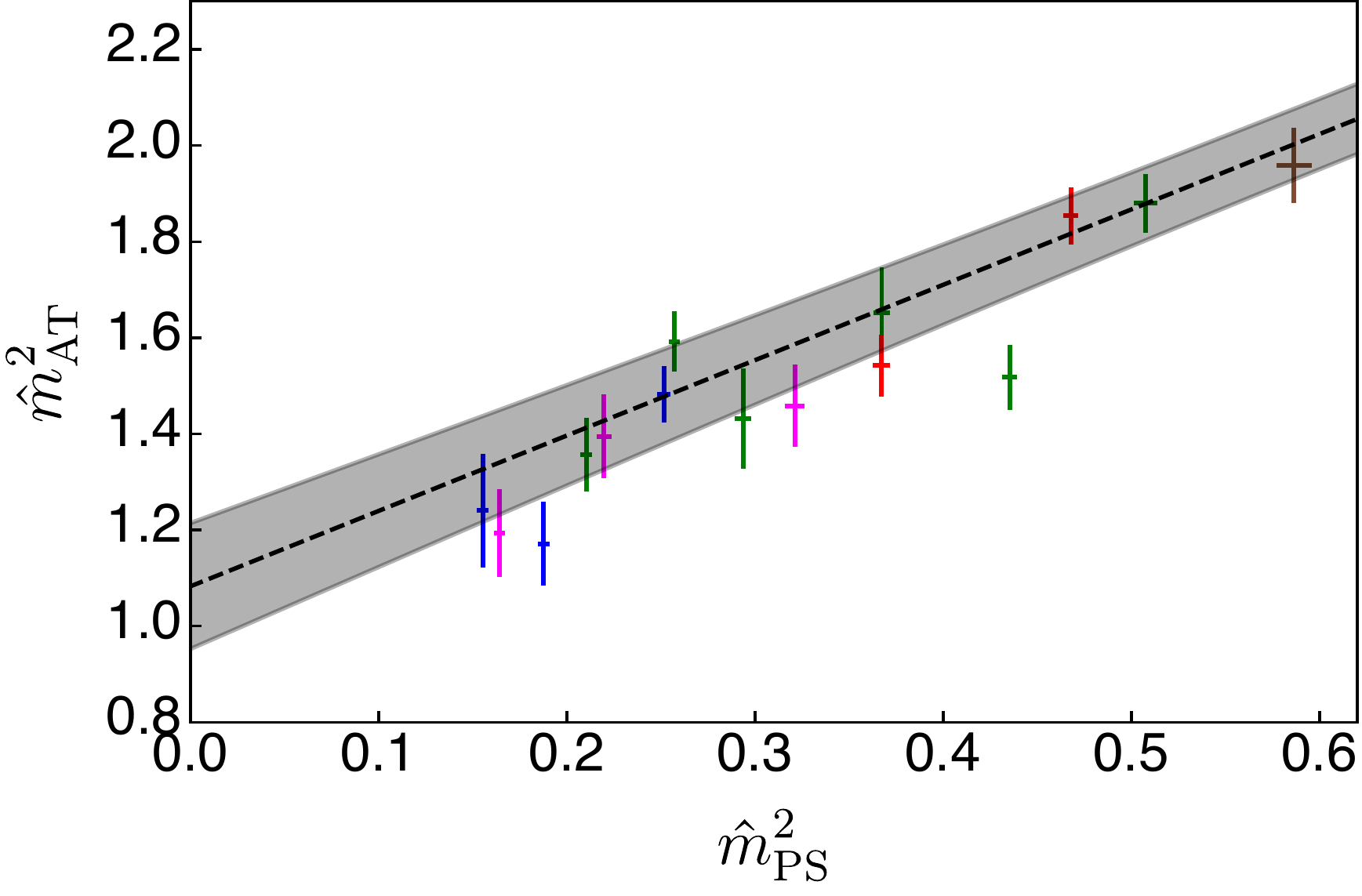}
\includegraphics[width=.49\textwidth]{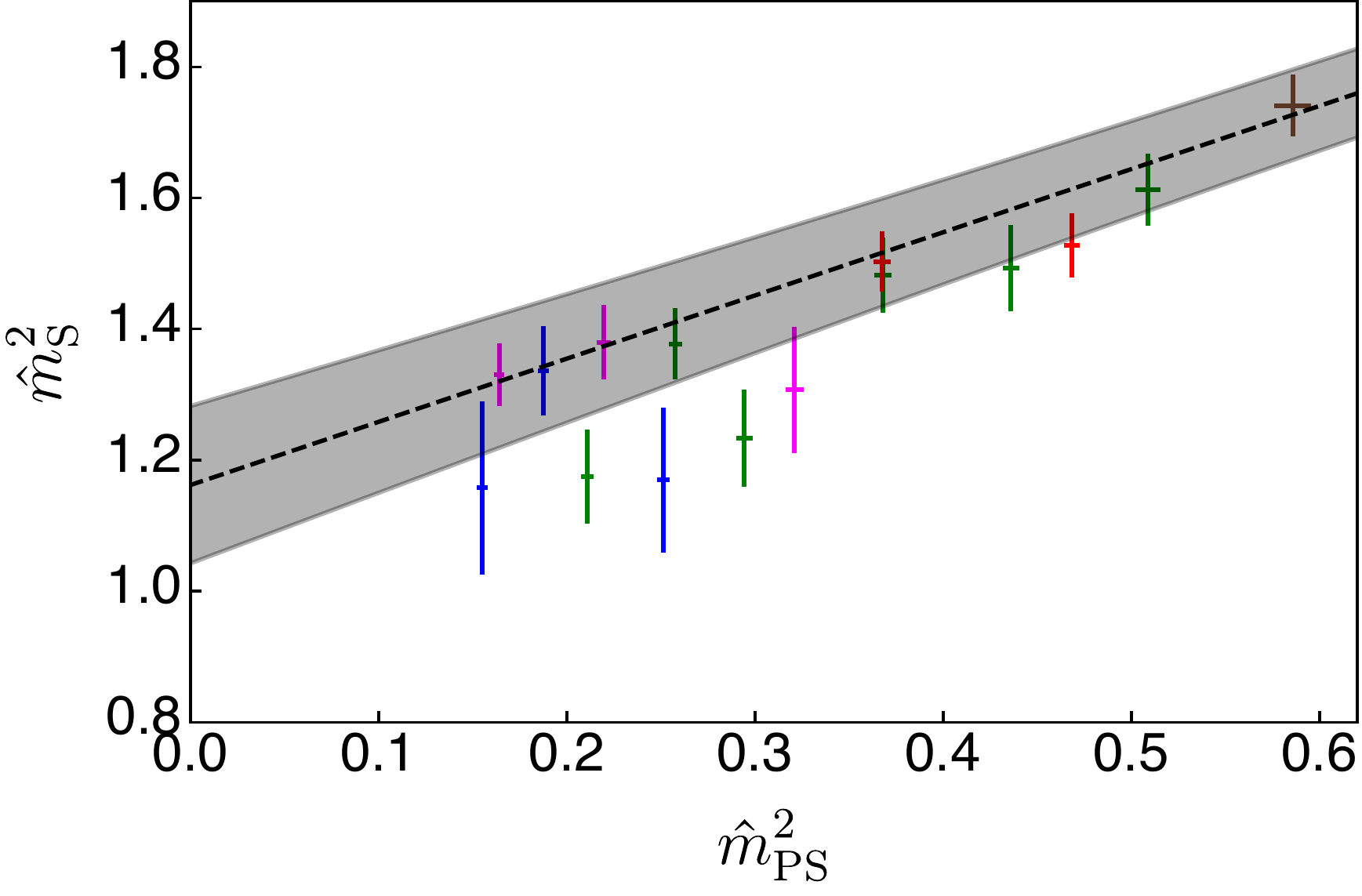}
\caption{%
\label{fig:m2_NLO_fit}%
Top to bottom, left to right: masses squared of the vector (V), 
tensor (T), axial-vector (AV), axial-tensor (AT), and scalar (S) mesons, 
as a function of the pseudoscalar mass squared $\hat{m}_{\rm PS}^2$.
Various colours denote different lattice couplings: $\beta=6.9$ 
(blue), $7.05$ (purple), $7.2$ (green), $7.4$ (red), and $7.5$ (brown). 
The grey bands (the width of which indicates the statistical error)
show the results of the continuum and massless extrapolations
after subtracting the discretisation artefact, as
 discussed in the text.  
}
\end{center}
\end{figure}

Among the underlying assumptions of  W$\chi$PT is the requirement that
 the measurements it describes satisfy 
\beq
\frac{m_{\rm PS}^2}{\Lambda_\chi^2} \sim a\Lambda_\chi < 1,
\label{eq:power_counting}
\eeq 
where  $\Lambda_\chi$ is the symmetry breaking scale,
for which we take the estimate $\Lambda_\chi=4\pi f_{\rm PS}$.
In \Sec{w0}, we found  the NLO EFT to describe the 
measurements of $\hat{w}_0$ up to $\hat{m}_{\rm PS}^2 \lsim 0.4$. 
The numerical results for the pseudoscalar decay constant squared in Fig.~\ref{fig:f2} 
shows linear mass dependence over the range of $0.15\lsim \hat{m}_{\rm PS}^2\lsim 0.40$. 
Using the (bare) results in \Tab{ensembles}, restricted to this range, 
 the two mass scales associated with the power counting for the tree-level NLO W$\chi$PT 
 are estimated to be
\beq
0.13\lsim \frac{m_{\rm PS}^2}{\Lambda_\chi^2}\sim 0.20\,,~~~{\rm and}~~~
0.60\lsim a\Lambda_\chi \sim 1.4.
\label{eq:power_counting_data}
\eeq
The first  estimate is roughly consistent with the scale separation in \Eq{power_counting}. 
But we find that  $a\Lambda_\chi$ evaluated on some of the ensembles is greater than unity. 
We further constrain the analysis to  the ensembles that satisfy the condition $\hat{a}\lesssim 1$, 
which is satisfied by DB1M5$-$7, DB2M1$-$3, DB3M5$-$8, and DB4M2.
Only these ensembles will be used in the continuum and massless extrapolations
that employ the tree-level NLO W$\chi$PT. 
We will report the values of the $\chi^2/N_{\rm d.o.f.}$ in the analysis, 
and we anticipate here that the quality of the  fits of the data supports
the results of  this exclusion process, otherwise based upon a set of estimates.

As the linear dependence of  $\hat{f}_{\rm PS}^{\rm NLO}$ on both $\hat{m}_{\rm PS}^2$ and $\hat{a}$ 
can be recast into linear behaviour of $\hat{f}_{\rm PS}^{2,{\rm NLO}}$
and because---on 
the basis of the EFT described in~\cite{Bennett:2017kga}---at NLO 
also the mass squared and decay constant squared
of spin-1 mesons have  leading corrections of $\mathcal{O}(m_{\rm PS}^2)$, 
we consider the following linear ansatz:
\bea
\hat{f}_M^{2,{\rm NLO}}&=&\hat{f}_M^{2, \chi}\left(1 + L_{f,M}^0 \hat{m}_{\rm PS}^2\right) + W_{f,M}^0 \hat{a}
\label{eq:f2_chipt}
\eea
for the decay constants squared of the mesons $M=$ PS, V, and AV, and 
\bea
\hat{m}_M^{2,{\rm NLO}}&=&\hat{m}_M^{2, \chi}\left(1 + L_{m,M}^0 \hat{m}_{\rm PS}^2\right) + W_{m,M}^0 \hat{a}
\label{eq:m2_chipt}
\eea
for the masses squared of the mesons $M=$ V, AV, S, T, and AT. 

We restrict the fit to the eleven ensembles identified earlier for  $\hat{f}_{\rm PS}^2$. 
Yet, the results for V, AV, S, T and AT mesons in Figs.~\ref{fig:f2}, \ref{fig:m2v} and~\ref{fig:m2} exhibit 
 linear dependence on $\hat{m}_{\rm PS}^2$ also in heavier ensembles,
so that in the fits of their properties we included additional ensembles with $\hat{m}_{\rm PS}^2\sim 0.6$ 
and $\hat{a} \lesssim 1$. The fit results are presented 
in Figs.~\ref{fig:f2_NLO_fit} and~\ref{fig:m2_NLO_fit}. 
In the figures, the grey bands denote the continuum extrapolated results 
obtained by setting $\hat{a}=0$
in Eqs.~(\ref{eq:f2_chipt}) and~(\ref{eq:m2_chipt}), and by using the fit parameters summarised
 in \Tab{LECs}. 
In the table, the numbers in the two parentheses denote the statistical and systematic uncertainties
associated with the numerical fits.
The latter is estimated by varying the fitting range to include or exclude the coarsest or the heaviest ensemble. 
We find acceptable values of $\chi^2/{\rm d.o.f}$ for all the fits, 
in support of the applicability of the tree-level NLO EFT to describe our data.

\section{Lattice results: summary}
\label{Sec:summary1}

We provide here a list of lattice highlights (from Sections~\ref{Sec:latticemodel}, 
\ref{Sec:systematics}, and \ref{Sec:mesons}).
We will be very schematic: precise definitions and detailed discussions
can be found in the main text.

\begin{itemize}
\item[L1.] The lattice ensembles used in the dynamical study, and the values 
of the lattice parameters that characterise them,
are discussed in Section~\ref{Sec:latticemodel}, and listed in Table~\ref{tab:ensembles},
 including the average plaquette
and the gradient flow scale $w_0$.
\item[L2.]  The gradient flow scale $w_0$ is defined and studied in 
Section~\ref{Sec:GF}. Its mass dependence is 
illustrated by Fig.~\ref{fig:GF_scales}, \ref{fig:GF_scale_fit}, and~\ref{fig:latt_spacing} and 
Table~\ref{tab:GF_LECs}. 
Throughout this work we employ the gradient flow scale to set the physical scale of the lattice calculations.
\item[L3.]  The topological charge is defined and studied in Sec.~\ref{Sec:topology}, 
and examples of topological charge histories
 are depicted in Fig.~\ref{fig:topcharge}. The ensembles used in the numerical 
 analysis do not show clear evidence of topological freezing.
\item[L4.]  Lattice finite size effects are studied in Section~\ref{Sec:FV}.
They are negligibly small provided the condition $m_{\rm PS} L \gsim 7.5$ 
is satisfied---see also Table~\ref{tab:FV_ensembles}
and Fig.~\ref{fig:mps_FV}.
In the analysis presented in this paper, all ensembles satisfy this condition.
\item[L5.]  In Section~\ref{Sec:mesons}, we define the operators and  the two-point functions
used for the measurement of decay constants and masses of the mesons---see Table~\ref{tab:mesons}. 
The  decay constants are perturbatively renormalised according to
Eqs.~(\ref{eq:f_con}) and~(\ref{eq:zfactor}).
\item[L6.]  The masses and (renormalised) decay constants (in units of the lattice spacing $a$)
of the spin-0 mesons are reported in Table~\ref{tab:meson_spec_spin0},
and those of the spin-1 mesons in Table~\ref{tab:meson_spec_spin1}. 
\item[L7.]  In Sec.~\ref{Sec:spectroscopy} we introduce the 
notation $\hat{m}\equiv m_Mw_0$, and equivalent for
all dimensional quantities expressed in terms of the gradient 
flow scale. We discuss the GMOR relation and illustrate it in Figure~\ref{fig:LO_relation}.
Linearity with the fermion mass holds for $\hat{m}_{\rm PS}^2\lsim 0.4$.
From this point on, we eliminate the dependence on the fermion mass, and study our observables
only as a function of the mass (squared) of the PS state.
\item[L8.]  Masses and decay constants, expressed in units of the gradient flow
scale, are plotted for all ensembles 
in Figs.~\ref{fig:f2}, \ref{fig:m2v}, and \ref{fig:m2}. We notice that the V and T 
masses are compatible with each other,
and so are the AV and AT---although the latter two states are physically distinct. 
We do not further discuss the
 T and AT states  in the paper.
\item[L9.]  We apply tree-level W$\chi$PT via Eqs.~(\ref{eq:f2_chipt}) and (\ref{eq:m2_chipt}), that we use 
to perform the continuum and massless extrapolations. We impose the conditions $\hat{a}\lsim 1$,
 $m_{\rm PS}L \gsim 7.5$, and $m_{\rm PS}^2\lsim 0.4$, which identify a subset of eleven ensembles
 to be used for the continuum extrapolation of $\hat{f}_{\rm PS}^2$. For the other observables we relax 
 the condition on the mass to $m_{\rm PS}^2\lsim 0.6$. The results are shown in Table~\ref{tab:LECs} and in 
Figs.~\ref{fig:f2_NLO_fit} and \ref{fig:m2_NLO_fit}.
\item[L10.] 
As reflected in the fit results for $W^0$ presented in
\Tab{LECs}, $\hat{f}_{\rm PS}^2$ and $\hat{m}_{\rm V}^2$ (and similarly $\hat{m}_{\rm T}^2$)
are affected by  discretisation effects, the sizes of which are about $4\sim 13\%$ 
and  $10\sim 35\%$, respectively, over the considered ranges of bare parameters. 
In all other observables,   discretisation effects are no larger than the statistical errors.

\end{itemize}

\section{Low-energy phenomenology: EFT and sum rules}
\label{Sec:result_con}

In this section, we study some  of the long distance properties of the $Sp(4)$ gauge theory
that may be useful inputs for phenomenological studies performed
  in the context of composite Higgs models based upon the  $SU(4)/Sp(4)$ coset.
We restrict our attention to the PS, V and AV states, and consider only 
ensembles with $\hat{a}\lsim 1$ and $\hat{m}_{\rm PS}^2\lsim 0.4$.
We perform a simplified continuum extrapolation, by subtracting from their masses
and decay constants the last term of Eqs.~(\ref{eq:f2_chipt}) and~(\ref{eq:m2_chipt}), 
with the coefficients $W^0_{f,M}$ and $W^0_{m,M}$
obtained by a fit similar to the one presented in Sec.~\ref{Sec:continuum}, 
but restricted to include only the eleven ensembles that satisfy all the constraints.
We then use the resulting eleven independent measurements of the three masses and three decay constants,
that we report in Table~\ref{tab:datacont},
to perform a simplified global fit based upon the NLO EFT in Ref.~\cite{Bennett:2017kga}.
The results of the fit allow us to provide an extrapolation to the massless limit, 
and to assess the validity of the EFT itself by providing a first estimate of the size of the $g_{VPP}$ coupling
between a vector state V and two pseudoscalar states PS.
We also discuss several phenomenological quantities of general interest, connected with
classical current algebra results, 
such as the aforementioned GMOR relation~\cite{GellMann:1968rz}, 
and the saturation of the Weinberg sum rules~\cite{Weinberg:1967kj}
by the lightest spin-1 states.

\begin{table}
\begin{center}
\begin{tabular}{|c|c|c|c|c|c|c|c|}
\hline\hline
Ensemble & $10~\hat{a}~$ & $10~\hat{m}_{\rm PS}^2~$ & $100~\hat{f}^2_{\rm PS}~$
& $10~\hat{m}_{\rm V}^2~$ & $100~\hat{f}^2_{\rm V}~$
& $~~\hat{m}_{\rm AV}^2~~$ & $100~\hat{f}^2_{\rm AV}$\cr

\hline
DB1M5 & 9.603(17) & 2.512(22) & 1.092(21) & 6.19(13) & 3.24(14) & 1.44(16) & 2.8(11) \cr
DB1M6 & 8.932(11) & 1.873(19) & 0.969(23) & 5.59(12) & 3.15(14) & 1.38(13) & 3.5(8) \cr
DB1M7 & 8.607(9) & 1.549(17) & 0.903(25) & 5.40(13) & 3.13(17) & 1.24(17) & 2.5(10) \cr
DB2M1 & 7.728(12) & 3.20(3) & 1.230(30) & 6.71(15) & 3.35(13) & 1.39(13) & 2.3(7) \cr
DB2M2 & 7.068(11) & 2.192(26) & 0.993(23) & 5.90(17) & 3.17(16) & 1.35(15) & 2.8(8) \cr
DB2M3 & 6.740(6) & 1.637(17) & 0.914(23) & 5.36(15) & 3.05(15) & 1.23(17) & 2.3(10) \cr
DB3M5 & 6.109(11) & 3.68(3) & 1.270(21) & 7.09(17) & 3.25(13) & 1.65(10) & 3.4(6) \cr
DB3M6 & 5.820(11) & 2.94(3) & 1.153(22) & 6.57(16) & 3.35(13) & 1.51(14) & 3.1(8) \cr
DB3M7 & 5.669(6) & 2.574(21) & 1.106(20) & 6.20(16) & 3.25(14) & 1.50(12) & 3.5(7) \cr
DB3M8 & 5.522(7) & 2.105(21) & 1.014(21) & 5.81(17) & 3.11(15) &1.37(14) & 3.1(8) \cr
DB4M2 & 4.466(7) & 3.67(3) & 1.312(20) & 7.24(14) & 3.49(12) & 1.56(10) & 3.1(5) \cr
\hline\hline
\end{tabular}
\end{center}
\caption{
\label{tab:datacont}
Input data used in the continuum EFT global fit.
We consider only ensembles with $\hat{a}\lsim 1$ and $\hat{m}_{\rm PS}^2\lsim 0.4$.
For each ensemble, we subtracted the finite lattice spacing effect,
according to Eqs.~(\ref{eq:f2_chipt}) and~(\ref{eq:m2_chipt}), with the coefficients $W^0_{f,M}$ and $W^0_{m,M}$
obtained by a fit restricted to these ensembles.
}
\end{table}

\subsection{Global fit and low-energy constants}
\label{Sec:global_fit}

The EFT presented in Ref.~\cite{Bennett:2017kga} describes the low-energy 
behaviour of the lightest PS, V, and AV mesons,
by  adopting the ideas of 
HLS~\cite{Bando:1984ej,Casalbuoni:1985kq,Bando:1987br,Casalbuoni:1988xm,Harada:2003jx}. 
The extension of the chiral Lagrangian is achieved by enhancing the $SU(4)$ global symmetry to 
$SU(4)_A\times SU(4)_B$, weakly gauging the $SU(4)_A$ group factor, 
and implementing the spontaneous breaking
to the global $Sp(4)$ by the VEVs of two spin-0 fields subject
to non-linear constraints, and  transforming one on the 
bi-fundamental of $SU(4)_A\times SU(4)_B$ and the other on the
antisymmetric of $SU(4)_A$, respectively. 
In this way, besides the $5$ pNGBs identified with the PS states of the theory, 
the spectrum also contains $15$ massive spin-1 particles, identified with the $10$ V and $5$ AV states.
Explicit breaking of the global symmetry due to the fermion mass $m_f$ 
is implemented by the familiar spurion analysis. The results for physical quantities 
in terms of the $12$ free parameters ($f$, $F$, $b$, $c$, $g_V$, $\kappa$, 
$v$, $v_1$, $v_2$, $v_5$, $y_3$, $y_4$) of the EFT are summarised in Eqs.~(2.19)-(2.24), together with
Eqs.~(2.29), (2.30) and~(2.33) of Ref.~\cite{Bennett:2017kga}.\footnote{We notice an inconsequential typo in Eq.~(2.16), of Ref.~\cite{Bennett:2017kga},
in which the last term should appear with a $+$ sign, 
rather than a $-$ sign, in order to be consistent with Eq.~(2.30).}

The simplified global fit in Ref.~\cite{Bennett:2017kga} 
suffered from uncontrolled systematics 
such as quenching effects and discretisation artefacts.
In this paper, we overcome these limitations with the continuum extrapolations 
of the data obtained by dynamical simulations,
as discussed in previous sections.
A further technical  difficulty of the global fit is due to the large parameter space 
and the limited number of observables available. 
Because we employ Wilson fermions, we have one more unknown fit
 parameter, the critical bare fermion mass $m_0^c$, 
if we use the bare mass $m_0$ for the fermion mass in the EFT. 
A better way to determine the parameters in the EFT might be to consider a 
more sophisticated definition of the fermion mass, 
such as the one calculated via partially conserved axial current (PCAC). 
However, this would require to carry out a more involved computation of the renomalisation factors 
associated with the pseudoscalar operators, which is beyond the scope of this work. 

\begin{figure}
\begin{center}
\includegraphics[width=.49\textwidth]{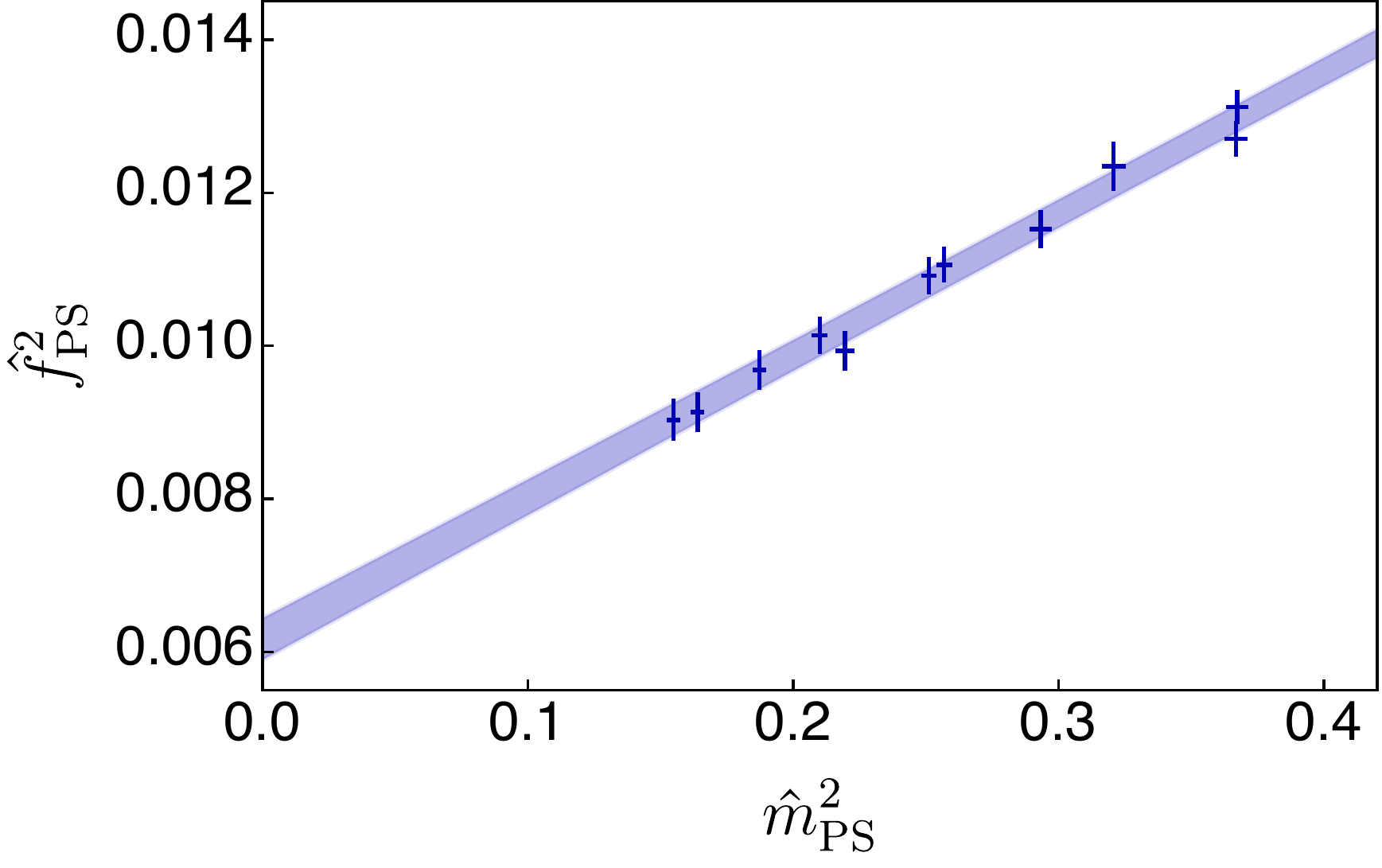}
\includegraphics[width=.49\textwidth]{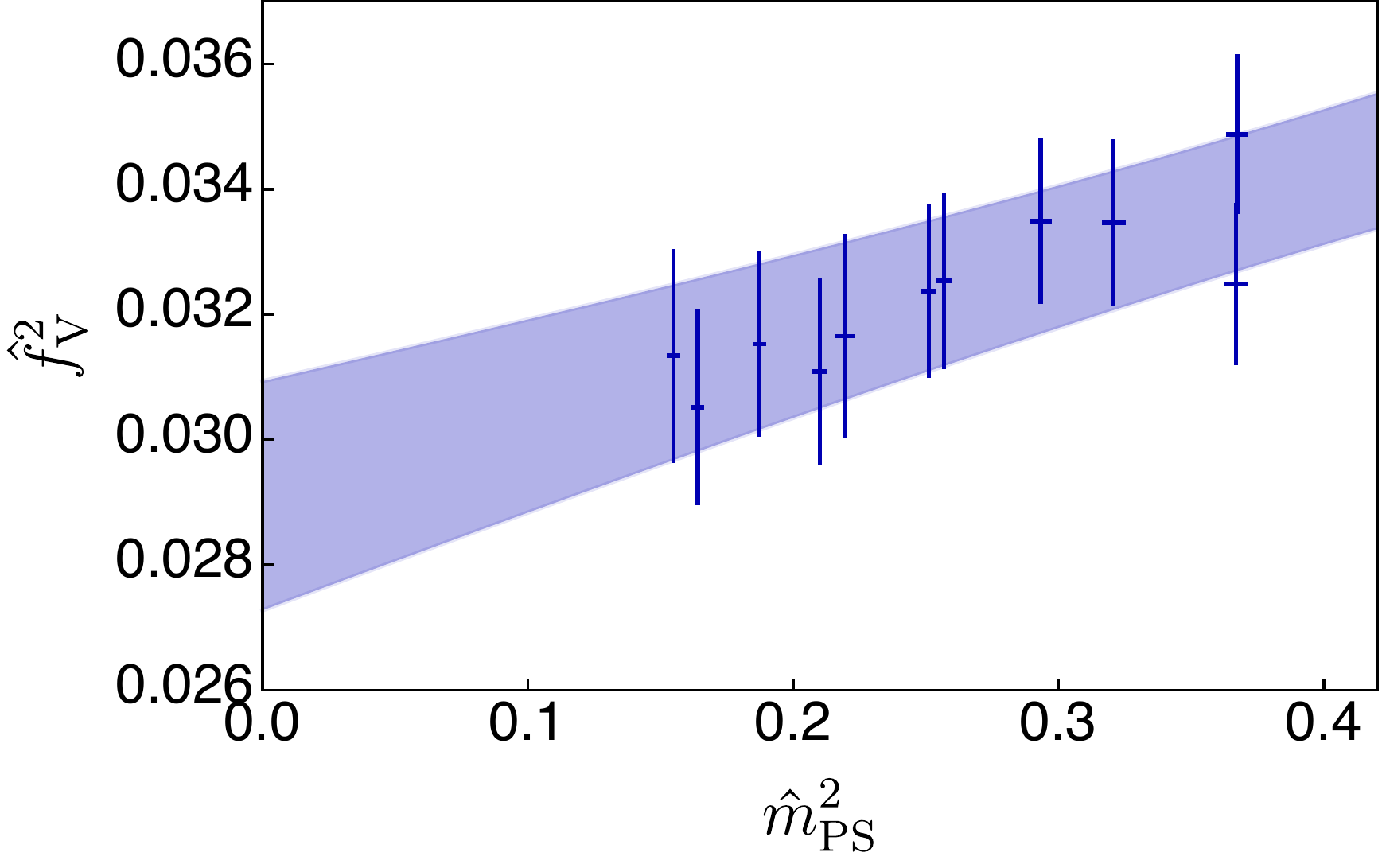}
\includegraphics[width=.49\textwidth]{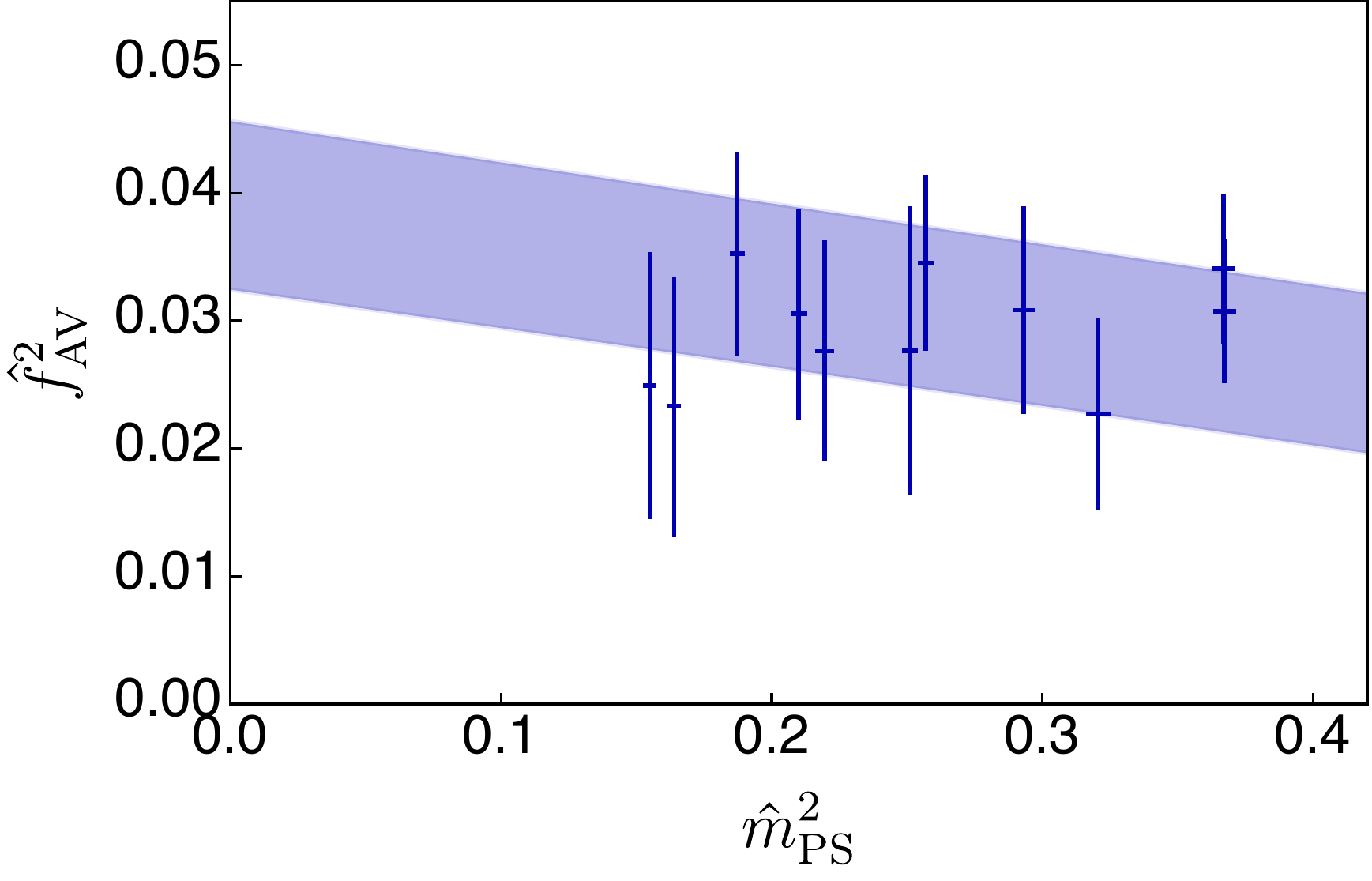}
\caption{%
\label{fig:f_gfit}%
Decay constants squared in the continuum limit---subtracting lattice artefacts due 
to a finite lattice spacing (see Table~\ref{tab:datacont})---as a function of the 
pseudoscalar mass $\hat{m}_{\rm PS}^2$.
The bars represent the statistical uncertainties.
The global fit results are denoted by shaded bands whose widths indicate the statistical errors.
}
\end{center}
\end{figure}

Along the lines discussed in~\Sec{spectroscopy}, 
here we take a different approach.
We eliminate from the analysis any direct reference to the fermion mass, 
effectively  replacing its role
 with the mass of the PS state. Because the linear relation
$m_{\rm PS}^2=2 B m_f$ holds only at leading-order, we restrict our
attention to ensembles for which $\hat{m}_{\rm PS}^2 \lesssim 0.4$ (see Table~\ref{tab:datacont}).
Furthermore, we expand the dependence of the other observable masses 
and decay constants~\cite{Bennett:2017kga}
on $\hat{m}_{\rm PS}^2$, truncating at  linear order.
(We express all quantities in units of $w_0$.)
Up to $\mathcal{O}(\hat{m}_{\rm PS}^4)$, we find 
\beqs
\label{eq:gfit_func1}
\hat{m}_{\rm V}^2&=&\frac{g_{\rm V}^2 (b \hat{f}^2+\hat{F}^2)}{4 (1+\kappa)}+
\frac{2 \hat{v}_1(\kappa+1)-\hat{y}_3(b\hat{f}^2+\hat{F}^2)}{4(\kappa+1)^2}g_{\rm V}^2 \hat{m}_{\rm PS}^2,\\
\label{eq:gfit_func2}
\hat{m}_{\rm AV}^2&=&\frac{(b+4)\hat{f}^2+\hat{F}^2}{4(1-\kappa)}g_{\rm V}^2
+\frac{\left((b+4)\hat{f}^2+\hat{F}^2\right) \hat{y}_4-2(1-\kappa)(\hat{v}_1-2\hat{v}_2)}{4 (1-\kappa)^2} g_{\rm V}^2 \hat{m}_{\rm PS}^2,\\
\label{eq:gfit_func3}
\hat{f}_{\rm V}^2&=&\frac{1}{2}(b\hat{f}^2+\hat{F}^2)+\hat{v}_1 \hat{m}_{\rm PS}^2,\\
\label{eq:gfit_func4}
\hat{f}_{\rm AV}^2&=&\frac{(\hat{F}^2-b \hat{f}^2)^2}{2((b+4)\hat{f}^2+\hat{F}^2)}
-\frac{((3b+8)\hat{v}_1-4(b+2)\hat{v}_2)\hat{f}^2
+\hat{F}^2\hat{v}_1}{((b+4)\hat{f}^2+\hat{F}^2)^2}(\hat{F}^2-b\hat{f}^2)\hat{m}_{\rm PS}^2,\\
\label{eq:gfit_func5}
\hat{f}_{\rm PS}^2&=&\frac{2 \hat{f}^2 ((b+4c+bc) \hat{f}^2+(1+b+c)\hat{F}^2)}{(4+b)\hat{f}^2+\hat{F}^2}
+\\
\nonumber
&&-\frac{4(2+b)\hat{f}^2((2+b)\hat{f}^2\hat{v}_1
+(\hat{F}^2-b\hat{f}^2)\hat{v}_2)}{((4+b)\hat{f}^2+\hat{F}^2)^2}\hat{m}_{\rm PS}^2\,,
\eeqs
having make use of the redefinitions
\beqs
\hat{f}\equiv fw_0,~\hat{F}\equiv F w_0,~\hat{y}_3\equiv \frac{y_3}{2 \hat{B} w_0},~\hat{y}_4\equiv\frac{y_4}{2 \hat{B} w_0},
~\hat{v}_1\equiv\frac{v_1 w_0}{2 \hat{B}},~\hat{v}_2\equiv\frac{v_2 w_0}{2 \hat{B}},
\eeqs
and
\beqs
 \hat{B}\equiv  B w_0\,,~~~~~~~~B\equiv v^3/(2 f_{\rm PS}^2)\,.
 \eeqs 
This linearised ansatz involves $10$ unknown parameters ($\hat{f}$, $\hat{F}$, ${b}$, ${c}$, ${g}_V$, ${\kappa}$, 
 $\hat{v}_1$, $\hat{v}_2$, $\hat{y}_3$, $\hat{y}_4$) to be determined from $5$ measurements. 
The remaining two parameters ($v$ and $v_5$) are associated with the 
 GMOR relation, as they enter at NLO  
the definition of  $\hat{m}_{\rm PS}^2$.
We will return to this point later on.

\begin{figure}
\begin{center}
\includegraphics[width=.49\textwidth]{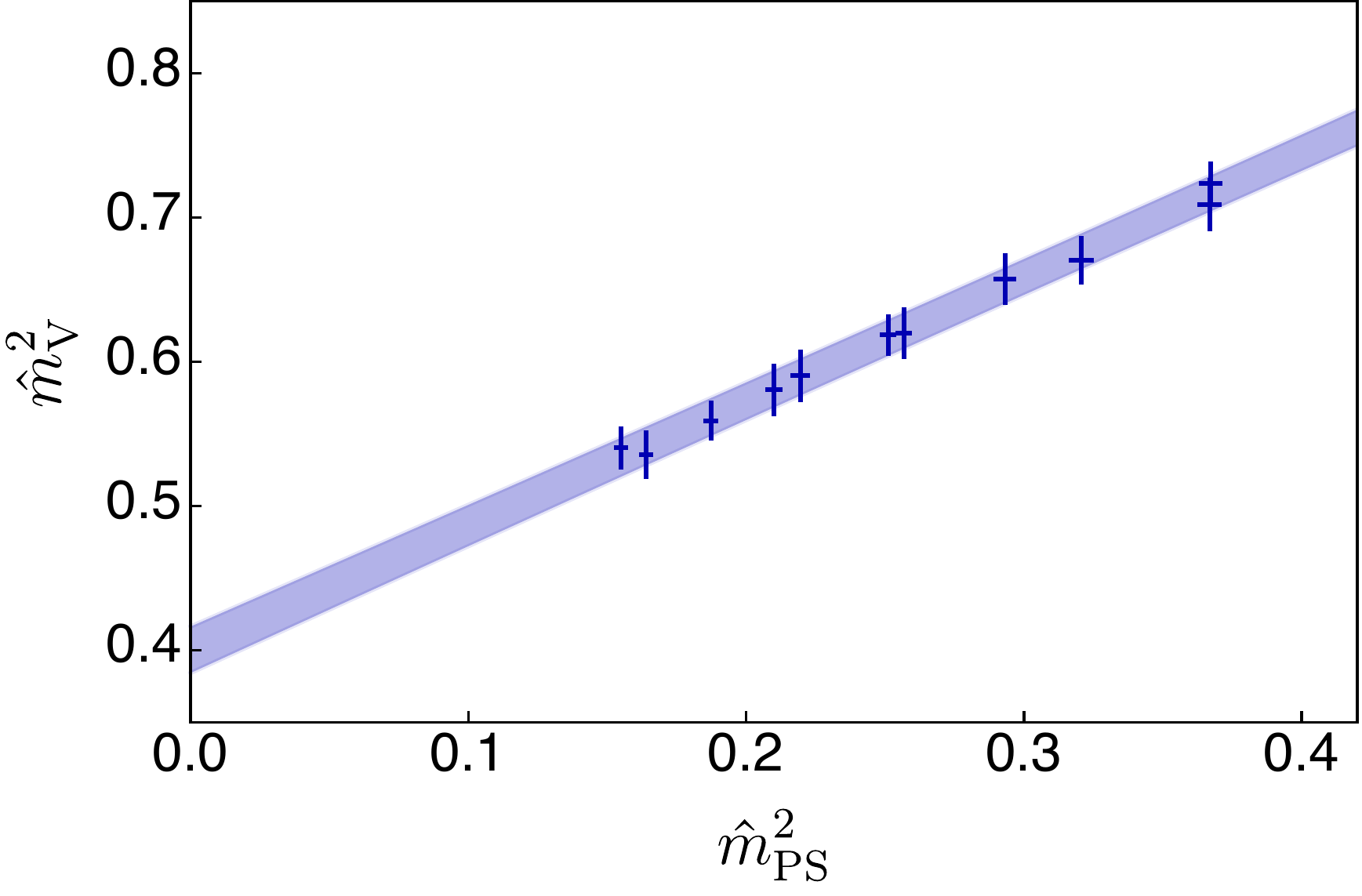}
\includegraphics[width=.49\textwidth]{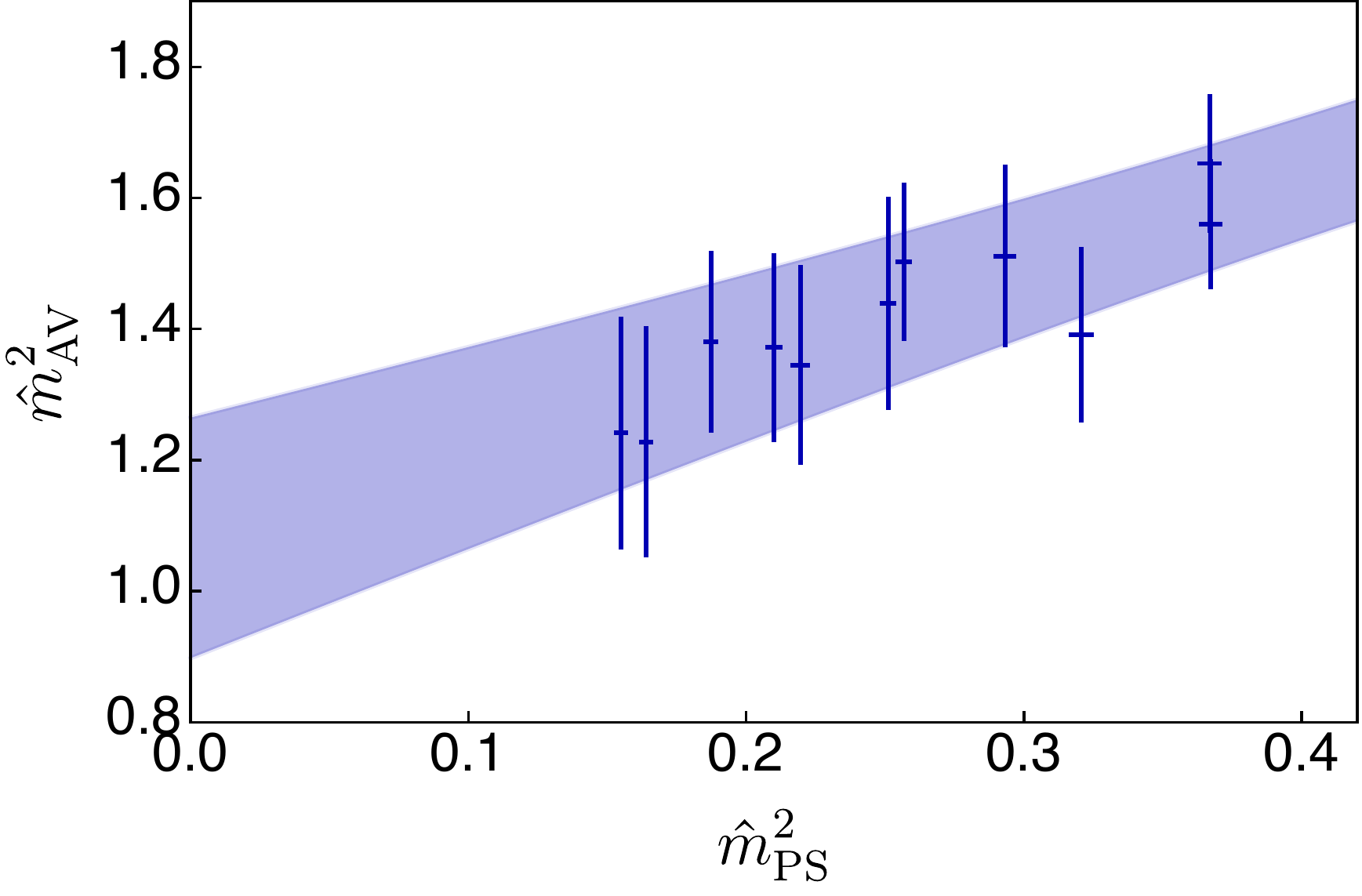}
\caption{%
\label{fig:m_gfit}%
Masses squared in the continuum limit---subtracting lattice artefacts due to a finite lattice spacing
 (see Table~\ref{tab:datacont})---as a function of the pseudoscalar mass $\hat{m}_{\rm PS}^2$.
The bars represent the statistical uncertainties.
The global fit results are denoted by shaded bands whose widths indicate the statistical errors. 
}
\end{center}
\end{figure}

We perform a global fit using the functions in Eqs.~(\ref{eq:gfit_func1})-(\ref{eq:gfit_func5}).
The $\chi^2$ function is defined in a simplified manner, by summing the individual 
$\chi^2$ obtained from the five 
independent fit equations, and hence ignoring correlations among the fit equations.
Besides subtracting the discretisation effects $W^0_{m,M}\hat{a}$ and $W^0_{f,M}\hat{a}$
 from the original data,
and restricting consideration to $\hat{m}_{\rm PS}^2 \lesssim 0.4$ and $\hat{a} \lesssim 1$, as anticipated,
we also constrain the fit by implementing the following conditions,
 which are the consequences of  unitarity~\cite{Bennett:2017kga}:
\beqs
1 &>& \kappa+\hat{m}_{\rm PS}^2 \hat{y}_4, \nn\\
-1 &<& \kappa+\hat{m}_{\rm PS}^2 \hat{y}_3, \nn \\
0 &<& b\hat{f}^2+\hat{F}^2+2 \hat{m}_{\rm PS}^2 \hat{v}_1, \nn \\
0 &<& 2+b+c+(b+4c)\frac{\hat{f}^2}{\hat{F}^2}-2\hat{m}_{\rm PS}^2 \hat{v}_1, \\
0 &<& b\left((c+1)\hat{f}^2+\hat{F}^2-2 \hat{m}_{\rm PS}^2 \hat{v_1} +2 \hat{m}_{\rm PS}^2 \hat{v}_2\right)+ \nn \\
&&+c\left(4\hat{f}^2+\hat{F}^2-2 \hat{m}_{\rm PS}^2 \hat{v_1}+4\hat{m}_{\rm PS}^2 \hat{v}_2\right) 
-\frac{\hat{m}_{\rm PS}^4\hat{v}_2^2}{\hat{f}^2}+\hat{F}^2-2 \hat{m}_{\rm PS}^2 \hat{v}_1. \nn
\eeqs

\begin{table}
\begin{center}
\begin{tabular}{|c|c|c|c|c|}
\hline\hline
 & $~~\hat{f}^{2,\,\chi}_M~~$ & $~~~L^0_{f,M}~~~$
 & $~~\hat{m}^{2,\,\chi}_M~~$ & $~~~L^0_{m,M}~~~$ \cr \hline
PS & $0.00617(28)(36)$ & $3.02(22)(35)$ 
&  &  \cr
V & $0.0291(18)(11)$ & $0.45(16)(14)$ 
& $0.400(16)(10)$ & $2.16(15)(9)$ \cr
AV & $0.039(7)(2)$ & $-0.82(15)(8)$ 
& $1.07(19)(8)$ & $1.42(6)(3)$ \cr
\hline\hline
\end{tabular}
\end{center}
\caption{
\label{tab:global_LECs}
Coefficients appearing in Eqs.~(\ref{eq:f2_chipt}) and (\ref{eq:m2_chipt}), 
determined by using the results of the  global fit
described in the main text. The results are compatible with those obtained  
with the alternative fit process in Table~\ref{tab:LECs},
except for the coefficient $L^0_{f, {\rm AV}}$.
}
\end{table}

In Figs.~\ref{fig:f_gfit} and~\ref{fig:m_gfit} we present the continuum 
values of the masses and decay constants 
from Table~\ref{tab:datacont}, along with the results of the global fits, represented by blue bands, 
obtained 
through a constrained bootstrapped $\chi^2$ minimisation with 10 parameters, 
with the constraints guided by an initial minimisation of the full data set.
The fit functions describe the data well,  with $\chi^2/N_{\rm d.o.f}\sim 0.4$. 
But we also find that some of the EFT parameters are not well 
determined (see discussion in \App{LECs_hist},
in particular the histograms of $200$ resampled data obtained by bootstrapping, 
 for the LO parameters in \Fig{LEC_LO_hist} and 
for the NLO ones in \Fig{LEC_NLO_hist}), indicating the presence of flat directions in the $\chi^2$.

The fact that the EFT yields a good quality fit of the masses and decay constants 
justifies using its results to calculate other physical quantities. 
The  coupling $g_{\rm VPP}$
is responsible for the decay of the vector meson into two pseudoscalar mesons.
The NLO EFT expression can be found in~\cite{Bennett:2017kga}
 (where $g_{\rho\pi\pi}$ denotes $g_{\rm VPP}$)
and  in the massless limit we have that the $g_{\rm VPP}^{\chi}$ coupling is
\beq
g_{\rm VPP}^\chi=\frac{g_{\rm V} (b+2) (2\hat{f}^2+\hat{F}^2)(b\hat{f}^2+\hat{F}^2)}
{((b+4)\hat{f}^2+\hat{F}^2)((b+(b+4)c)\hat{f}^2+(b+c+1)\hat{F}^2)\sqrt{1+\kappa}}\,.
\label{eq:gvpp_massless}
\eeq
We find that  $g_{\rm VPP}^\chi=6.0(4)(2)$---see also \App{LECs_hist}, in particular 
 the histogram of the coupling in \Fig{gpp_hist}, which shows a nice gaussian distribution.

As a consistency check, we also calculate the relevant coefficients 
in Eqs.~(\ref{eq:f2_chipt}) and~(\ref{eq:m2_chipt}),  by using the results of the global fit. 
We present the results in \Tab{global_LECs}: 
they are consistent with the ones in \Tab{LECs}, with the only exception of  
$L^0_{f, {\rm AV}}$ which is over-constrained by Eqs.~(\ref{eq:gfit_func1})-(\ref{eq:gfit_func5}),
but also affected by larger systematic as well as statistical uncertainties.

The results of this exercise have to be taken with caution, for a number of reasons.
The fermion masses considered are not small enough to fully justify the linearisation 
of the fit equations, nor the truncation of the EFT itself to exclude higher-order terms.
This is also clear from the fact that the masses considered are such that the V mesons
are stable, because the 2-PS channel is closed.
These objections can be addressed in the future, by lowering the fermion mass studied on the lattice.
Also, the coupling $g_{\rm V}$ (or alternatively the coupling $g_{\rm VPP}$)
turns out not to be small, bring into question the truncation of the EFT.
This coupling should be suppressed if we consider future lattice studies of $Sp(2N)$ theories with $N\geq3$.

\subsection{GMOR relation and Weinberg sum rules}
\label{Sec:GMORWein}

\begin{figure}
\begin{center}
\includegraphics[width=.65\textwidth]{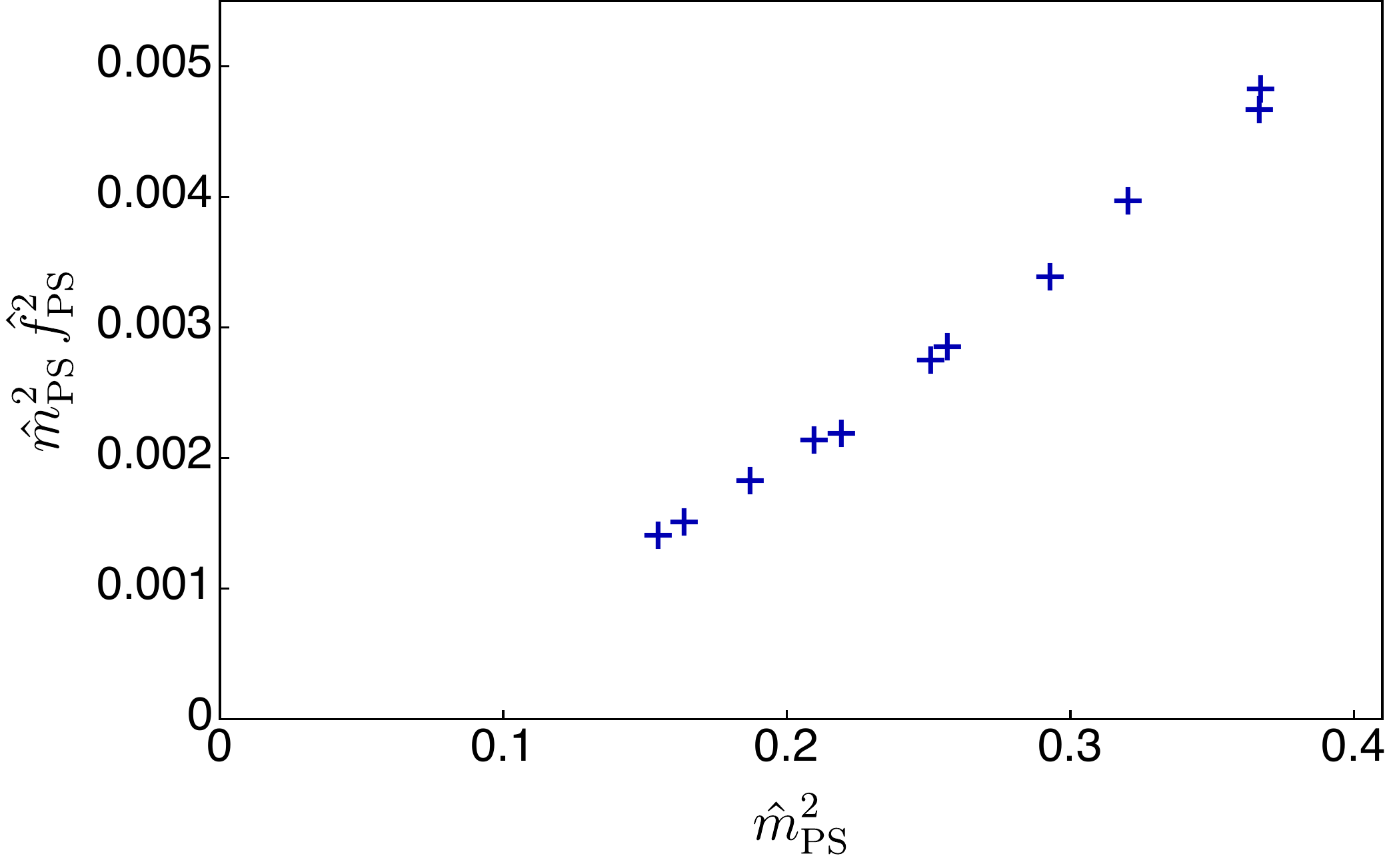}
\end{center}
\caption{
The quantity $\hat{m}_{\rm PS}^2 \hat{f}_{\rm PS}^2$ entering the GMOR relation,
 for the data set of $\hat{m}_{\rm PS}^2 \lesssim 0.4$ and $\hat{a} \lesssim 1$, in the continuum limit
(see Table~\ref{tab:datacont}).
}
\label{fig:GMOR}
\end{figure}

There are several consequences of the HLS EFT (and of current algebra) that can be tested on our lattice data.
If we first restrict our attention to the PS sector, 
we have the GMOR relation, which at NLO takes the corrected form 
\beq
m_{\rm PS}^2 f_{\rm PS}^2=m_f (v^3+m_f v_5^2),
\label{eq:GMOR}
\eeq
where $v$ and $v_5$ are the two low-energy constants defined in
 Ref.~\cite{Bennett:2017kga} that we removed from our global fit. 
As noted in~\Sec{spectroscopy}, we cannot extract $v$ and $v_5$ 
without renormalisation of the fermion mass.
In \Fig{GMOR} we plot the numerical results of 
$\hat{m}_{\rm PS}^2 \hat{f}_{\rm PS}^2$ with respect to $\hat{m}_{\rm PS}^2$,
 as an illustration of Eq.~(\ref{eq:GMOR}).

\begin{figure}
\begin{center}
\includegraphics[width=.49\textwidth]{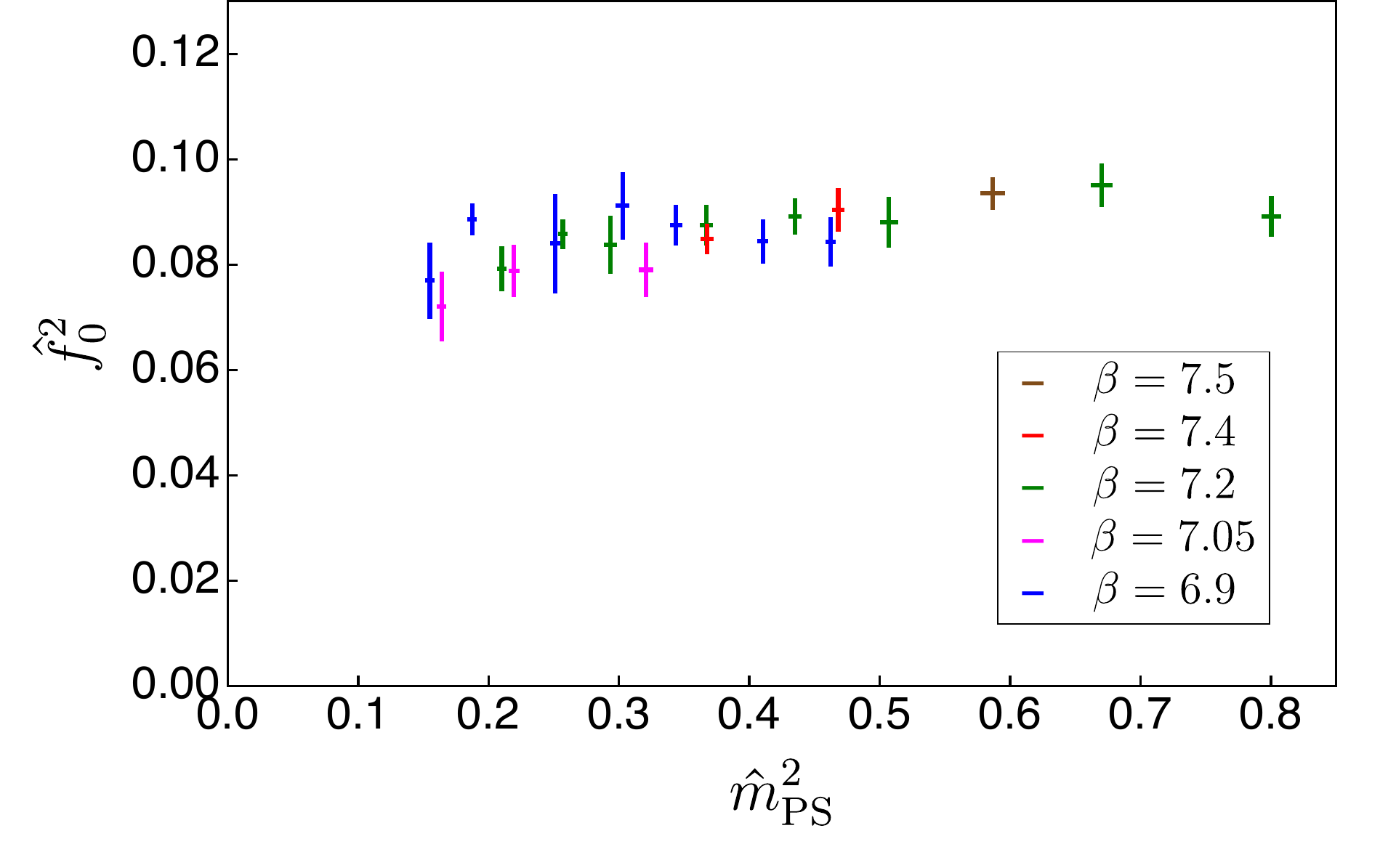}
\includegraphics[width=.49\textwidth]{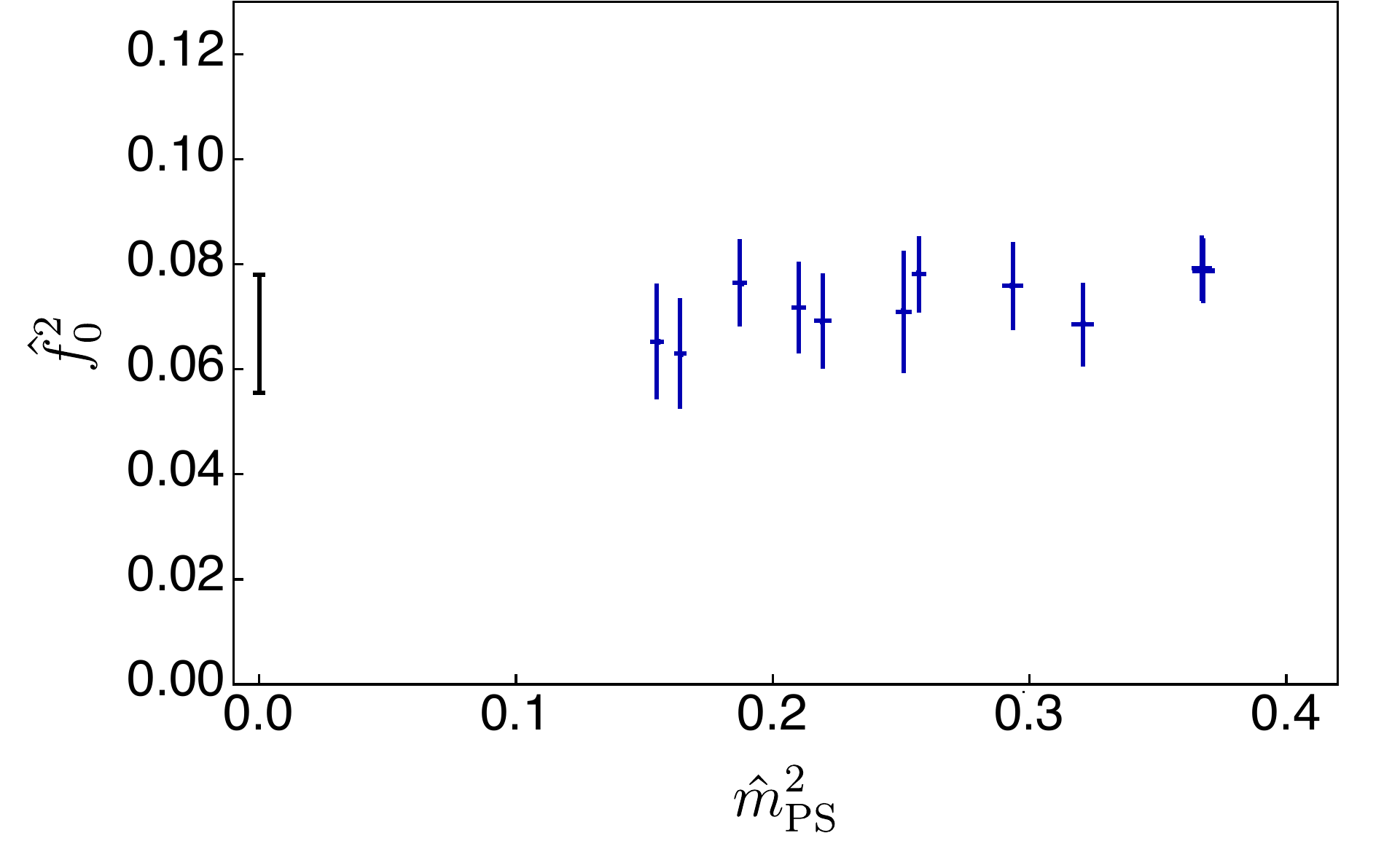}
\includegraphics[width=.49\textwidth]{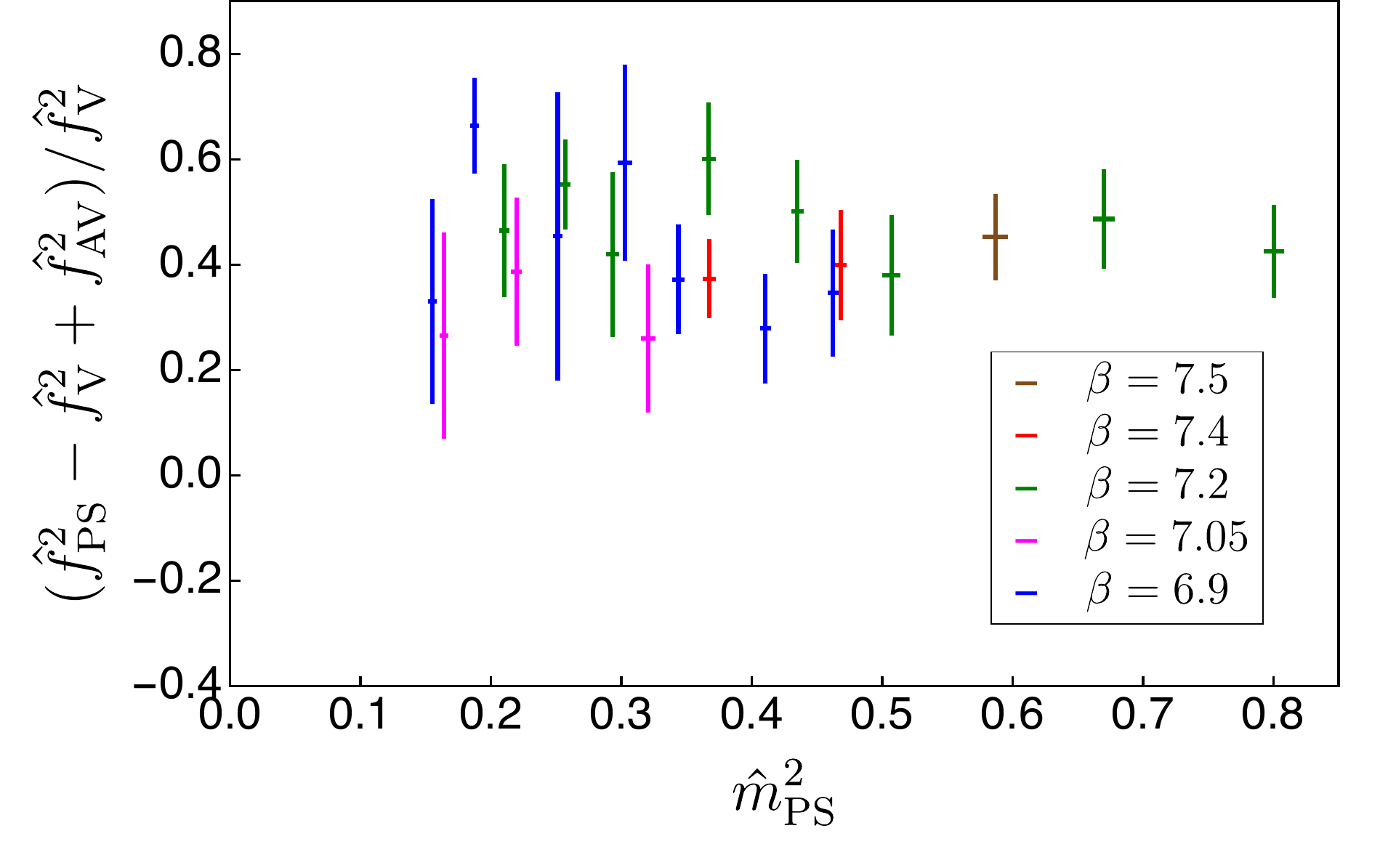}
\includegraphics[width=.49\textwidth]{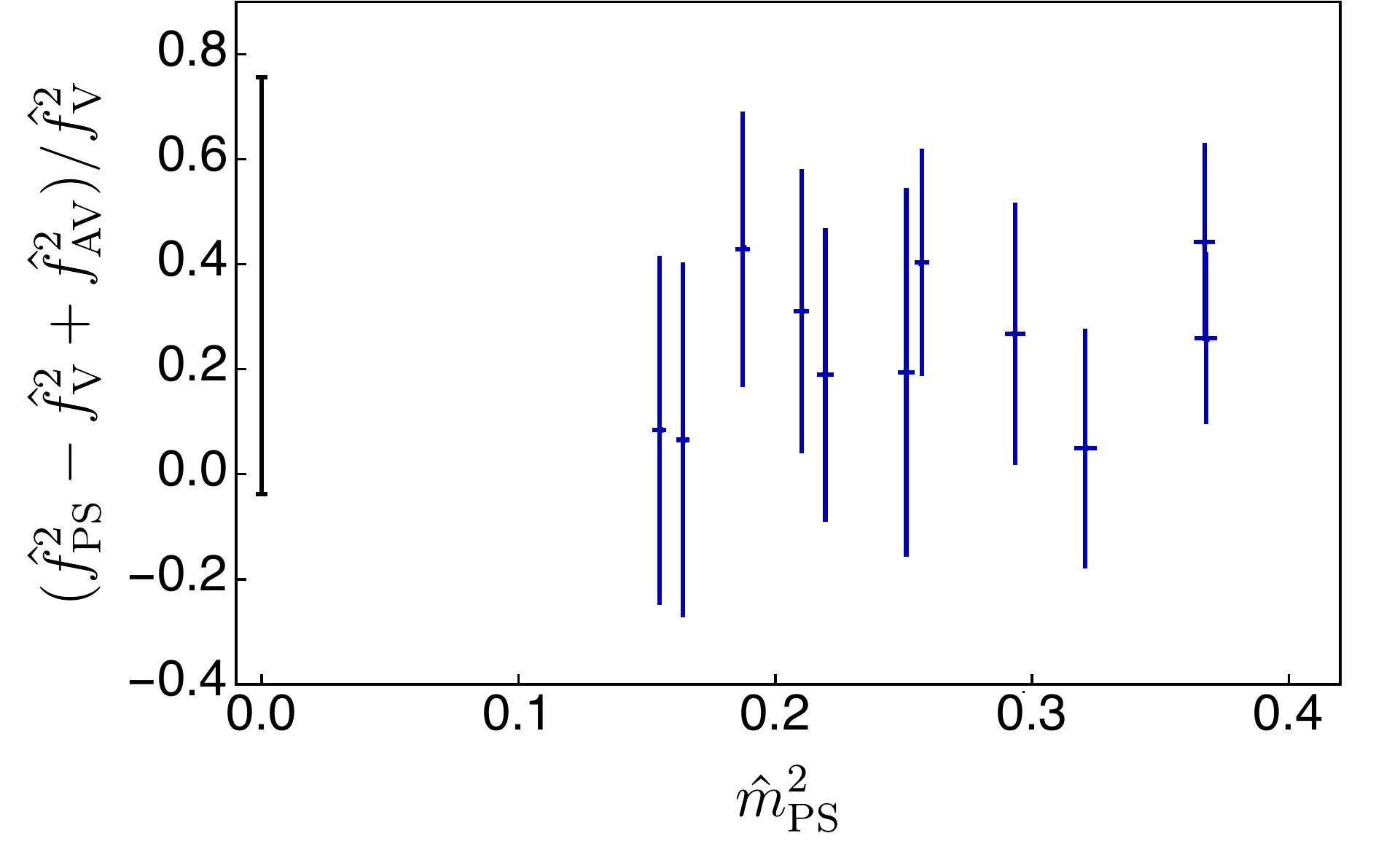}
\includegraphics[width=.49\textwidth]{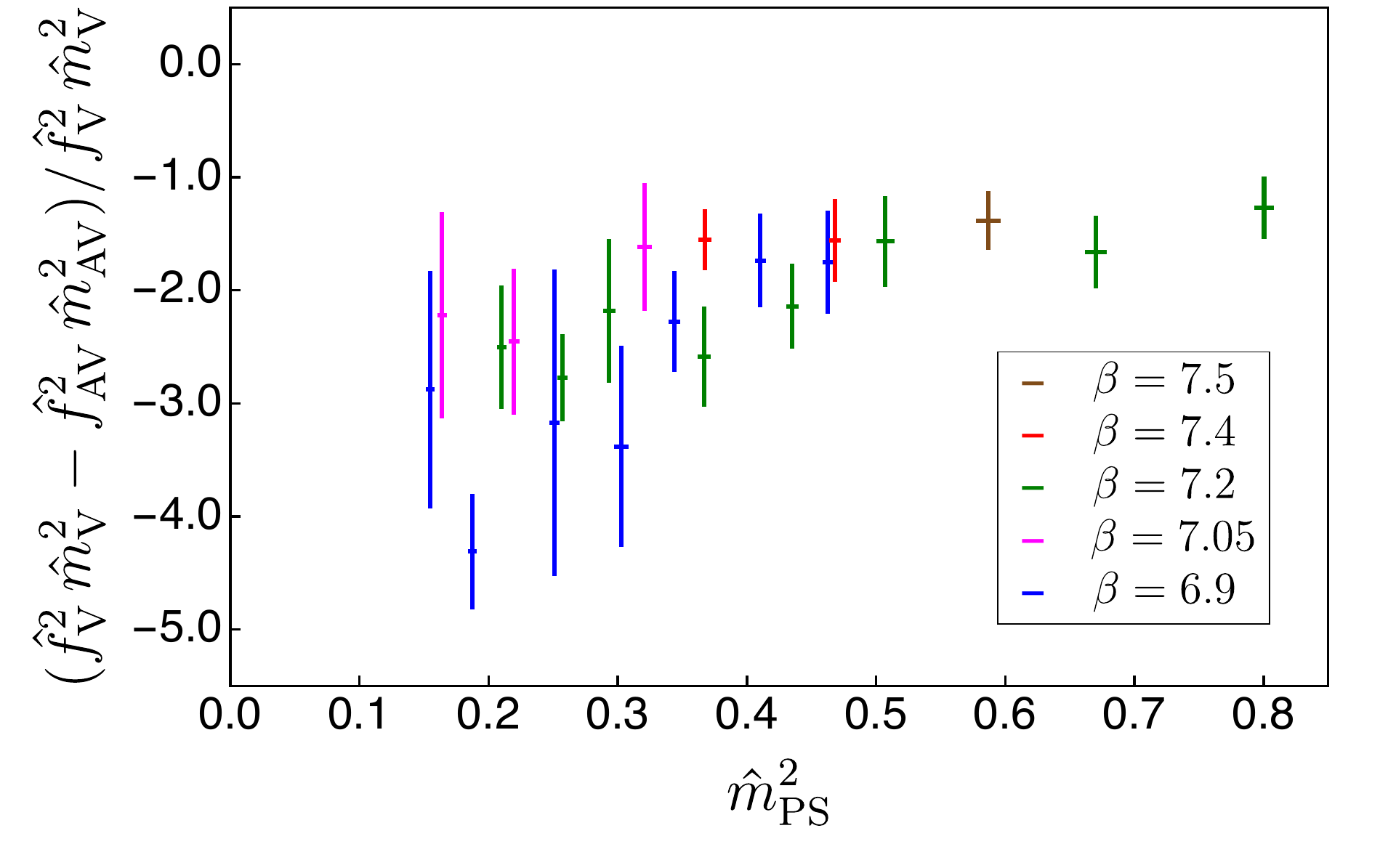}
\includegraphics[width=.49\textwidth]{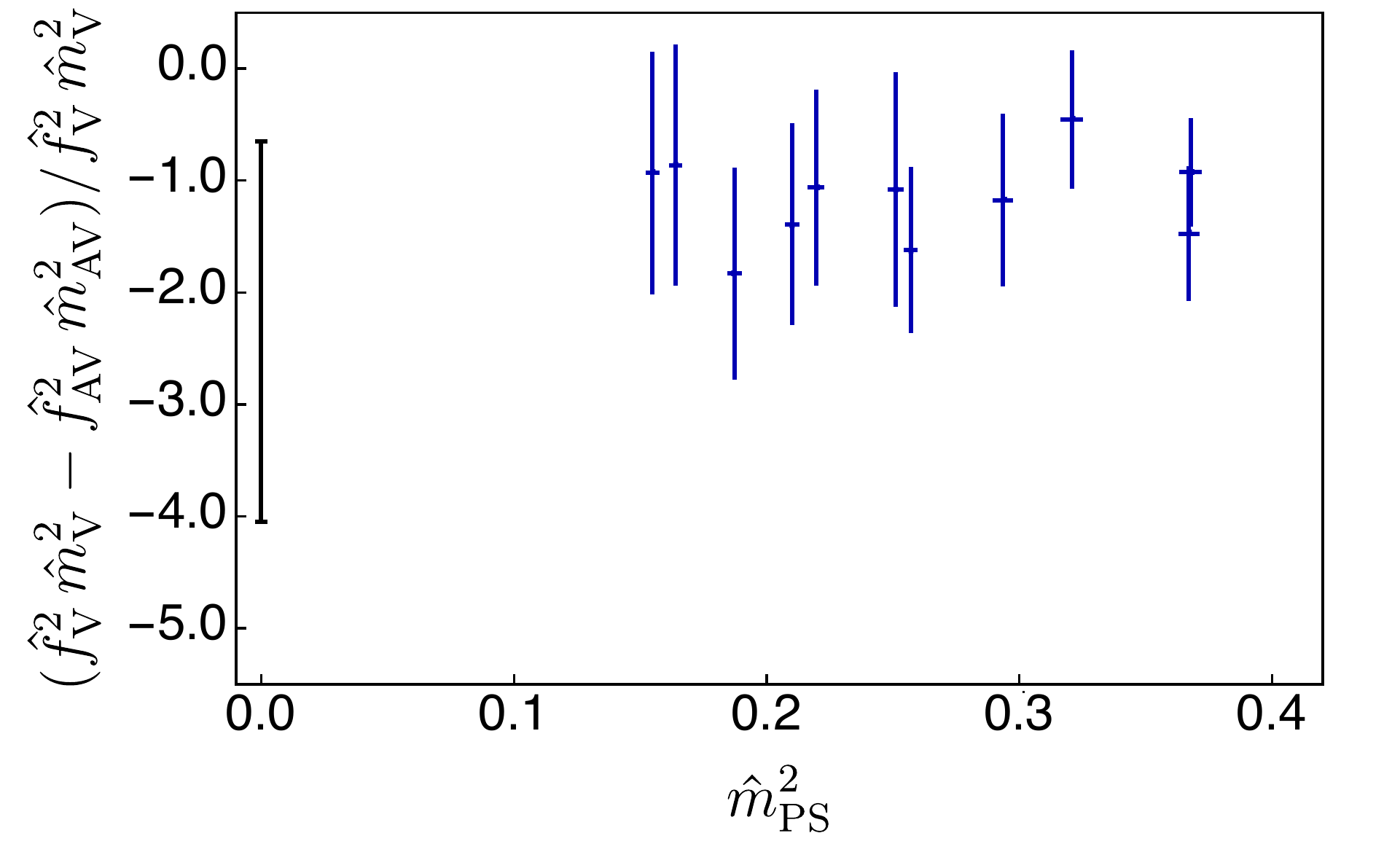}
\end{center}
\caption{
\label{fig:f0}
Top to bottom: $\hat{f}_0^2$, the first, and the second Weinberg sum rule,
saturated over the lightest spin-1 states, as a function of $\hat{m}_{\rm PS}^2$.
Left panels: we do not  perform the continuum extrapolation, colour coding represents 
the bare lattice coupling of the various ensembles.
Blue, purple, green, red and brown colours
have $\beta=6.9$, $7.05$, $7.2$, $7.4$ and $7.5$, respectively. 
Right panels: continuum-extrapolation  results as in Table~\ref{tab:datacont} (in blue),
and the results of  the continuum and massless extrapolation (in black). Statistical and systematic
errors are summed in quadrature.
}
\end{figure}

Going beyond the pseudoscalar sector, 
the first non-trivial prediction of the NLO EFT---related to
some reasonable assumptions for the truncation of the series of operators---is 
that the sum of the decay constants (at zero momentum) 
\beqs
f_0^2&\equiv& f_{\rm PS}^2+f_{\rm V}^2+f_{\rm AV}^2\,,
\eeqs
is expected to be independent on the fermion mass.
In the top-left panel of \Fig{f0} we plot the numerical results
of $\hat{f}_0^2$ with respect to $\hat{m}_{\rm PS}^2$,
with the continuum extrapolation shown in the top-right panel. 
As seen in the figure, at the level of the present precision  the numerical results 
support the mass independence of $f_0^2$ 
over the range of mass considered.

While the Weinberg spectral theorems, which involve integration over all momenta, are exact, 
because they reflect the properties of  the underlying condensates in
the theory~\cite{Weinberg:1967kj},  
the saturation of the resulting sum rules over a finite number among 
the lightest spin-1 states is an 
approximation. 
The  NLO EFT analysis captures such violations 
in the non-nearest-neighbour interactions appearing
in the HLS Lagrangian.
These terms are generated by integrating out heavier resonant spin-1 states 
that contribute to the dispersion relations.
Saturated over the lightest resonances, the first and second sum rules are 
\beqs
f_{\rm AV}^2-f_{\rm V}^2+f_{\rm PS}^2&=&0\,,
\eeqs 
and 
\beqs
f_{\rm V}^2 m_{\rm V}^2 - f_{\rm AV}^2 m_{\rm AV}^2&=&0\,,
\eeqs
respectively. The aforementioned corrections give finite contributions 
 to the right-hand-side of these two relations.
The numerical results for these quantities, normalised 
by $f_{\rm V}^2$ and $f_{\rm V}^2 m_{\rm V}^2$, respectively,
are shown in the middle and bottom panels of \Fig{f0}. 
As  the figures show, after  continuum extrapolation the results are closer to saturating 
the Weinberg sum rules. The conservative estimate of the uncertainties
includes both statistical and systematic error, 
and with its current size this comparison is somewhat inconclusive.

\section{Comparison to other gauge theories}
\label{Sec:discussion}

\begin{table}
\begin{center}
\begin{tabular}{|c|c|c|c|c|c|c|c|}
\hline\hline
Ensemble & $\hat{m}_{\rm V}/\hat{m}_{\rm PS}$ & $\hat{m}_{\rm PS}/\hat{f}_{\rm PS}$ & 
$\hat{m}_{\rm V}/\hat{f}_{\rm PS}$ 
& $\hat{m}_{\rm AV}/\hat{f}_{\rm PS}$ & $\hat{m}_{\rm S}/\hat{f}_{\rm PS}$ & $\hat{f}_{\rm V}/\hat{f}_{\rm PS}$ & $\hat{f}_{\rm AV}/\hat{f}_{\rm PS}$ \\
\hline
DB1M5 & 1.569(18) & 4.80(5) & 7.53(13) & 11.5(6) & 10.9(6) & 1.72(4) & 1.6(3)\\
DB1M6 & 1.728(20) & 4.40(6) & 7.60(14) & 11.9(6) & 12.3(6) & 1.80(4) & 1.90(21)\\
DB1M7 & 1.867(26) & 4.14(6) & 7.74(15) & 11.7(8) & 11.9(9) & 1.86(5) & 1.6(4)\\
DB2M1 & 1.446(18) & 5.10(6) & 7.37(13) & 10.6(5) & 10.7(5) & 1.65(3) & 1.34(23)\\
DB2M2 & 1.640(26) & 4.70(6) & 7.71(16) & 11.6(7) & 12.2(5) & 1.78(5) & 1.65(26)\\
DB2M3 & 1.807(28) & 4.24(5) & 7.66(16) & 11.6(8) & 12.5(6) & 1.83(5) & 1.6(4)\\
DB3M5 & 1.390(17) & 5.38(5) & 7.47(12) & 11.4(4) & 11.1(4) & 1.60(3) & 1.63(14)\\
DB3M6 & 1.497(19) & 5.05(5) & 7.55(13) & 11.4(5) & 10.7(5) & 1.70(3) & 1.62(21)\\
DB3M7 & 1.553(21) & 4.82(5) & 7.49(13) & 11.6(5) & 11.5(5) & 1.72(4) & 1.76(17)\\
DB3M8 & 1.662(25) & 4.55(5) & 7.57(15) & 11.6(6) & 11.1(6) & 1.75(4) & 1.72(23)\\
DB4M2 & 1.403(15) & 5.29(4) & 7.43(10) & 10.9(3) & 10.9(4) & 1.63(3) & 1.53(14)\\
\hline
Massless & N/A & N/A & 8.08(32) & 13.4(1.5) & 14.2(1.7) & 2.15(8) & 2.3(4)\\
\hline\hline
\end{tabular}
\end{center}
\caption{%
\label{tab:meson_ratio}%
Ratios of meson masses and decay constants, extrapolated to the continuum limit
by the subtraction method explained in Section~\ref{Sec:continuum}.
We restrict the data to the eleven ensembles for which the subtraction of  effects 
due to the finite lattice spacing can be done for all measurable quantities.
In the last row, we report the  results of the continuum and massless extrapolation, 
obtained by applying Eqs.~(\ref{eq:f2_chipt}) and~(\ref{eq:m2_chipt}) to the eleven ensembles.
Statistical and systematic uncertainties have been added in quadrature.
}
\end{table}

\begin{figure}
\begin{center}
\includegraphics[width=.49\textwidth]{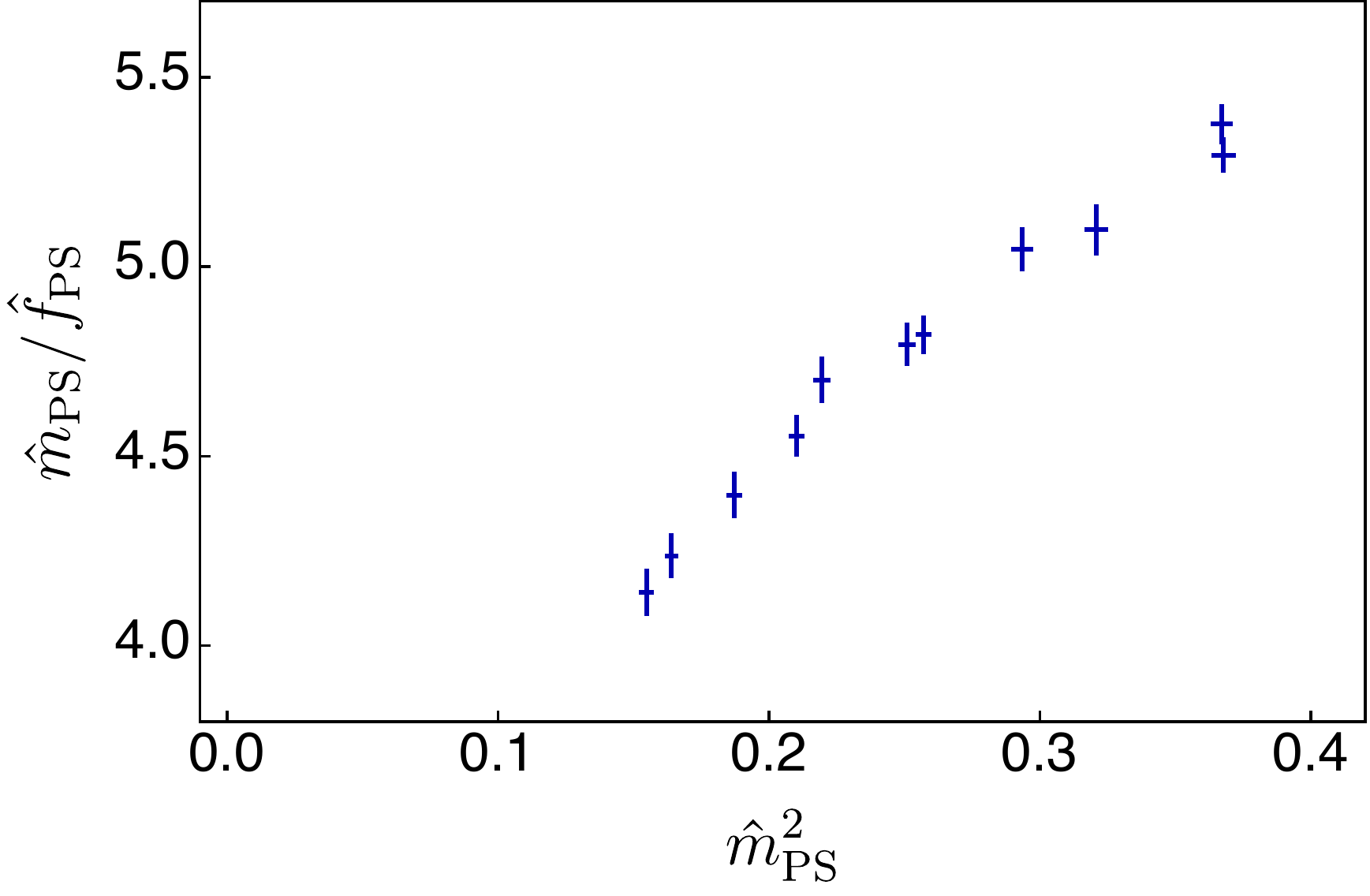}
\includegraphics[width=.49\textwidth]{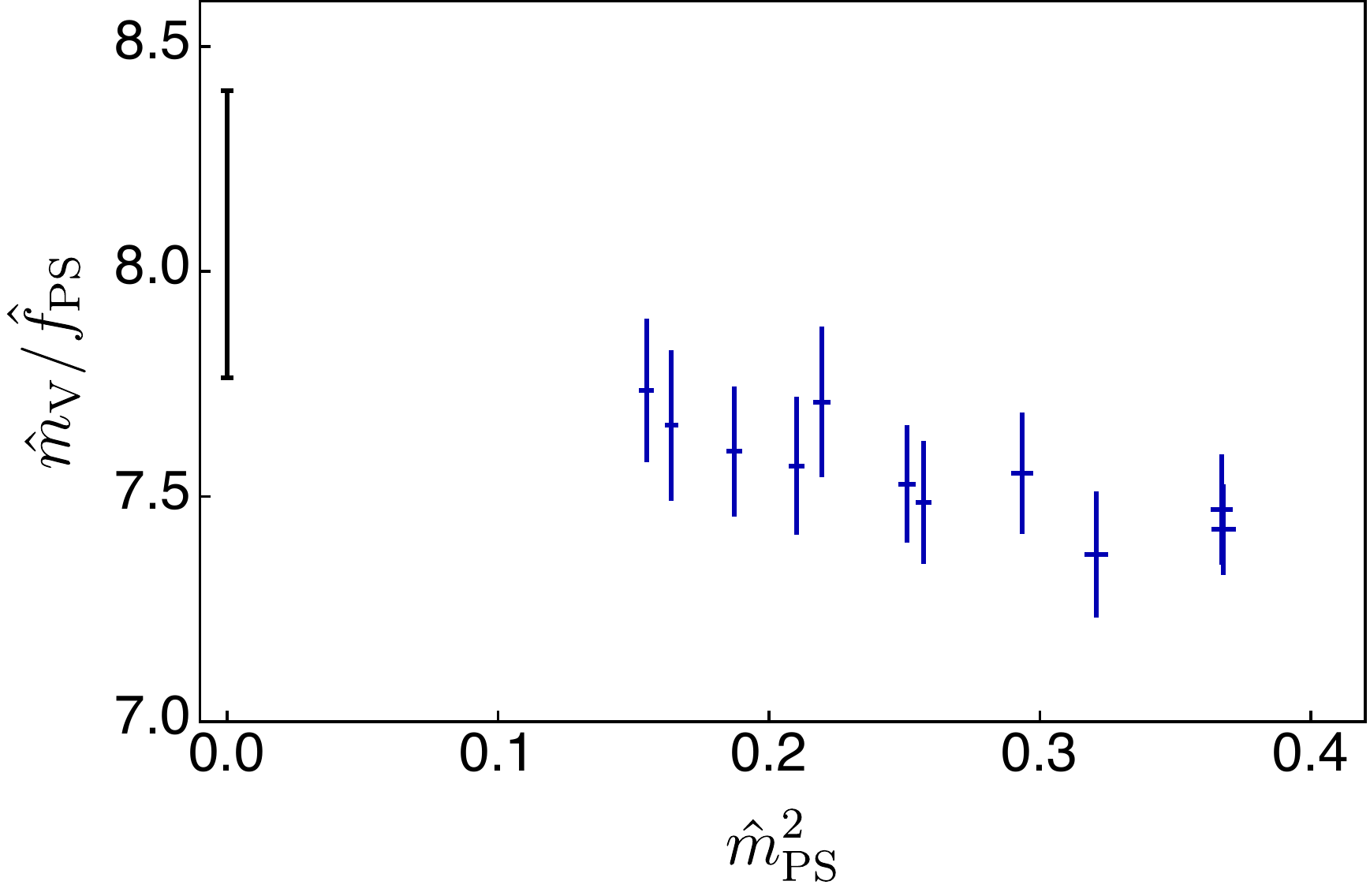}
\includegraphics[width=.49\textwidth]{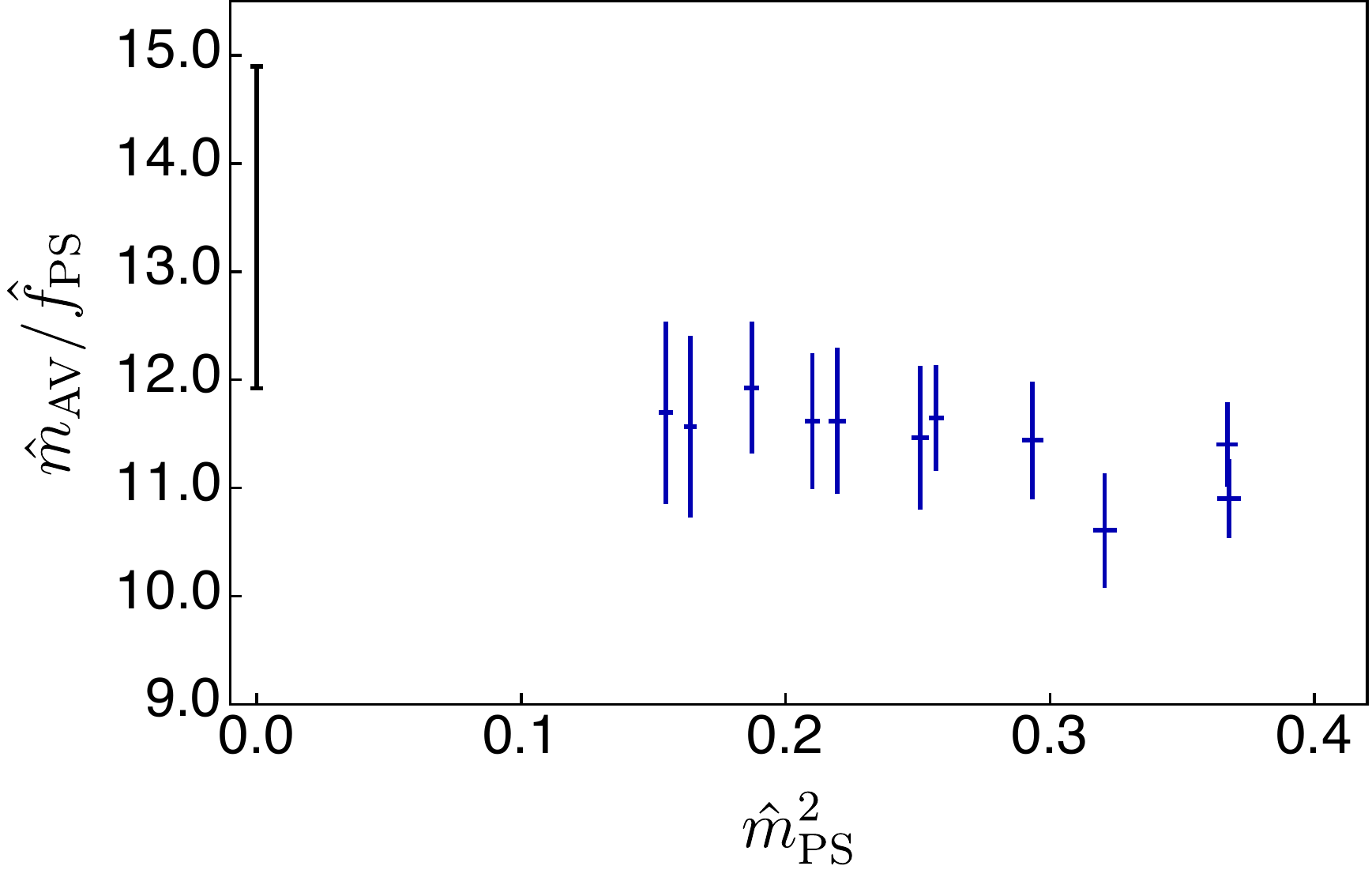}
\includegraphics[width=.49\textwidth]{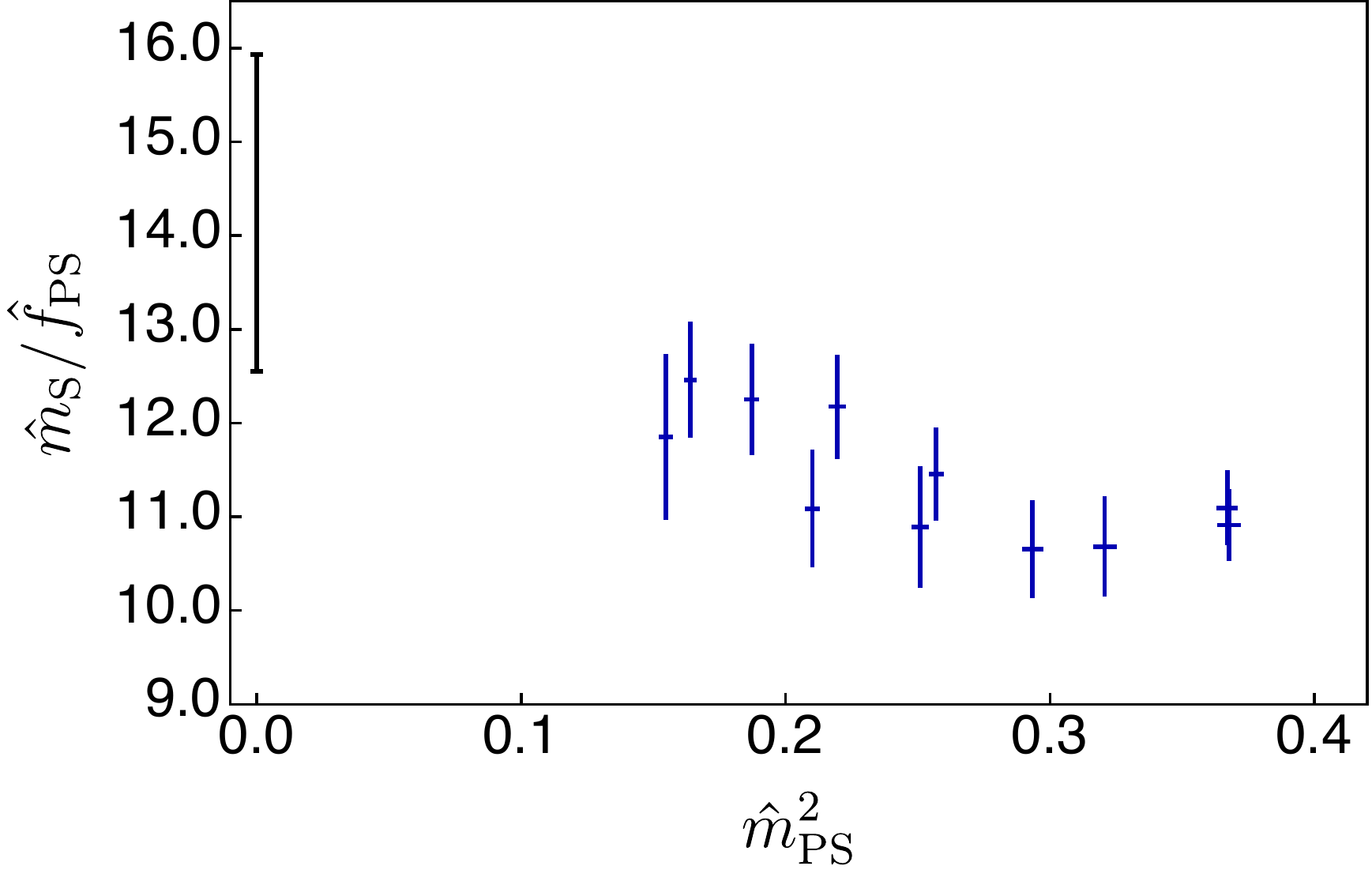}
\includegraphics[width=.49\textwidth]{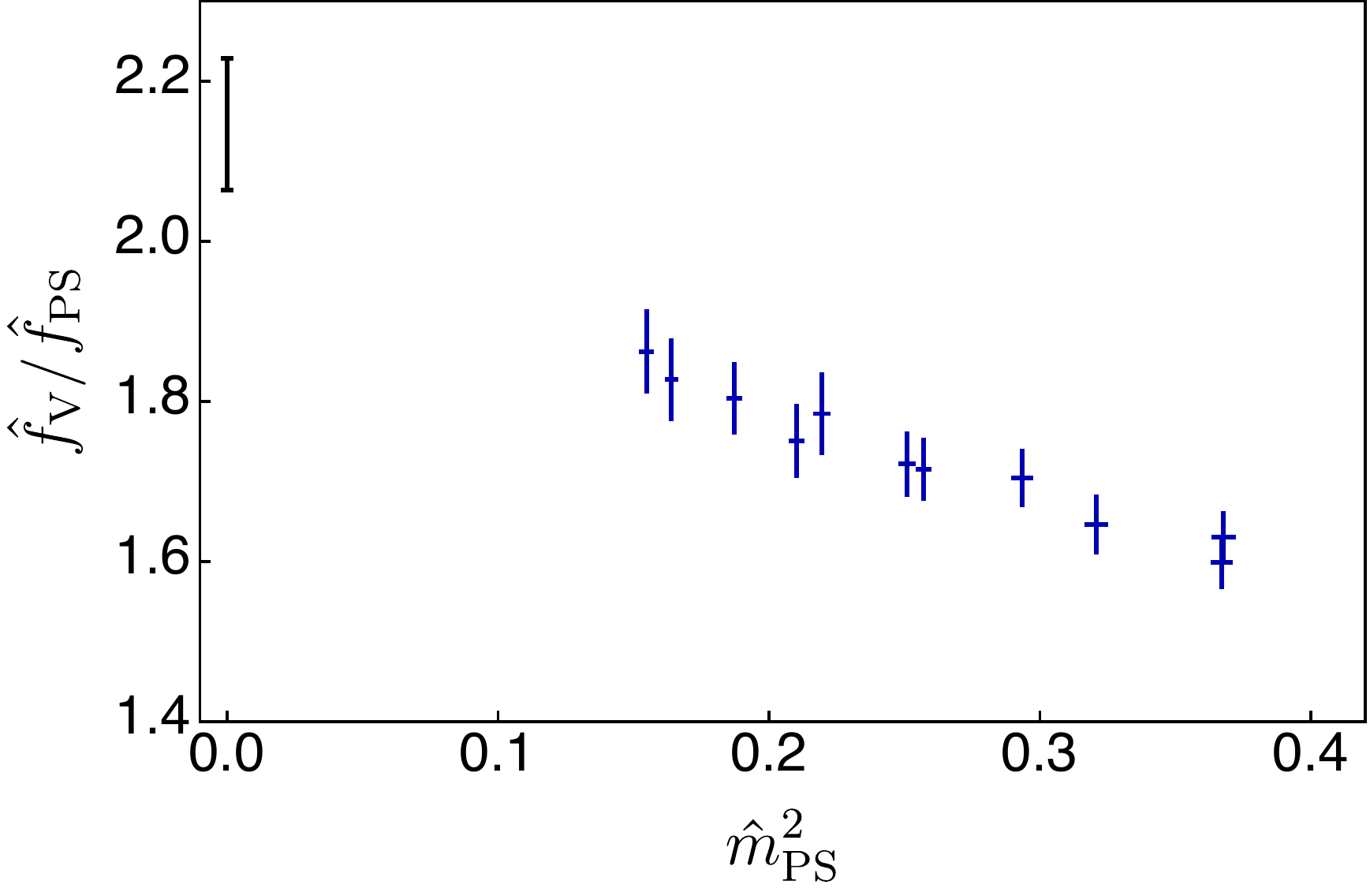}
\includegraphics[width=.49\textwidth]{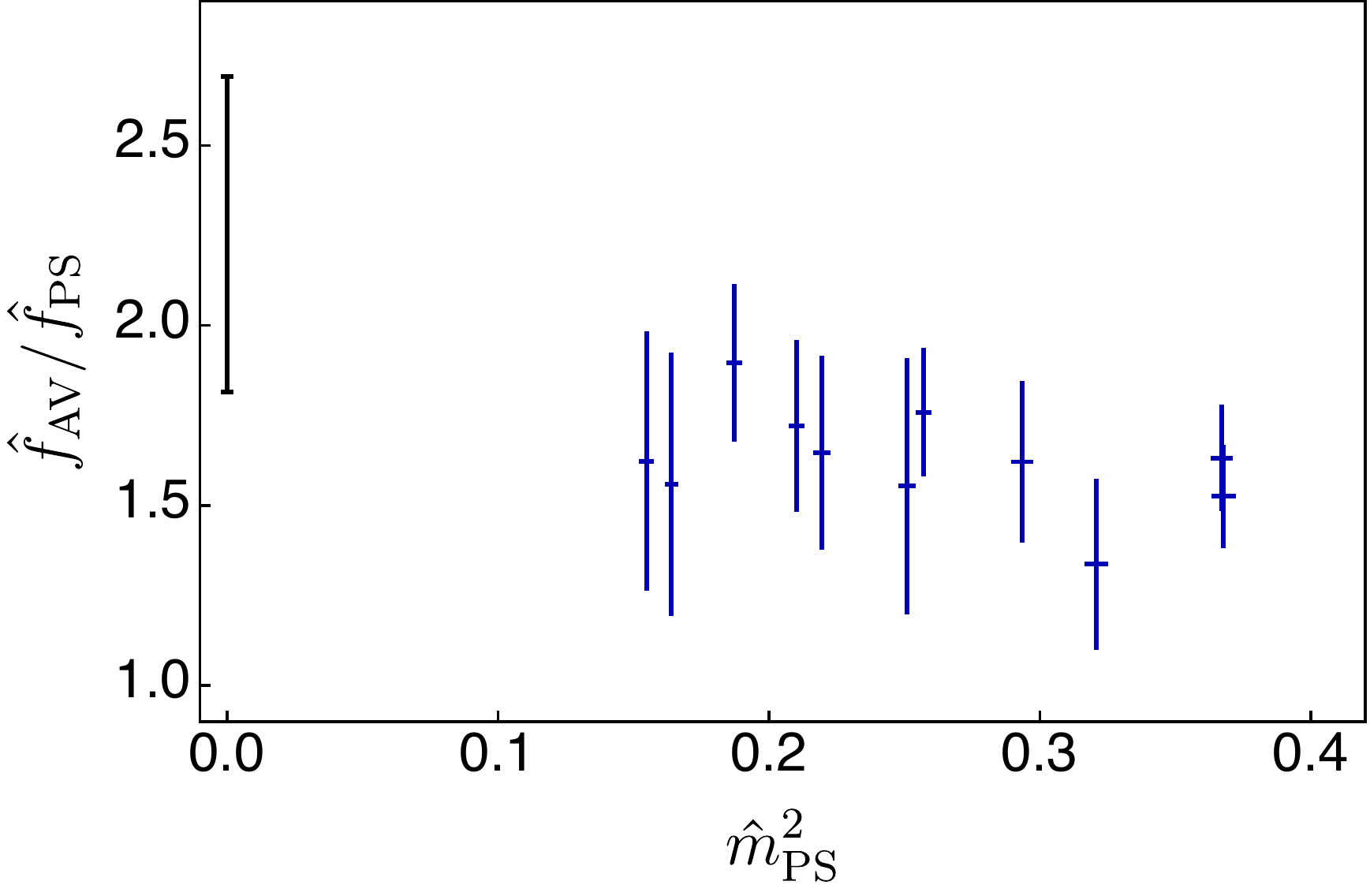}
\caption{%
\label{fig:ratio1}%
Ratios of masses and decay constants, in the continuum limit, expressed 
in units of the decay constant of the PS meson (as in Table~\ref{tab:meson_ratio}),
shown in blue. We also show the extrapolation to the continuum  and massless limit in black,
except for $\hat{m}_{\rm PS}$, which vanishes in the limit of massless fermions.
}
\end{center}
\end{figure}

This section is devoted to comparing our results with those obtained in a few other lattice gauge theories
with two Dirac fermions. We start by presenting our results in a different way, 
by normalising all masses and decay constants
 to  the decay constant $f_{\rm PS}$ of the lightest PS state. 
By doing so, we remove direct reference to the 
gradient flow scale $w_0$, and make the comparison to other theories more transparent.
We report the results of this exercise in Table~\ref{tab:meson_ratio}
and Fig.~\ref{fig:ratio1}. We restrict attention to the eleven ensembles for which 
we are able to take the continuum limit for all the observables,
 as described in Section~\ref{Sec:continuum}.
We then compare our results to those obtained in gauge theories 
with  $SU(2)$~\cite{Arthur:2016dir}, $SU(3)$~\cite{Jansen:2009hr}, and
$SU(4)$~\cite{Ayyar:2017qdf} with matter field content consisting 
of $N_f=2$ Dirac fermions in the fundamental.\footnote{
In the literature more lattice results are available for the $SU(2)$ theory with two fundamental Dirac fermions, see Refs. \cite{Detmold:2014kba,DeGrand:2016pur}.
However, we note that in these references continuum extrapolations of the numerical data computed at finite lattice spacing have not been carried out. 
Hence, we only use the results from \cite{Arthur:2016dir} for the comparison. 
} 
We do so by
testing the validity of the two phenomenological 
Kawarabayashi-Suzuki-Riazuddin-Fayyazuddin (KSRF) 
relations~\cite{Kawarabayashi:1966kd,Riazuddin:1966sw}.
We conclude this section with a comparison to the continuum extrapolation 
of the quenched $Sp(4)$ theory;
details of additional quenched computations performed beyond 
those presented in Ref.~\cite{Bennett:2017kga} to facilitate this analysis
are presented elsewhere~\cite{Sp4ASFQ}.

The fermion masses in the selected eleven ensembles are moderately heavy, 
with $1.39 \lsim \hat{m}_{\rm V}/\hat{m}_{\rm PS} \lsim 1.87$, 
so that both  V and AV mesons are stable, as the decay channels to  two and
 three PS mesons, respectively, are
kinematically forbidden. 
Bearing  this in mind, 
we nevertheless apply Eqs.~(\ref{eq:f2_chipt}) and~(\ref{eq:m2_chipt}) to the eleven ensembles,
and we perform the  massless (and continuum) extrapolations.
In the last row of the Table~\ref{tab:meson_ratio} we  report the resulting 
ratios of  the masses and decay constants to $\hat{f}_{\rm PS}$, 
 which are rendered in black colour in Figure~\ref{fig:ratio1} .
Notice that the ratio 
are independent of the gradient flow scale $w_0$, so that
 $f_{\rm V}/f_{\rm PS}=\hat{f}_{\rm V}/\hat{f}_{\rm PS}$
(and analogous for all dimensionless 
quantities in Table~\ref{tab:meson_ratio}).

The spectroscopy of the lightest meson states 
is captured to some approximation by the two KSRF
 phenomenological relations~\cite{Kawarabayashi:1966kd,Riazuddin:1966sw}.
The first such relation states that
\beq
f_{\rm V}=\sqrt{2}f_{\rm PS}\,,
\eeq
which is rather close to  real-world QCD, 
as taking  the experimental value of the rho meson decay constant to be 
$f_{\r}\simeq 148$ MeV, yields $f_\rho/f_\pi \sim 1.6$ \cite{Tanabashi:2018oca}. 
We can compare our results for $Sp(4)$ with the first KSRF relation by looking 
at $\hat{f}_{\rm V}/\hat{f}_{\rm PS}$ in Figure~\ref{fig:ratio1}. 
For $\hat{m}_{\rm PS}^2 < 0.4$ the ratio monotonically increases from
 $\hat{f}_{\rm V}/\hat{f}_{\rm PS}\sim 1.5$ 
as $\hat{m}_{\rm PS}^2$ decreases. A simple linear extrapolation yields the ratio in the massless and continuum limit to
be $\hat{f}_{\rm V}/\hat{f}_{\rm PS}\sim 2.1$. 
Therefore, our numerical results do not support the first KSRF relation: 
the resulting values  not only depend on the fermion mass, 
but also become larger in the massless limit.

The second KSRF relation involves $m_{\rm V}$, $f_{\rm PS}$, and
 the on-shell coupling constant $g_{\rm VPP}$ associated with the decay of a vector meson:
\beq
g_{\rm VPP}=\frac{m_{\rm V}}{\sqrt{2} f_{\rm PS}}. 
\eeq
In real-world QCD, the mass of $\rho$ meson $m_{\rho}\simeq 775$ MeV, expressed
in units of the pion decay constant $f_\pi$,
 yields roughly $m_\rho/\sqrt{2}f_\pi \sim 5.9$ \cite{Tanabashi:2018oca}. 
For comparison, we adopt the tree-level definition for the decay rate of  $\r$ meson, 
$\Gamma_{\rho}=\frac{g_{\rho\pi\pi}^2}{48\pi}m_{\r}\left(1-\frac{4m_{\pi}^2}{m_{\r}^2}\right)^{3/2}$, and the 
reference experimental values $\Gamma_{\rho}\simeq 150$ MeV and $m_{\pi}\simeq 140$ MeV.
We find $g_{\rho\pi\pi}\simeq 6.0$, which is in quite good agreement.
By evaluating the right-hand side of the second KSRF relation for the
$Sp(4)$ gauge theory, computed with 
the lightest ensemble and in the massless limit, we find  ${\hat{m}_{\rm V}}/{\sqrt{2} \hat{f}_{\rm PS}}=5.47(11)$ 
and ${\hat{m}_{\rm V}}/{\sqrt{2} \hat{f}_{\rm PS}}=5.72(18)(13)$, respectively. 
By comparing with the independent measurement of $g_{\rm VPP}=6.0(4)(2)$ from Section~\ref{Sec:result_con},
obtained from the global fit of the EFT, we conclude that the second KSRF relation holds.

\begin{figure}
\begin{center}
\includegraphics[width=.75\textwidth]{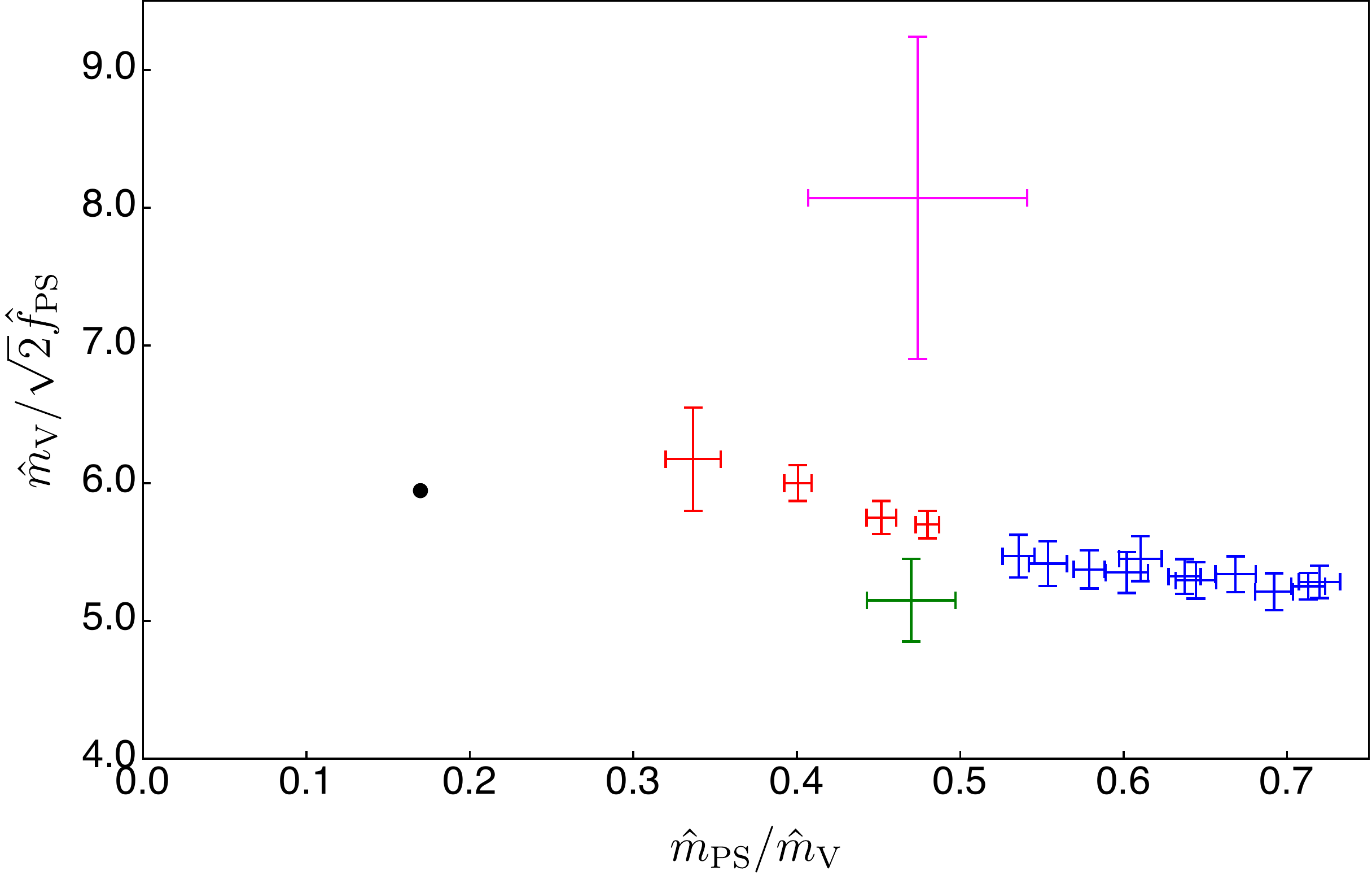}
\caption{%
\label{fig:gvpp_comparison}%
The ratios of the vector mass and pseudoscalar decay constant $m_{\rm V}/\sqrt{2}f_{\rm PS}$ 
in several lattice gauge theories with $N_f=2$ fundamental Dirac flavours. 
Purple, red, green and blue colours are for $SU(2)$~\cite{Arthur:2016dir}, $SU(3)$~\cite{Jansen:2009hr}, 
$SU(4)$~\cite{Ayyar:2017qdf} and $Sp(4)$ gauge groups, respectively. 
The $Sp(4)$ result obtained in this work
 are obtained in the continuum limit for the eleven ensembles identified in 
Section~\ref{Sec:continuum}.
The black circle is the experimental value of the coupling in the real world of QCD. 
Notice that for the $SU(2)$ and $SU(4)$ theories we show only 
the lightest data points, extracted approximately from the plots in the respective publications.
}
\end{center}
\end{figure}

It is interesting to compare the right-hand side of the second KSRF relation
 with other lattice results available in the literature on $SU(N)$ gauge theories 
with two fundamental Dirac fermions.
We show the comparison in \Fig{gvpp_comparison}.
For the lightest ensembles available, it is found that in the continuum limit $m_{\rm V}/\sqrt{2}f_{\rm PS}\sim 8.1(1.2)$ 
for $SU(2)$~\cite{Arthur:2016dir}
and $m_{\rm V}/\sqrt{2}f_{\rm PS}\sim 5.2(3)$ for $SU(4)$~\cite{Ayyar:2017qdf}, respectively. 
The general trend in $SU(N)$ theories is that the value of $m_{\rm V}/\sqrt{2}f_{\rm PS}$ decreases as 
$N$ increases, which complies with the expectation that 
$g_{\rm VPP}$ decreases in the large-$N$ limit.\footnote{
It is also interesting to investigate the flavour dependence of the ratio $m_{\rm V}/\sqrt{2}f_{\rm PS}$. 
A recent lattice study for $SU(3)$ gauge theory coupled to $N_f$ fundamental fermions finds that 
the ratio is statistically independent on $N_f$ up to $N_f=6$, 
for which all  theories considered are expected to behave in a way resembling QCD~\cite{Nogradi:2019iek}. 
On the other hand, the ratio could depend on the group representation of the fermion matter fields. 
For instance, large-$N_c$ arguments suggest that
$g_{\rm VPP} \propto1/\sqrt{N_c}$ and $g_{\rm VPP} \propto1/N_c$ for single index and two-index 
fermion representations, respectively.
Pioneering lattice results in $SU(4)$ are 
consistent with this scaling~\cite{Ayyar:2017qdf}. 
}
Near the threshold of $m_{\rm PS}/m_{\rm V} \sim 0.5$, the vector meson mass we find
for $Sp(4)$ in the continuum limit lies between the values for $SU(3)$ and $SU(4)$.

\begin{figure}
\begin{center}
\includegraphics[width=.49\textwidth]{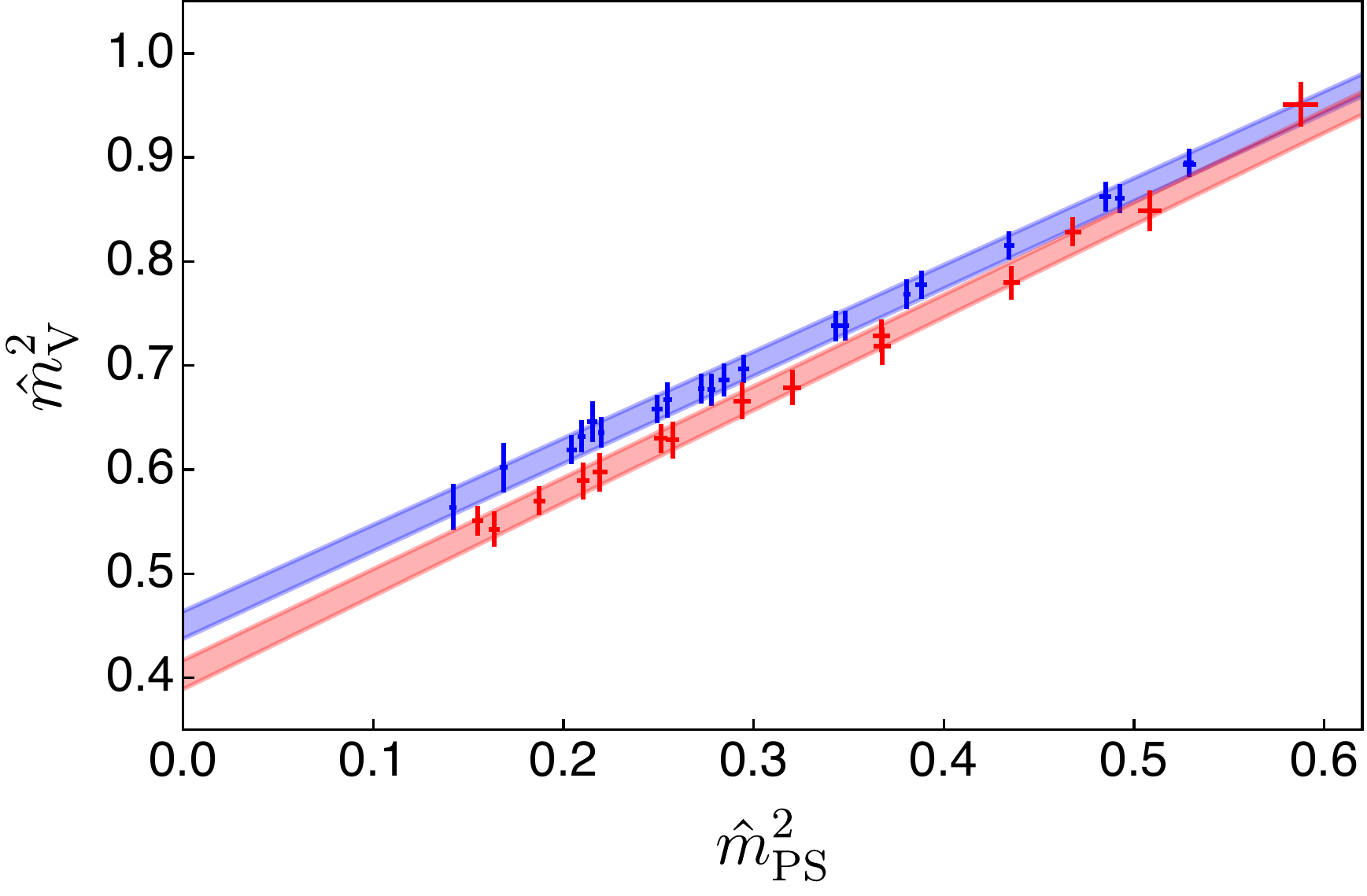}
\includegraphics[width=.49\textwidth]{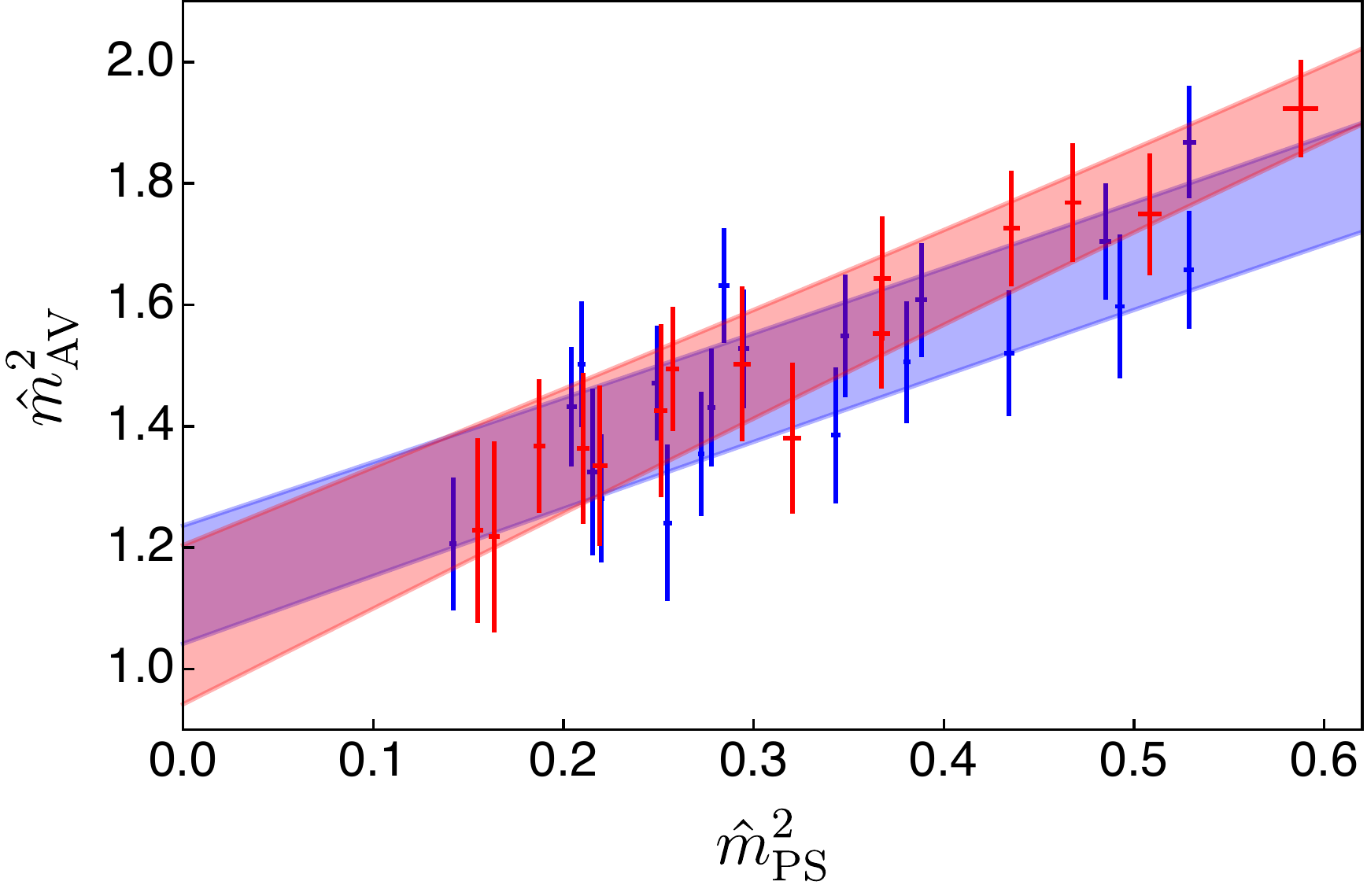}
\includegraphics[width=.49\textwidth]{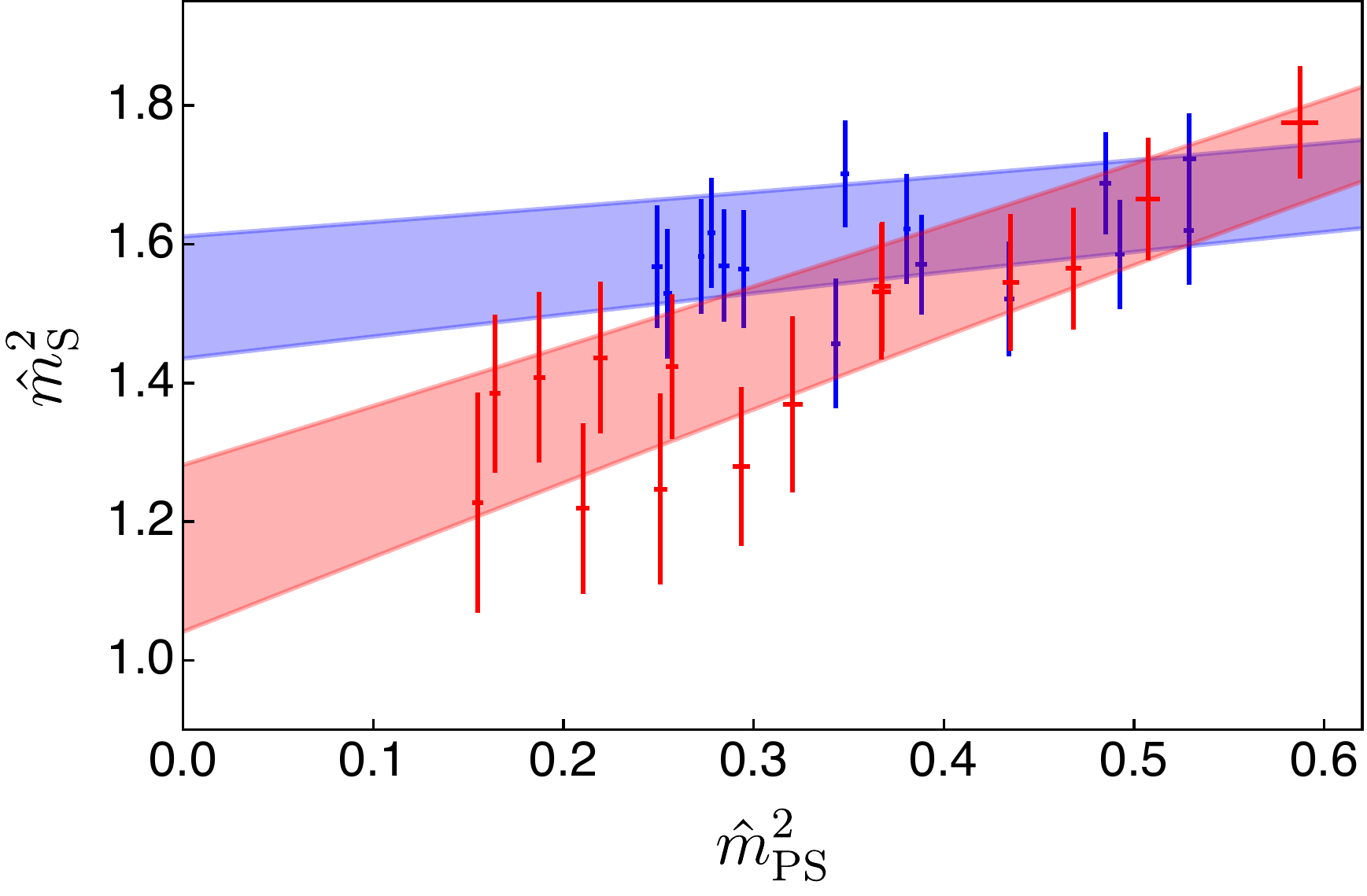}
\caption{%
\label{fig:m_QvsD}%
Meson masses squared from quenched (blue) and dynamical (red) calculations, in the continuum limit
obtained by considering all the ensembles with 
$\hat{m}_{\rm PS}^2\lsim 0.6$, as in Section~\ref{Sec:continuum}.
The coloured bands illustrate the fit of the measurements 
used in the massless extrapolations, with the width of the bands 
representing the statistical error in the fit.
}
\end{center}
\end{figure}

\begin{figure}
\begin{center}
\includegraphics[width=.49\textwidth]{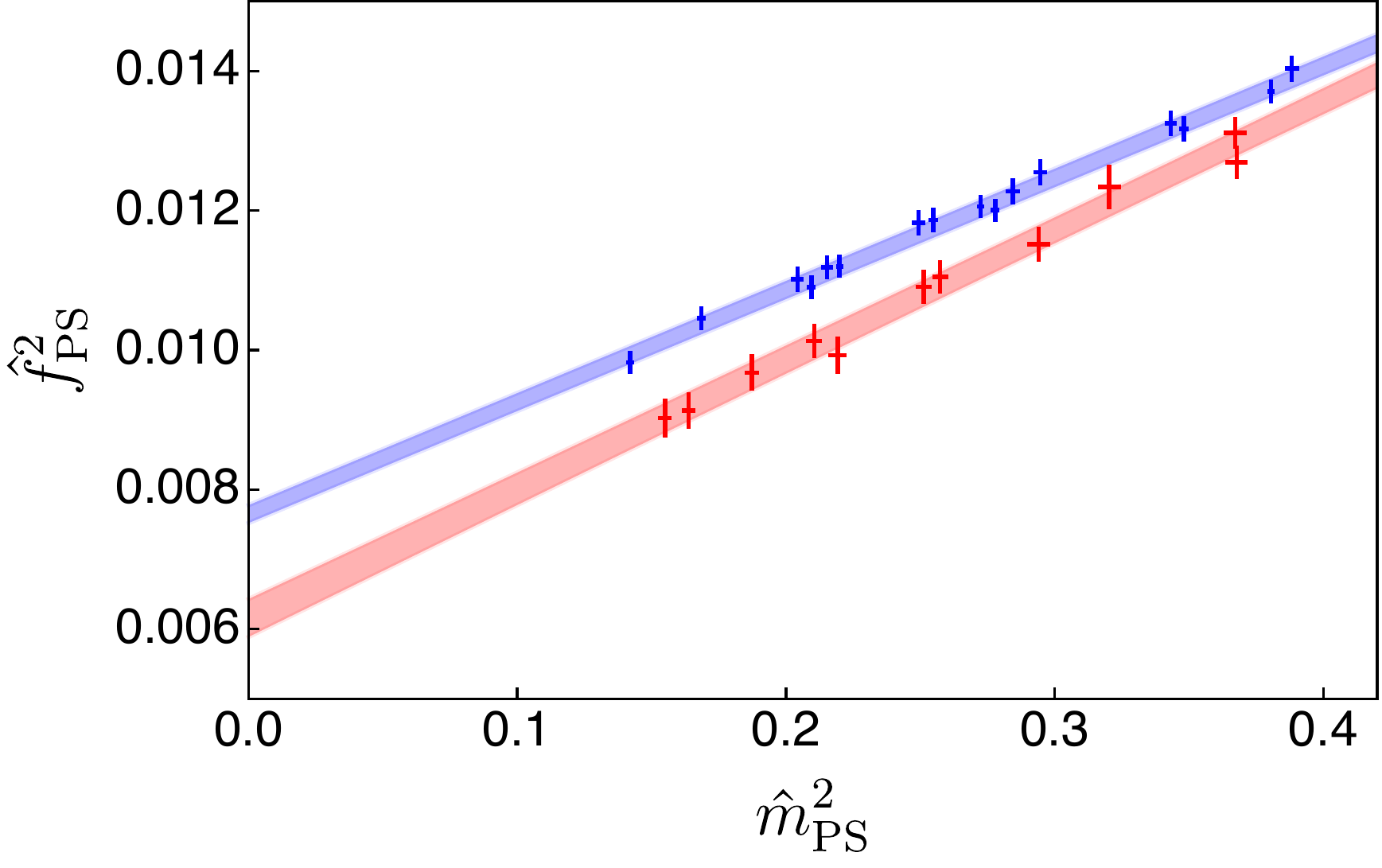}
\includegraphics[width=.49\textwidth]{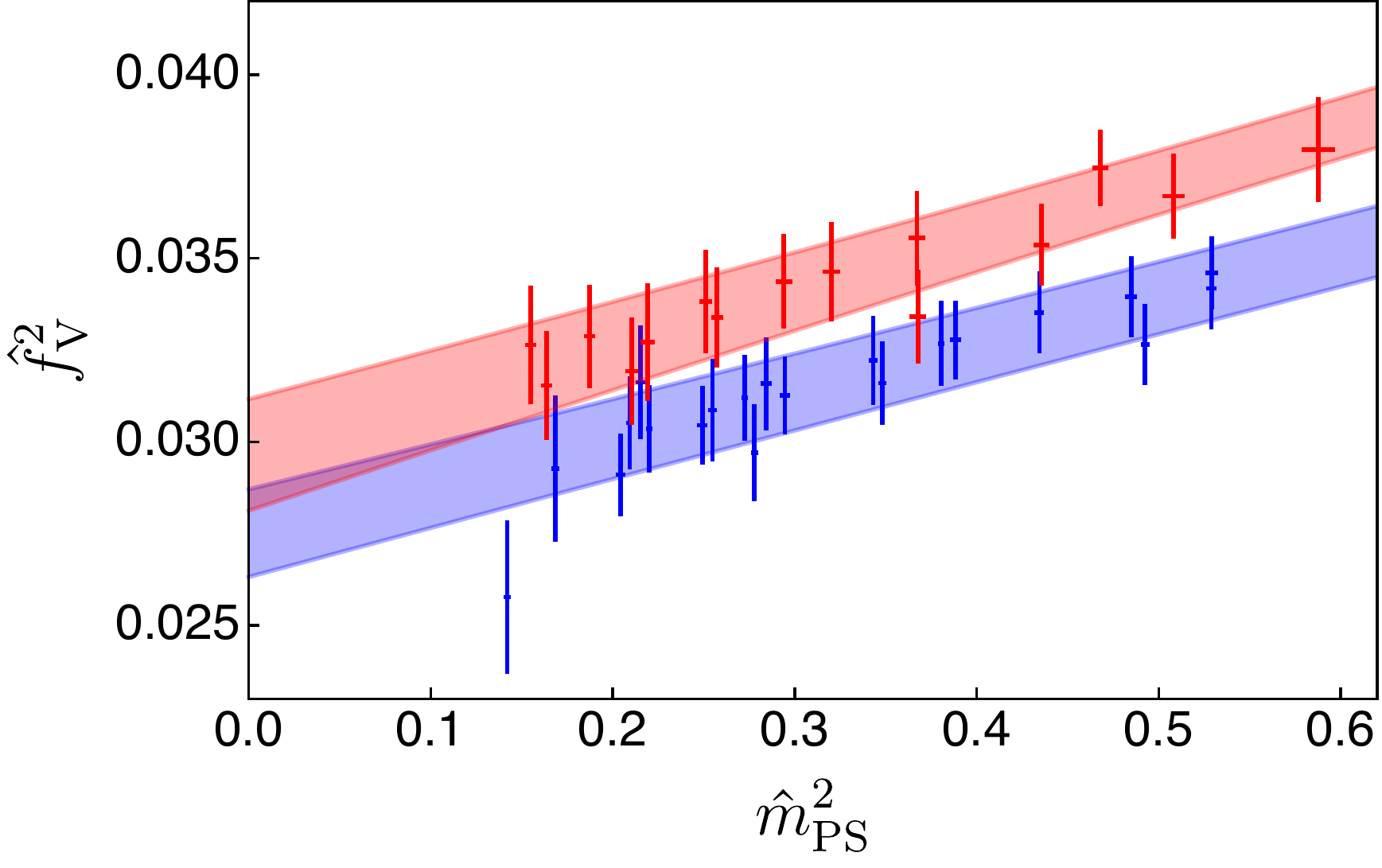}
\includegraphics[width=.49\textwidth]{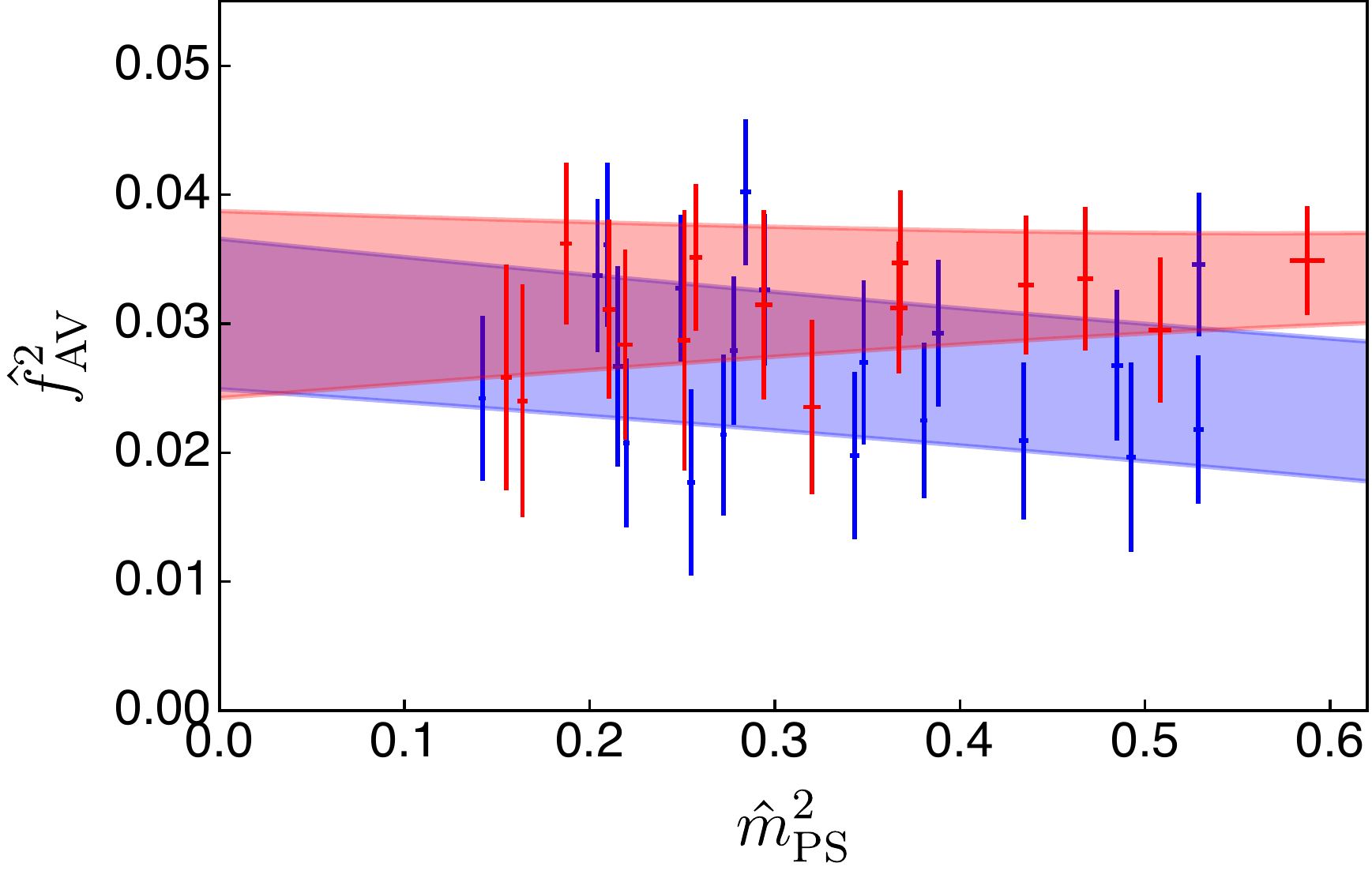}
\caption{%
\label{fig:f_QvsD}%
Meson decay constants squared from quenched (blue) 
and dynamical (red) calculations, in the continuum limit
obtained by considering all the ensembles with 
$\hat{m}_{\rm PS}^2\lsim 0.6$ for $\hat{f}_{\rm V}^2$ and $\hat{f}_{\rm AV}^2$, 
but restricting to $\hat{m}_{\rm PS}^2\lsim 0.4$ for $\hat{f}_{\rm PS}^2$, 
as in Section~\ref{Sec:continuum}.
The coloured bands illustrate the fit of the measurements 
used in the massless extrapolations, with the width of the bands 
representing the statistical error in the fit.
}
\end{center}
\end{figure}

We  close this section by comparing the dynamical results to the quenched ones. 
Experience from lattice QCD suggests that the quenched results, 
which require only a small amount of computing resources, 
capture well the qualitative features of the dynamical results 
for masses and decay constants,
in spite of the fact that the associated systematic uncertainties 
are not under analytical control.
Quenching effects are expected to be smaller as the number of colours increases with a fixed number of fundamental flavours.

In order to make the comparison possible,
 our preliminary exploration of quenched simulations~\cite{Bennett:2017kga}
 had to be extended 
by considering larger lattice volumes and various lattice couplings~\cite{Sp4ASFQ}. 
As in the dynamical case, the quenched ensembles satisfy the
 condition $m_{\rm PS} L \gtrsim 7.5$,
that allows one to neglect finite volume effects.
Continuum extrapolations use  ensembles 
constructed for $\beta=7.62$, $7.7$, $7.85$, $8.0$ and $8.2$. 
In contrast to the case of dynamical results---in which different ensembles are characterised by
the choices of $\beta$ as well as $a m_0$---only 
five independent ensembles generated at different $\beta$ 
values are used for all the quenched measurements. 
The results are hence affected by correlations among the data for various values of the fermion mass, 
which are obtained from the same ensemble, 
as well as from the continuum extrapolation carried out by using only
 five different values of the lattice spacing. 
The uncertainties associated with these systematic effects were estimated
by varying the fitting intervals in the continuum and massless extrapolations.
The numerical details and results of this quenched calculations are presented in Ref.~\cite{Sp4ASFQ}.

In Figs.~\ref{fig:m_QvsD} and \ref{fig:f_QvsD}, 
 we show together the continuum 
extrapolated data both for quenched and dynamical fermions,
restricted to the linear-mass regime---to $\hat{m}_{\rm PS}^2 \lsim 0.4$ for the
pseudoscalar decay constant and to $\hat{m}_{\rm PS}^2 \lsim 0.6$ for 
masses and decay constants of all other mesons. 
As seen in the figures, $\hat{f}_{\rm PS}^2$ and $\hat{m}_{\rm S}^2$ 
are significantly affected by quenching, 
and the differences become more substantial as the fermion mass decreases.
We estimate the discrepancies to be  $\delta_{\hat{f}_{\rm PS}^2}/\hat{f}_{\rm PS}^2 \sim 20\%$ 
and $\delta_{\hat{m}_{\rm S}^2}/\hat{m}_{\rm S}^2 \sim 25\%$, in the massless limit. 
The mass of the V meson shows a somewhat milder discrepancy, at the level of $\sim 10\%$.
For other quantities, quenching effects are not visible:
 the corresponding discrepancies are
 smaller than the uncertainties associated with the fits. 
Interestingly, the resulting values of $\hat{m}_{\rm V}/\sqrt{2} \hat{f}_{\rm PS}$
 for the dynamical and quenched simulations, 
which may be used to estimate the coupling $g_{\rm VPP}$ via the second KSRF relation, 
are found  to be consistent with each other in the massless limit~\cite{Sp4ASFQ}.  
The general conclusion of the comparison with the quenched results is quite encouraging,
although at present we do not know whether this conclusion is an indication  that the quenched 
approximation adequately captures the information encoded in the two-point functions---possibly
because of the proximity to large-$N$---or whether it is 
just a trivial consequence of the large fermion masses we studied.

\section{Continuum results: summary}
\label{Sec:summary2}

 In this section, we briefly summarise  the continuum extrapolation results for the dynamical theory, 
 presented in Section~\ref{Sec:result_con}  and~\ref{Sec:discussion}.

\begin{itemize}
\item[C1.]  Our continuum results for the 
decay constants and masses  of PS, V, and AV states,
 for the ensembles satisfying $\hat{m}_{\rm PS}^2 \leq 0.4$,
 are reported,
 in units of the gradient flow scale,  in Table~\ref{tab:datacont}.
\item[C2.] The global fit based on the EFT describing PS, V and AV states using hidden local symmetry
yields the results in Table~\ref{tab:global_LECs}, illustrated in Figs.~\ref{fig:f_gfit} and~\ref{fig:m_gfit}.
\item[C3.] Section~\ref{Sec:global_fit} discusses the GMOR relation and three sum rules in the continuum limit.
The main results are shown in Figs.~\ref{fig:GMOR} and \ref{fig:f0}. 
\item[C4.] In Section~\ref{Sec:discussion} the continuum limit results are discussed in units of the
 decay constant $f_{\rm PS}$ of the PS states, that are summarised in Table~\ref{tab:meson_ratio},
 and illustrated in Fig.~\ref{fig:ratio1}.  We include also the mass of the scalar flavoured state S.
 All our measurements are in the range $1.39 \lsim \hat{m}_{\rm V}/\hat{m}_{\rm PS} \lsim 1.87$,
 in which the V  states cannot decay to states containing two PS particles. 
 (Analogous considerations apply to  the 3-body decay of the AV mesons.)
 This range may be of direct relevance in the context of dark matter models.
 \item[C5.]  We perform the extrapolation to the massless limit. The masses of V, AV, and S states are
 $m_{\rm V}/f_{\rm PS}=8.08(32)$, 
 $m_{\rm AV}/f_{\rm PS}=13.4(1.5)$, and $m_{\rm S}/f_{\rm PS}=14.2(1.7)$.
 The decay constants of V and AV states in the continuum and massless limit 
 are $f_{\rm V}/f_{\rm PS}=2.15(8)$
 and $f_{\rm AV}/f_{\rm PS}=2.3(4)$, respectively---see Table~\ref{tab:meson_ratio}.
 \item[C6.]  We find that  $f_{\rm V}/f_{\rm PS}=2.15(8)$ is larger than
 expected on the basis of the first KSRF relation, which would  yield $f_{\rm V}=
 \sqrt{2}\,f_{\rm PS}$---see Section~\ref{Sec:discussion}.
 \item[C7.]  The second KSRF relation is satisfied, with $g_{\rm VPP}=6.0(4)(2)$ from the global fit, and 
 $m_{\rm V}/(\sqrt{2}f_{\rm PS})=5.72(18)(13)$  obtained from the massless limit extrapolation.
 \item[C8.]  We compare  the continuum and massless
 limit results to the literature on theories with two Dirac fermions on the fundamental.
 Fig.~\ref{fig:gvpp_comparison} shows that the VPP coupling is smaller than in the $SU(2)$ theory,
 but comparable to $SU(3)$ and $SU(4)$.
 \item[C9.] We close Section~\ref{Sec:discussion} by comparing our results, 
 obtained with dynamical fermions, to
 quenched calculations~\cite{Sp4ASFQ}. (See Figures~\ref{fig:m_QvsD} and~\ref{fig:f_QvsD}.) 
 We find that, in the massless limit,
 the decay constant squared of the PS state $\hat{f}_{\rm PS}^2$ is $\sim 20\%$ lower 
 than the quenched result.
 The mass squared of the (flavoured) scalar is $\sim 25\%$ lower than in the quenched result,
 while that of the vector is lower by $\sim 10\%$. In the other observables, the 
 dynamical and quenched results are compatible with one another, given the current uncertainties.
  
\end{itemize}

\section{Conclusions and outlook}
\label{Sec:conclusion}

Following along the programme outlined in Ref.~\cite{Bennett:2017kga}, 
with this paper we have made a substantial step
forwards in the  study of the gauge theory with $Sp(4)$ group and $N_f=2$ (Dirac) fundamental fermions,
that at long distance yields pNGBs describing the  $SU(4)/Sp(4)$ coset,
of relevance in the CHM context. 
We have adopted the Wilson-Dirac lattice action for gauge and fermion degrees of freedom, 
and performed numerical studies via the HMC method, with dynamical fermions. We have repeated the 
calculations at several values of the lattice bare parameters (fermion mass and gauge coupling).
Our main result is the  first continuum extrapolation
of the $Sp(4)$ measurements for the  
masses and decay constants of the lightest, flavoured, spin-0 and spin-1 mesons.
While numerical studies are performed in a regime of fermion masses large enough
to preclude the decay of heavier mesons to the pNGBs,  we have presented  also a preliminary 
massless extrapolation, and compared the results to those of the quenched approximation.
Summaries of the lattice and continuum limit results can be found 
in the brief Sections~\ref{Sec:summary1} and \ref{Sec:summary2}, respectively.
Other results of relevance to the broad research programme 
of lattice gauge theories with $Sp(2N)$ group will be
 reported in Refs.~\cite{Sp4ASFQ} and~\cite{Sp2Nglue} 
(see also Ref.~\cite{Lee:2018ztv}).

In the CHM context, while it is not strictly  necessary to have massless fermions, it would be important
 in the future to extend this study to reach closer to the massless regime, for pNGB
 masses that allow the heavier mesons to decay. This would allow to test whether the moderate departure of
 our results from the quenched ones is due to the large fermion mass in the action, to the approach to
 the large-$N$ limit, to dynamical properties of the theory or to a combination of the above.
 Of great importance in the CHM context is also the problem of vacuum 
 alignment. A first step towards addressing this point,
 without specifying the full set of couplings to SM fermions, would be
  to compute the coefficients of higher-order operators in
 the EFT obtained by retaining only the pNGB degrees of freedom.
  This information could be used as input in model building, 
 to compute the (perturbative) Higgs potential responsible for EWSB.
 In general, it would also be interesting to track the dependence on $N$ 
 of the spectra of mesons in $Sp(2N)$ theories with the
 same fermion field content, and understand how closely it follows the 
 expectations from the large-$N$ analysis.
 
The mass regime we studied is suitable for direct application in the context 
of models with strongly-coupled origin
of dark matter (along the lines of 
Refs.~\cite{Hochberg:2014dra,Hochberg:2014kqa,Berlin:2018tvf}, for example).
The information we provide here is a necessary first step towards adopting
 this theory as a candidate for the origin of dark matter.
Further information about the dynamics feeds into the relevant cross-sections,
 the calculation of which requires knowing the interactions.
The first coupling one would like to measure in the future is the $g_{\rm VPP}$ 
coupling between one vector state V and two pseudoscalar states PS.
We have performed a preliminary determination of this coupling based 
upon the EFT constructed to include also V and AV mesons,
but the intrinsic limitations of this process imply that the results should be used with some caution.

Finally, CHM constructions often involve advocating for (partial) compositeness of some of the SM fermions,
 the top quark in particular.
To this purpose, model building requires the existence of new composite
 fermion states with special properties, that can be realised in the $Sp(4)$
gauge theory by introducing $n_f=3$ (Dirac) fermions transforming in the 
2-index antisymmetric representation 
of the group~\cite{Barnard:2013zea,Ferretti:2013kya}. 
Some results about the (quenched) spectrum of mesons in the theory with 
this field content will be presented in Ref.~\cite{Sp4ASFQ}, 
but the implementation of fully dynamical calculations with 
multiple fermion representations will require a significant amount of development, with only few
examples of calculations of this type existing in the current
literature~\cite{DeGrand:2016pgq,Ayyar:2017qdf,Ayyar:2018zuk,Ayyar:2018glg,Cossu:2019hse}.

\vspace{1.0cm}
\begin{acknowledgments}

We acknowledge useful discussions with Axel Maas and Roman Zwicky.
We would like to thank Michele Mesiti and Jarno Rantaharju for their assistance on 
the modification and improvement of the HiRep code for this project. 

The work of EB has been funded in part by the Supercomputing Wales project, 
which is part-funded by the European Regional Development Fund (ERDF) via Welsh Government. 

The work of DKH was supported by Basic Science Research Program 
through the National Research Foundation of Korea (NRF) funded by 
the Ministry of Education (NRF-2017R1D1A1B06033701).  

The work of JWL is supported in part by the National Research Foundation of Korea grant funded 
by the Korea government(MSIT) (NRF-2018R1C1B3001379) and 
in part by Korea Research Fellowship programme funded 
by the Ministry of Science, ICT and Future Planning through the National 
Research Foundation of Korea (2016H1D3A1909283).

The work of CJDL is supported by the Taiwanese MoST grant 105-2628-M-009-003-MY4. 

The work of BL and MP has been supported in part by the STFC 
Consolidated Grants ST/L000369/1 and ST/P00055X/1.
The work of BL is further supported in part by the Royal Society Wolfson Research Merit Award WM170010.

DV acknowledges support from the INFN HPC-HTC project.

Numerical simulations have been performed on the Swansea SUNBIRD 
system, on the local HPC
clusters in Pusan National University (PNU) and in National Chiao-Tung University (NCTU),
 and on the Cambridge Service for Data Driven Discovery (CSD3). The Swansea SUNBIRD 
system is part of the Supercomputing Wales project, which is part-funded by the European Regional
Development Fund (ERDF) via Welsh Government. CSD3 is operated in part by
 the University of Cambridge Research Computing on
behalf of the STFC DiRAC HPC Facility (www.dirac.ac.uk). 
The DiRAC component of CSD3 was funded by BEIS capital funding via
STFC capital grants ST/P002307/1 and ST/R002452/1 and 
STFC operations grant ST/R00689X/1. DiRAC is part of the National e-Infrastructure.

\end{acknowledgments}

\vspace{1.0cm}
\appendix
\section{Lattice action and numerical calculations: additional details}
\label{Sec:technique}

We collect in this Appendix some supplementary details, pertaining the
lattice numerical study that ultimately leads to the results summarised in Tables~\ref{tab:meson_spec_spin0}
and \ref{tab:meson_spec_spin1}, 
which provide the data upon which we perform the continuum and massless extrapolations.
In Appendix~\ref{Sec:HMC} we discuss the treatment of the $Sp(2N)$ matrices within the HMC algorithm.
We briefly clarify the role of diquark operators in Appendix~\ref{Sec:diquark}. 
Finally, in Appendix~\ref{Sec:meff} we present details of the fitting procedure of the correlation functions
used to extract masses and decay constants.

\subsection{Hybrid Monte Carlo}
\label{Sec:HMC}

We perform numerical simulations using a variant of HiRep code~\cite{DelDebbio:2008zf},
which is designed to simulate $SU(N)$ and $SO(N)$ lattice gauge theories with 
fermions in higher representations. 
While a detailed description of the implementations of $Sp(2N)$ theories in the
 HiRep code is described  in Ref.~\cite{Bennett:2017kga} 
(see also Refs.~\cite{Bennett:2017tum} and~\cite{Bennett:2017kbp}), 
here we briefly summarise its main features and 
then report on some improvements we implemented for  the purposes of this and future projects.
This paper focuses on the spectroscopy of $Sp(4)$ with two fundamental fermions,
 higher-dimensional representations will be discussed elsewhere.\footnote{
Preliminary results for  $Sp(2N)$ theories with fermions in 
the anti-symmetric two-index irreducible representation 
are presented in Ref.~\cite{Lee:2018ztv}, and a more detailed study will
be discussed in Ref.~\cite{Sp4ASFQ}.} 
To study the $N_f=2$ theory we use the standard hybrid Monte Carlo algorithm, 
a well established technique for lattice QCD. 

In our preliminary study of the two-flavor $Sp(4)$ theory~\cite{Bennett:2017kga}, 
we used the specific form of group generators $T^A$ given in Ref.~\cite{Lee:2017uvl} 
with the normalisation of $\textrm{Tr}(T^A T^B)=\delta^{AB}/2$ 
for the molecular dynamics (MD) evolution. 
We also implemented a resymplectisation process, to ensure that 
 updated configurations 
stay inside the $Sp(4)$ group manifold, that requires 
performing a (normalised) group projection onto the quaternion basis.
This process works well for $Sp(4)$ theories, but in order
to enhance the capability of the software in  view of future studies of
 dynamical $Sp(2N)$ theories with arbitrary $N$, 
in this study we further improve and generalise the code in the following way. 

First, we remind the reader that the group elements $U$ of $Sp(2N)$ satisfy the condition 
\bea
U^* =\Omega U \Omega^\dagger, ~~~\textrm{with}~
\Omega&=&
\left(\begin{array}{cc}
0 & \mathbb{I}_{N\times N} \cr
-\mathbb{I}_{N\times N} & 0 \cr
\end{array}\right)\,.
\label{eq:spn_def}
\eea
 $U$ is a unitary matrix, and it can be written as  $U=\textrm{exp}(i a^A T^A)$, 
where $T^A$ are Hermitian traceless $2N\times 2N$ matrices, and $a^N$ real numbers. 
Combining  \Eq{spn_def} and unitarity, one can also write the matrix $U$ in the block-diagonal form:
\bea
U&=&
\left(\begin{array}{cc}
V & W \cr
-W^* & V^* \cr
\end{array}\right)\, ,
\eea
where $V$ and $W$ are complex $N\times N$ matrices. 
Because of its adaptability to any $N$, we 
perform the resymplectisation via a variant of the modified 
Gram-Schmidt algorithm, that has been tested for the pure gauge model with 
Heath Bath algorithm~\cite{Bennett:2017kbp,Bennett:2017kga}.
The basic idea is that the ($N+j$)-th column of $U$ 
can be obtained from $j$-th via $col_{j+N}=-\Omega col_j$, 
while the Gram-Schimt procedure is used to find the ($j+1$)-th column through the orthonormalisation 
with respect to the $j$-th and ($N+j$)-th. 
We also modified the code to save
 two $N\times N$ matrices ($V$ and $W$), instead of the full $2N\times 2N$ unitary matrix $U$,
 reducing  by half the size of each configuration. 

The MD update in HMC makes explicit use of the 
generators of the group, not just of the group elements.
We write  the group generators $T^A$ as follows.
\Eq{spn_def} can be rewritten in terms of the group generator $T^A$,  as the condition 
\beq
T^{A*}=-\Omega T^A \Omega^\dagger, 
\eeq
which allows to write $T^A$ in block-diagonal matrix form as
\bea
T^A&=&
\left(\begin{array}{cc}
X & Y \cr
Y^{*} & -X^{*}\cr
\end{array}\right)\, ,
\label{eq:spn_gen}
\eea
where of the two $N\times N$ matrices, $X$  is Hermitian and  $Y$ is complex symmetric. 
We use the definition in \Eq{spn_gen}, supplemented by the normalisation
 $\textrm{Tr}\,(T^AT^B)=\delta^{AB}/2$, 
for the generators implemented into the HiRep code.

We conclude by summarising some  technical details about the algorithm used
in  the $Sp(4)$ gauge theory with two Dirac fermions in the fundamental representation. 
Gauge configurations are generated using the HMC with a 
second order Omelyan integrator for the MD update. 
We use different lengths of MD time steps $\delta \tau_g$ and $\delta \tau_f$ 
for gauge and fermions actions, respectively, 
which are optimised to keep the acceptance rate of the Metropolis test,
performed at the end of each HMC update, in the range of $75-85\%$. 
Thermalisation and autocorrelation lengths are determined by monitoring 
the average value of the plaquette.

\subsection{Of diquarks}
\label{Sec:diquark}

In the theory studied in this paper, 
mesons and diquarks combine together  in the low-energy spectrum, to form irreducible representations 
after the spontaneous breaking of the enlarged global $SU(4)$ symmetry breaking to $Sp(4)$. 
Hence, we do not calculate the diquark correlators, 
as they are identical to the corresponding meson correlators. 
A general discussion of both real and pseudoreal representations can be 
found in Ref.~\cite{DeGrand:2015lna}, yet
we think it is useful to explicitly write the diquark operators and 
show the identity at the level of correlators by using our lattice action in \Eq{fermion_action}. 
Such an analysis for SU($2$) theory with two fundamental fermions can 
be  found in Ref.~\cite{Lewis:2011zb}. 

The diquark operators are defined by
\beq
\mathcal{O}_D(x)\equiv Q^T_i(x) (-\Omega) C \Gamma Q_j(x), 
\eeq
where $C$ is the  charge conjugation operator satisfying $\gamma^{\mu T}C=-C\gamma^\mu$,
$\Gamma$ is a generic matrix with spinor indices, 
spinor indices are understood and summed over, 
and the anti-symmetric matrix $\Omega$ defined in \Eq{spn_def} acts on the (understood) colour indices. 
Then, the diquark correlation function is 
\bea
C_{\mathcal{O}_D}(t)&=&\sum_{\vec{x}} 
\langle 0 | \mathcal{O}_D(\vec{x},t) \mathcal{O}_D^\dagger(\vec{0},0) | 0\rangle \nn \\
&=&\sum_{\vec{x}} 
\textrm{Tr}\left[
\Gamma S_j(x;0) \gamma^0 \Gamma^\dagger C^\dagger (-\Omega)^\dagger \gamma^{0T} S_i^T(x;0) (-\Omega)C
\right]\nn \\
&=&-\sum_{\vec{x}} 
\textrm{Tr}\left[
\Gamma S_j(x;0) \gamma^0 \Gamma^\dagger \gamma^0 C^\dagger (-\Omega)^\dagger S_i^T(x;0) (-\Omega)C
\right].
\label{eq:diquark_corr}
\eea
The inverse of a quark propagator $S^{-1}$---or, equivalently, the Wilson-Dirac 
operator---can be obtained from \Eq{fermion_action}: 
\beq
S^{-1}_i(x;y)=(4+a\,m_0)\delta_{x,y}-\frac{1}{2}
\sum_\mu \left((1-\gamma_\mu)U_\mu (x) \delta_{x+\hat{\mu},y}
+(1+\gamma_\mu)U^\dagger(x) \delta_{y,x-\hat{\mu}}\right).
\eeq
By recalling \Eq{spn_def} and the defining property of the charge-conjugation matrix 
$C^{-1}=C^\dagger=C^T=-C$, one can show that
\beq
C^{-1}(-\Omega)^{-1} (S^{-1}_i(x;y))^T (-\Omega) C =S^{-1}_i(y;x),
\eeq
which in turn leads to
\beq
C^\dagger(-\Omega)^\dagger S^T_i(x;y) (-\Omega) C = S_i(y;x).
\eeq
Combining this result with \Eq{meson_corr} and \Eq{diquark_corr}, we finally arrive at
\beq
C_{\mathcal{O}_D}(t)=C_{\mathcal{O}_M}(t)\,,
\eeq
where $C_{\mathcal{O}_M}(t)$ is the meson correlation function defined in Eq.~(\ref{eq:meson_corr}).

\begin{table}
\begin{center}
\begin{tabular}{|c|c|c|cc|cc|cc|}
\hline\hline
\multirow{2}{*}{Ensemble} & \multirow{2}{*}{$N_{\rm configs}$} & \multirow{2}{*}{$\delta_{\rm traj}$} 
& \multicolumn{2}{c|}{PS} & \multicolumn{2}{c|}{V} & \multicolumn{2}{c|}{AV} \\
 &  &  & 
$I_{\rm fit}$ & $\chi^2/N_{\rm d.o.f}$ & $I_{\rm fit}$ & $\chi^2/N_{\rm d.o.f.}$ & $I_{\rm fit}$ &
 $\chi^2/N_{\rm d.o.f.}$  \\
\hline
DB1M1 & 100 & 24 & 9-16 & 0.3 & 10-16 & 0.4 & 6-10 & 1.3 \\
DB1M2 & 100 & 24 & 11-16 & 0.8 & 9-16 & 0.2 & 6-9 & 0.1\\
DB1M3 & 100 & 24 & 10-16 & 0.5 & 10-16 & 1.0 & 6-10 & 1.1\\
DB1M4 & 74 & 32 & 10-16 & 1.1 & 10-16 & 0.2 & 6-9 & 0.9\\
DB1M5 & 183 & 12 & 10-16 & 0.6 & 10-15 & 1.0 & 7-10 & 0.6\\
DB1M6 & 80 & 28 & 10-16 & 0.7 & 10-16 & 0.3 & 6-9 & 0.5\\
DB1M7 & 71 & 12 & 10-16 & 0.5 & 10-16 & 1.3 & 7-10 & 0.5\\
\hline
DB2M1 & 100 & 20 & 12-18 & 0.8 & 11-18 & 0.5 & 8-11 & 1.1\\
DB2M2 & 100 & 24 & 12-18 & 0.5 & 13-18 & 0.6 & 8-14 & 0.6\\
DB2M3 & 102 & 20 & 13-18 & 1.0 & 12-16 & 0.4 & 9-11 & 0.1\\
\hline
DB3M1 & 120 & 20 & 14-18 & 0.5 & 14-18 & 0.2 & 8-12 & 0.3\\
DB3M2 & 100 & 24 & 12-18 & 0.6 & 12-18 & 0.8 & 8-14 & 1.0\\
DB3M3 & 200 & 12 & 13-18 & 1.0 & 13-17 & 1.0 & 9-15 & 0.4\\
DB3M4 & 200 & 12 & 14-17 & 0.7 & 14-18 & 0.7 & 9-15 & 0.9\\
DB3M5 & 150 & 12 & 15-18 & 0.3 & 15-18 & 0.8 & 9-15 & 1.3\\
DB3M6 & 200 & 12 & 13-18 & 0.8 & 13-17 & 0.1 & 10-15 & 0.6\\
DB3M7 & 199 & 12 & 14-18 & 0.6 & 14-18 & 0.8 & 9-14 & 1.4\\
DB3M8 & 150 & 12 & 15-20 & 0.6 & 15-20 & 0.9 & 10-16 & 0.2\\
\hline
DB4M1 & 150 & 12 & 19-24 & 1.0 & 15-24 & 1.9 & 12-17 & 0.6\\
DB4M2 & 150 & 12 & 19-23 & 0.4 & 17-24 & 0.6 & 12-15 & 0.2\\
\hline
DB5M1 & 105 & 12 & 16-24 & 0.6 & 18-24 & 0.2 & 11-19 & 0.6\\
\hline\hline
\end{tabular}
\end{center}
\caption{%
\label{tab:meson_measurement_1}%
Fitting intervals of the Euclidean time $I_{\rm fit}=[t_i,t_f]$ 
used in the  single-exponential fit of the measured correlators for PS, V and AV mesons. 
The number of configurations and the separation between the adjacent configurations 
are denoted by $N_{\rm configs}$ and $\delta_{\rm traj}$, respectively. 
We carry out a correlated fit using standard $\chi^2$ minimisation.
We show the resulting values of $\chi^2/N_{\rm d.o.f.}$, as a way to assess the quality of the fit itself. 
}
\end{table}

\begin{table}
\begin{center}
\begin{tabular}{|c|cc|cc|cc|}
\hline\hline
\multirow{2}{*}{Ensemble} & \multicolumn{2}{c|}{T} & \multicolumn{2}{c|}{AT} & \multicolumn{2}{c|}{S}\\
  & 
$I_{\rm fit}$ & $\chi^2/N_{\rm d.o.f}$ & $I_{\rm fit}$ & $\chi^2/N_{\rm d.o.f.}$ & $I_{\rm fit}$ &
 $\chi^2/N_{\rm d.o.f.}$\\
\hline
DB1M1 & 9-16 & 1.1 & 6-10 & 0.8 & 6-9 & 1.7\\
DB1M2 & 9-16 & 0.9 & 6-9 & 0.9 &  6-9 & 0.4\\
DB1M3 & 8-16 & 1.0 & 6-10 & 1.6 & 7-10 & 0.6\\
DB1M4 & 10-16 & 1.4 & 6-10 & 1.0 & 5-8 & 1.2\\
DB1M5 & 10-16 & 0.4 & 6-10 & 0.6 &  7-10 & 0.7\\
DB1M6 & 9-16 & 0.7 & 7-10 & 1.6 &  6-9 & 0.5\\
DB1M7 & 8-16 & 0.5 & 7-10 & 0.8 &  6-10 & 0.1\\
\hline
DB2M1 & 11-18 & 0.7 & 8-12 & 0.6 & 8-13 & 0.4\\
DB2M2 & 12-18 & 1.5 & 8-12 & 1.6 & 7-9 & 1.0\\
DB2M3 & 11-18 & 1.0 & 9-11 & 1.1 & 7-12 & 0.3\\
\hline
DB3M1 & 14-18 & 0.2 & 7-12 & 0.4 & 9-14 & 0.9\\
DB3M2 & 12-18 & 0.9 & 7-13 & 1.5 & 8-14 & 1.0\\
DB3M3 & 13-18 & 0.8 & 8-14 & 0.9 & 8-14 & 0.2\\
DB3M4 & 11-18 & 1.2 & 10-15 & 0.4 & 9-16 & 0.5\\
DB3M5 & 13-18 & 0.4 & 10-14 & 1.4 & 9-15 & 0.7\\
DB3M6 & 12-17 & 0.1 & 10-15 & 0.7 & 10-15 & 1.2\\
DB3M7 & 13-18 & 0.6 & 9-16 & 1.6 & 9-15 & 0.9\\
DB3M8 & 11-20 & 0.4 & 10-14 & 1.0 & 10-17 & 1.7\\
\hline
DB4M1 & 15-24 & 0.4 & 11-16 & 1.1 & 12-18 & 0.5\\
DB4M2 & 16-24 & 0.7 & 12-16 & 0.2 & 11-17 & 0.5\\
\hline
DB5M1 & 18-24 & 0.6 & 11-19 & 0.2 & 11-20 & 1.3\\
\hline\hline
\end{tabular}
\end{center}
\caption{%
\label{tab:meson_measurement_2}%
Fitting intervals of the Euclidean time $I_{\rm fit}=[t_i,t_f]$ 
used in the  single-exponential fit of the measured correlators for T, AT and S mesons, 
where the same $N_{\rm configs}$ and $\delta_{\rm traj}$ in \Tab{meson_measurement_1} are considered. 
We carry out a correlated fit using standard $\chi^2$ minimisation.
We show the resulting values of $\chi^2/N_{\rm d.o.f.}$, as a way to assess the quality of the fit itself. 
}
\end{table}

\begin{figure}
\begin{center}
\includegraphics[width=.79\textwidth]{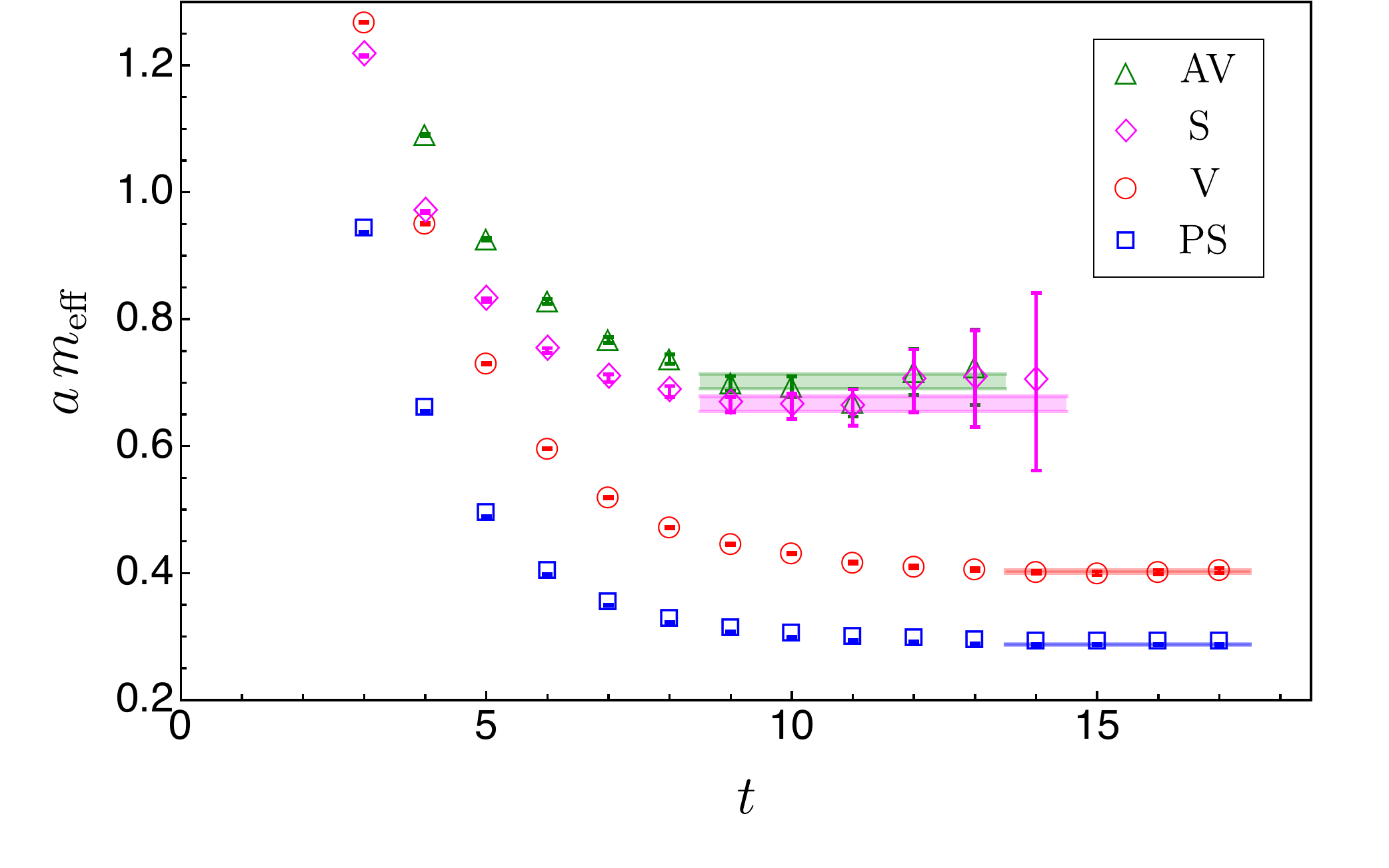}
\end{center}
\caption{
\label{fig:meff}
Example of effective mass plots of low-lying spin-$0$ and spin-$1$ mesons. 
The data is taken from the DB3M7 ensemble (see list in Table~\ref{tab:ensembles}),
which is characterised by the  lattice parameters $\beta=7.2$ and $a m_0=-0.794$. 
The individual fits that yield the masses of the PS, V, S and AV states are restricted to include only
the plateau regions, which are highlighted by the shaded bands. 
The width of each band represents for the statistical uncertainty.
}
\end{figure}

\subsection{Effective mass and fitting procedure}
\label{Sec:meff}

In order to extract the lattice measurements of masses and decay constants, 
we fit  the numerical data for the relevant two-point correlation functions. 
In all ensembles we produced, the Euclidean time is large enough 
to reach a plateau region in the plot of $m_{\rm eff}$ versus time $t$,
 in which the correlation functions 
are well approximated by a single exponential function,  with its decay rate identified by the mass of 
mesons---the relevant 2-point function asymptotically 
behaves as $C_{{\cal O}_M}(t)\propto e^{-m_{\rm M} t}$.
We use standard $\chi^2$ minimisation to this functional form
to determine the best fit parameters (mass and decay constant, 
in units of the lattice spacing) and their statistical uncertainties. 

This process requires choosing, for each ensemble and each meson,
 an interval $I_{\rm fit}=[t_i,t_f]$ of Euclidean time,  and restricting the fit to data collected
  between its minimum value
 $t_i$ and its maximum time $t_f$.
We choose the fitting intervals as follows: 
we first  look at the effective mass plots and identify the emergence of the plateau. 
Over the plateau region, we perform a single-exponential fit. 
We vary the choices of $t_i$ and $t_f$, and ultimately identify the best fit range 
with the choice that yields the smallest value for
$\chi^2/N_{\rm d.o.f}$, with the largest possible range $t_f-t_i$.
We provide the relevant details of this process in Tables~\ref{tab:meson_measurement_1}
and~\ref{tab:meson_measurement_2}.
As an illustrative example, in \Fig{meff} we show the effective mass plots for PS, V, AV and S  mesons 
at $\beta=7.2$ and $a m_0=-0.794$ (corresponding to ensemble DB3M7). 
The shaded bands represent the fit results, showing both
 the statistical uncertainties (width of the band) and the best fitting ranges (length of the band). 
For the PS meson, we perform a simultaneous fit of the two-point functions of 
 PS and AV operators in \Eq{corr_ground} and \Eq{axial_corr}. 
 
We notice that while the effective masses retained in the fit extend to the maximum 
length of the temporal directions $T_{\rm max}$ for PS and V mesons, 
those for AV and S mesons  typically cease at $t<T_{\rm max}$ due to severe numerical noise problems, 
which in practical terms reduce the fitting ranges. 
As a result, we expect a comparatively large systematic error  associated 
with the choice of the fitting range for AV and S states
(analogous arguments apply to the AT states).

\section{Low-energy constants and global fit}
\label{Sec:LECs_hist}

\begin{figure}
\begin{center}
\includegraphics[width=.49\textwidth]{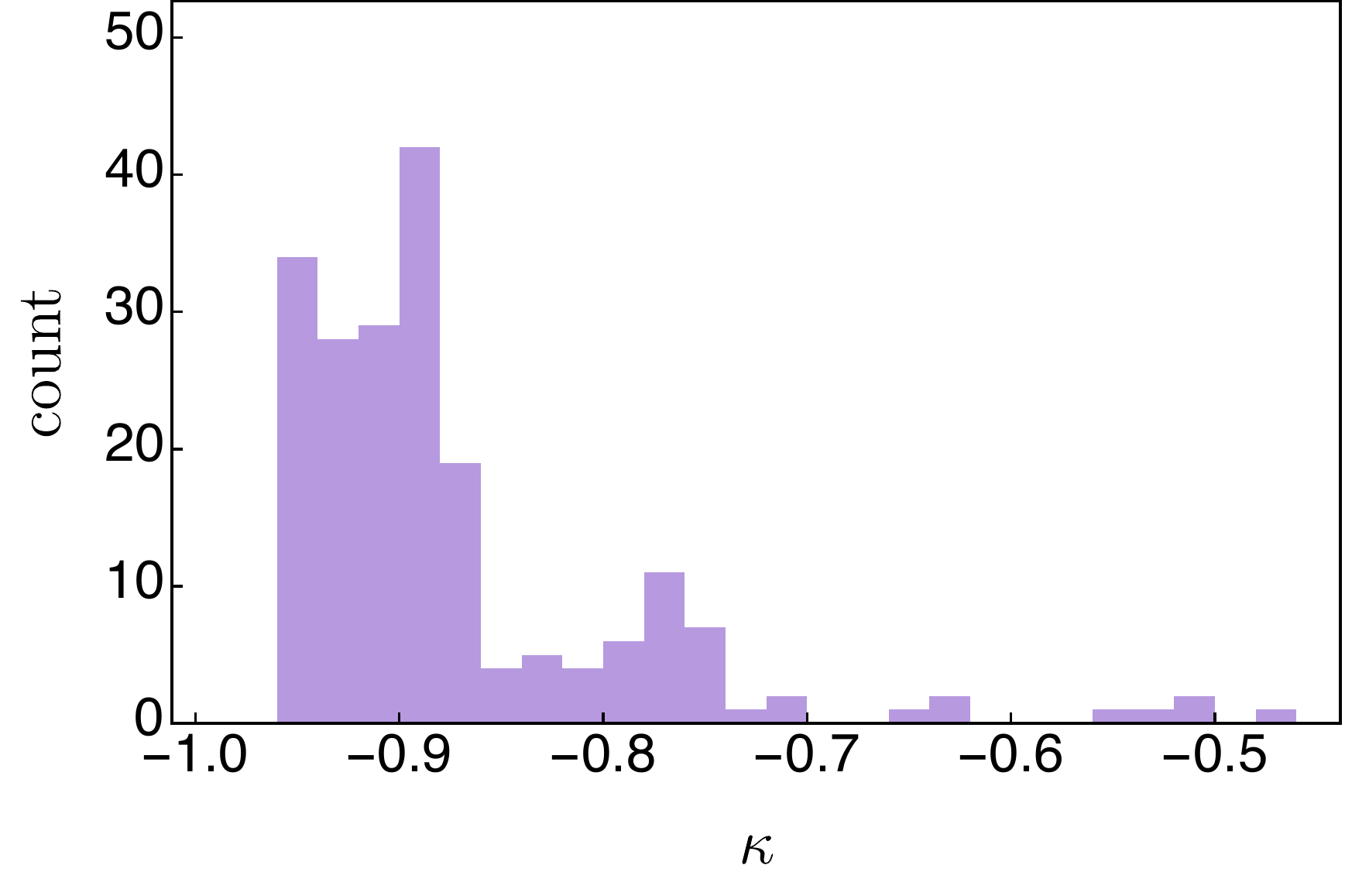}
\includegraphics[width=.49\textwidth]{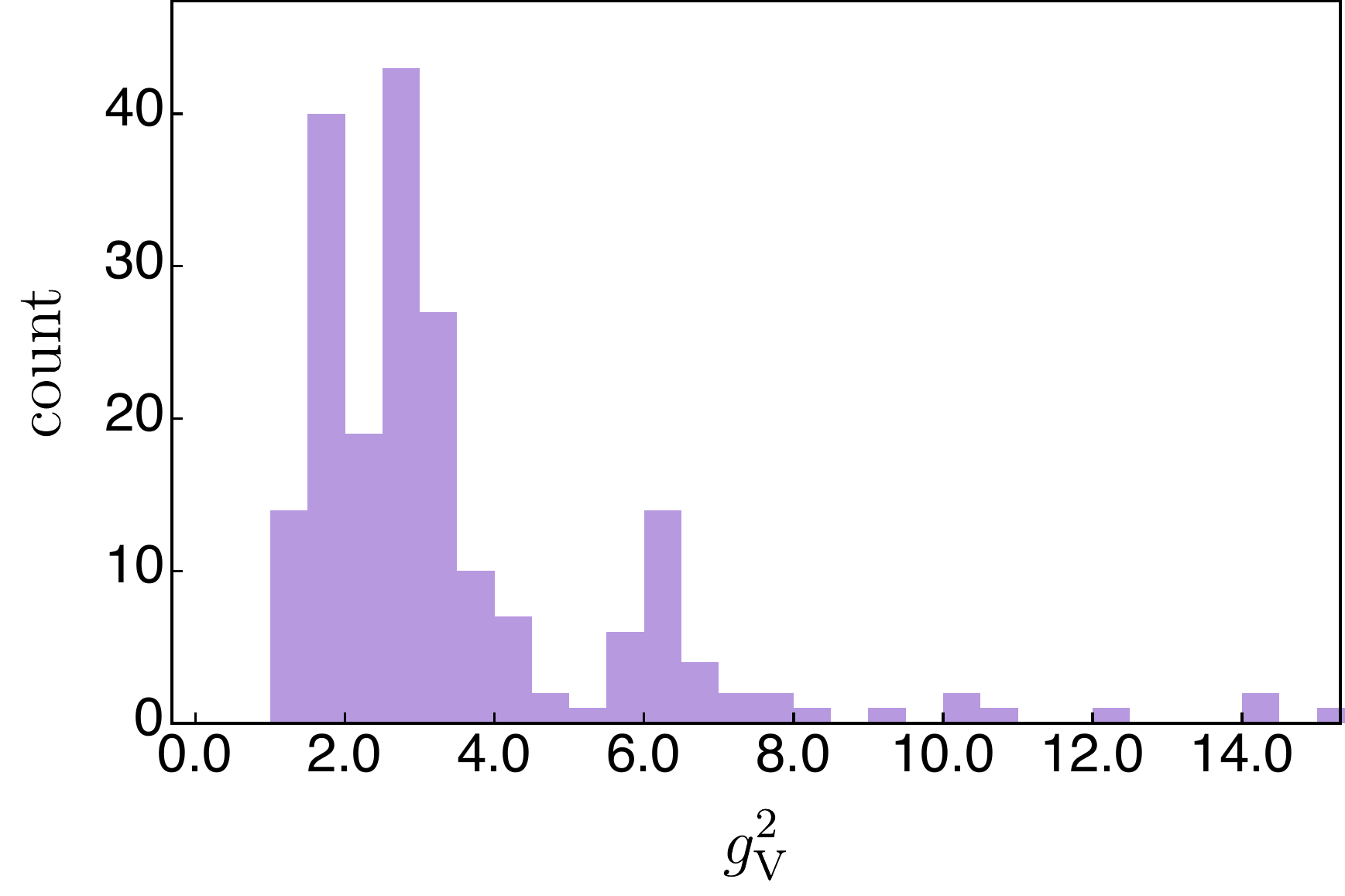}
\includegraphics[width=.49\textwidth]{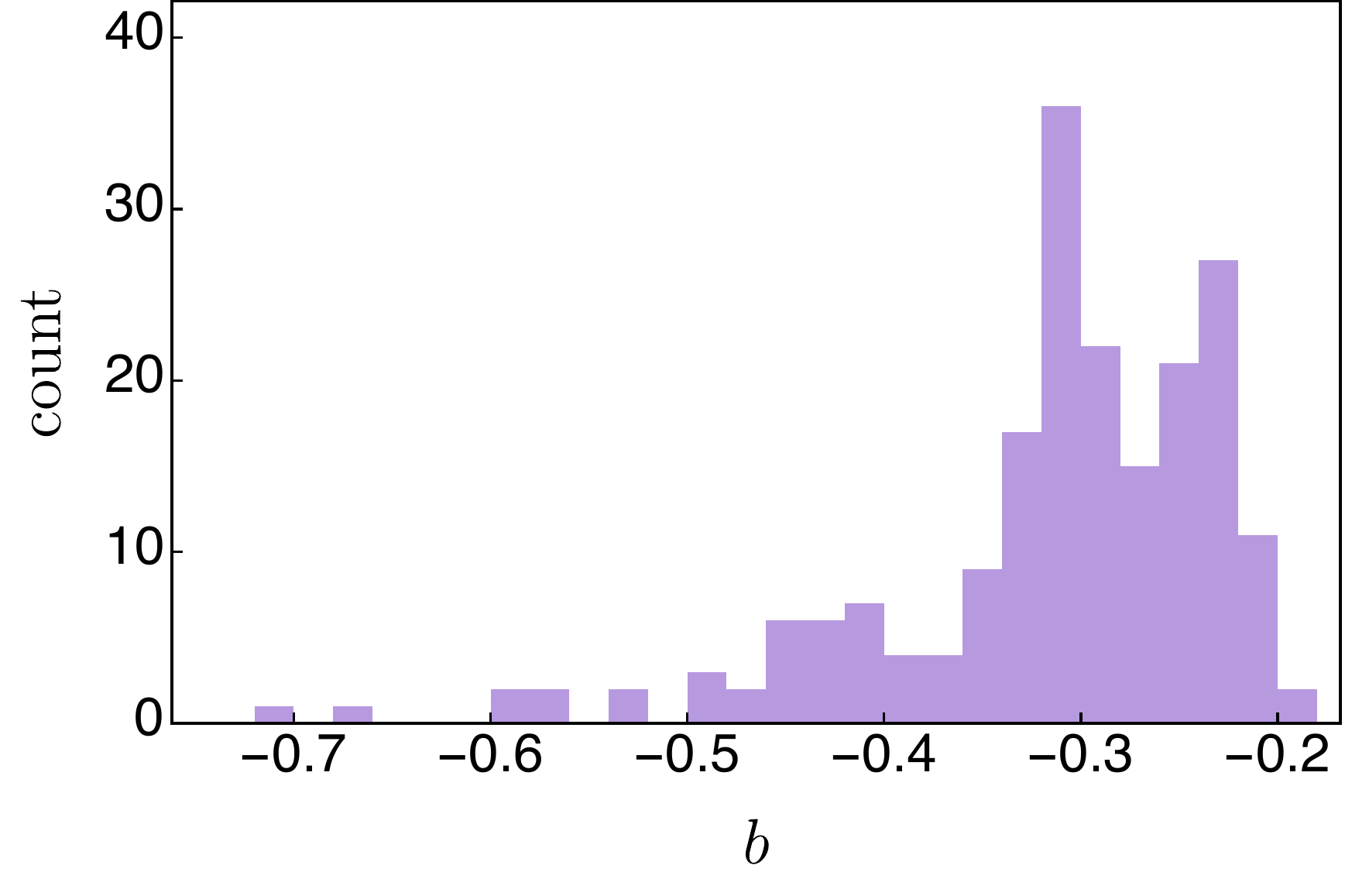}
\includegraphics[width=.49\textwidth]{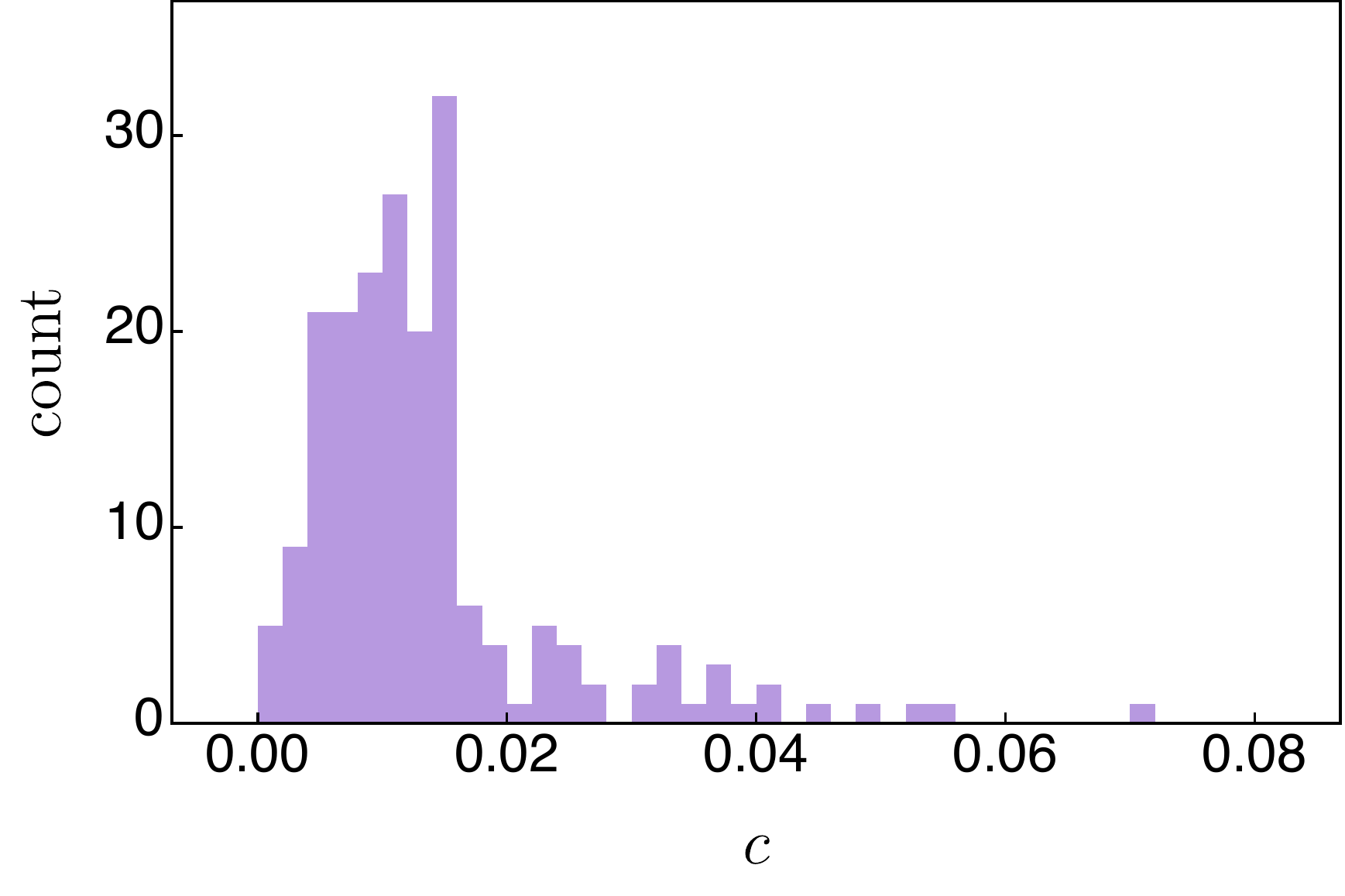}
\includegraphics[width=.49\textwidth]{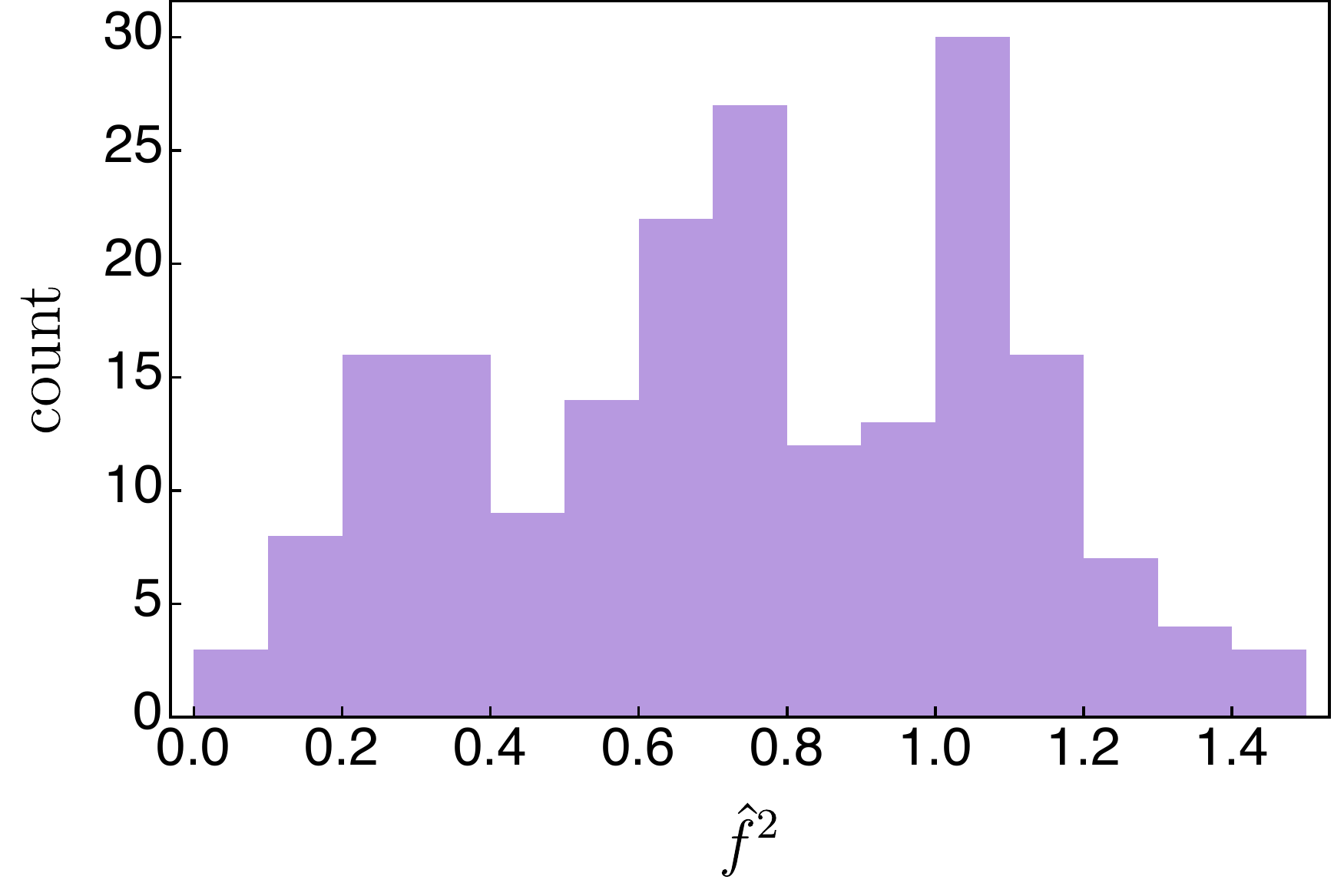}
\includegraphics[width=.49\textwidth]{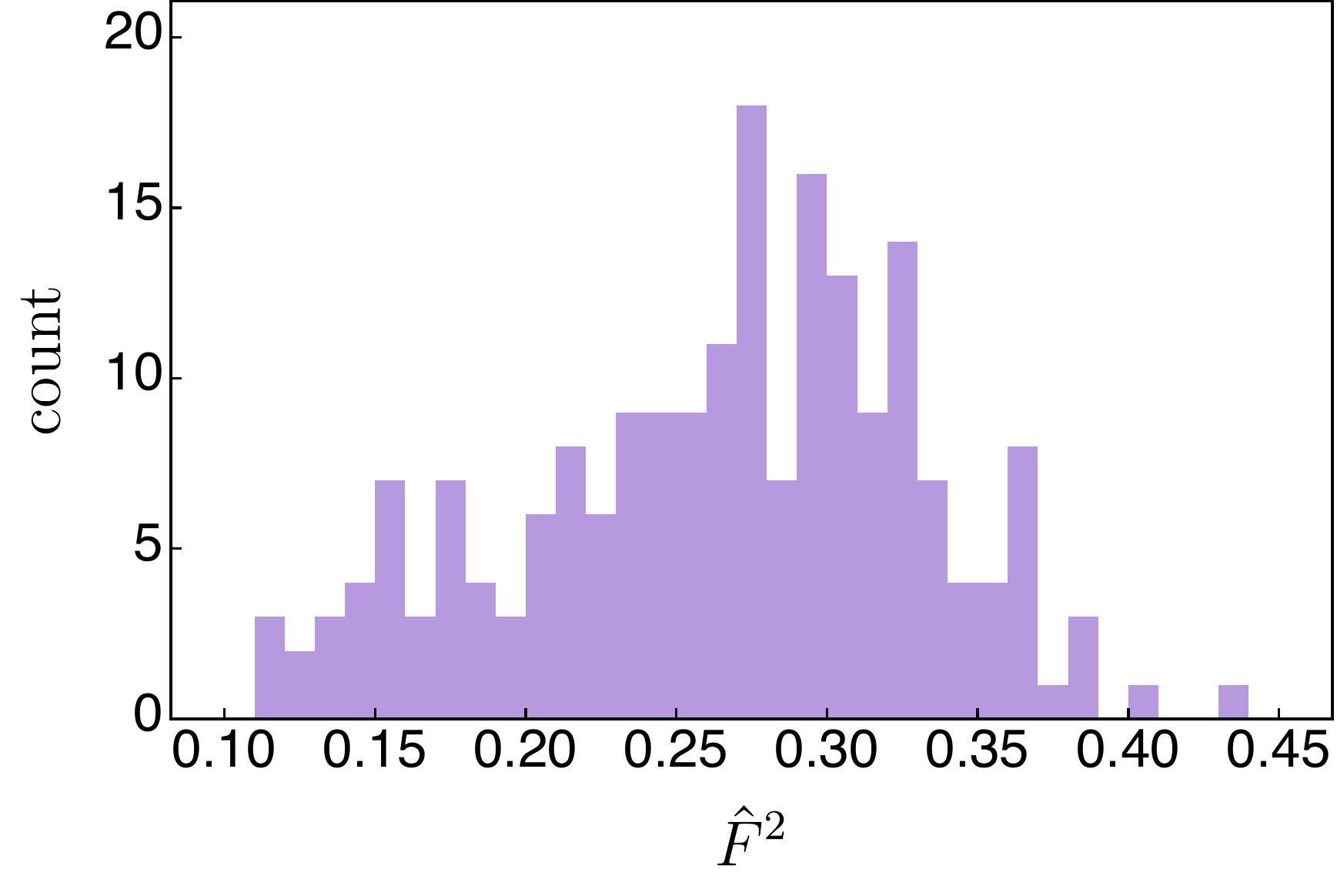}
\caption{%
\label{fig:LEC_LO_hist}%
Histograms of the distribution of the six low-energy constants 
appearing in the EFT of Section~\ref{Sec:global_fit} at the leading order, 
as determined from the global fit.
}
\end{center}
\end{figure}

\begin{figure}
\begin{center}
\includegraphics[width=.49\textwidth]{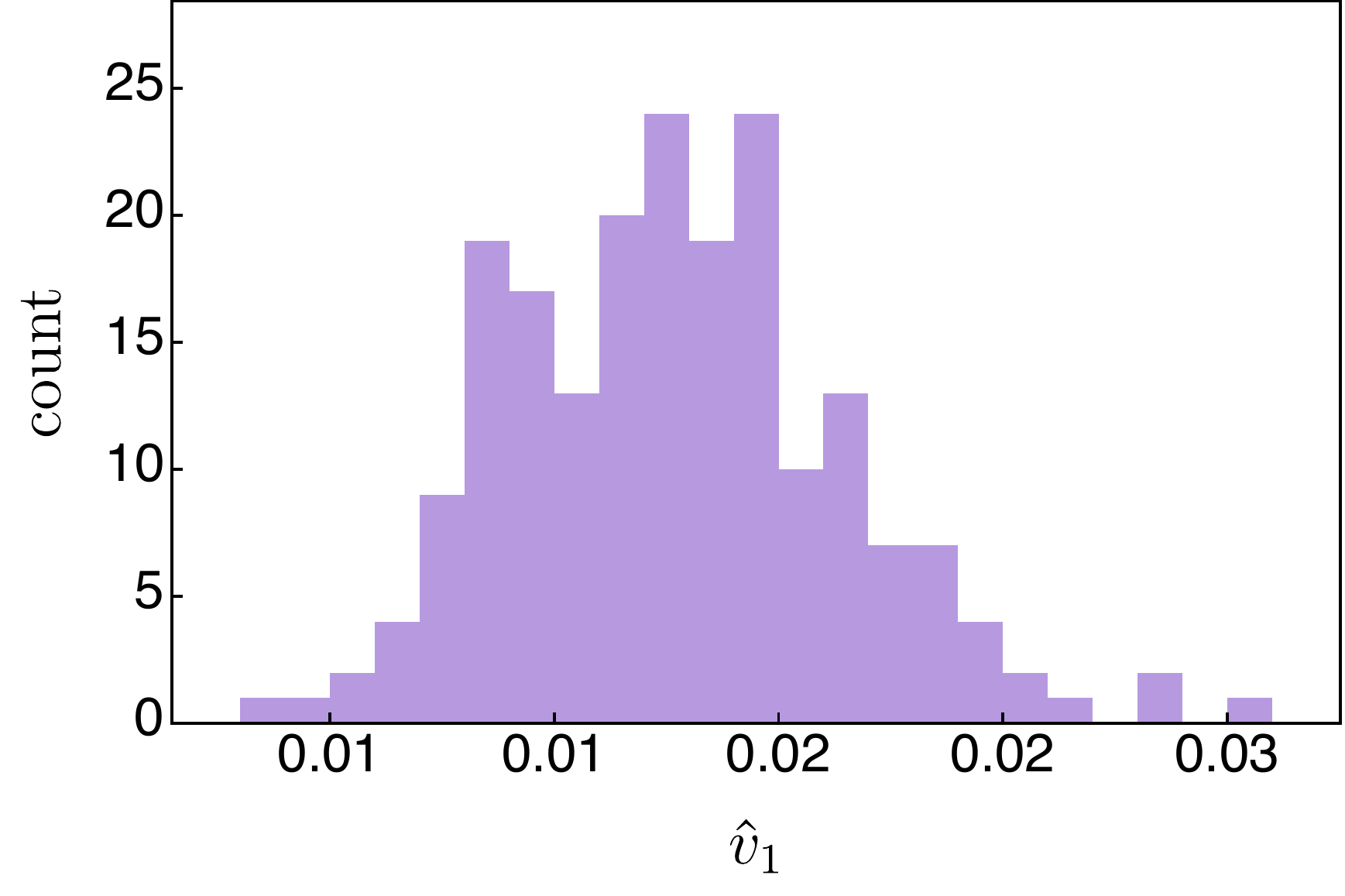}
\includegraphics[width=.49\textwidth]{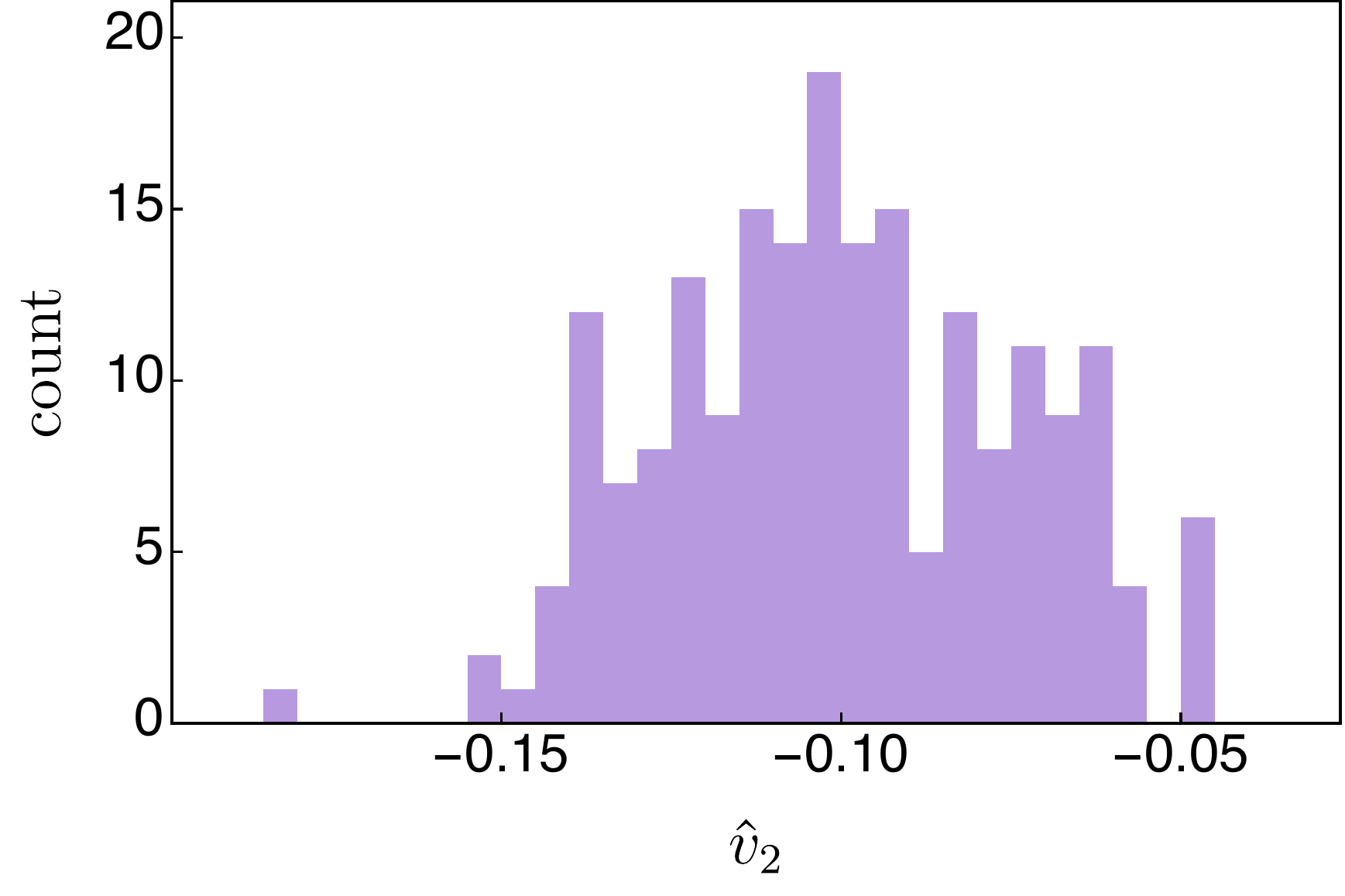}
\includegraphics[width=.49\textwidth]{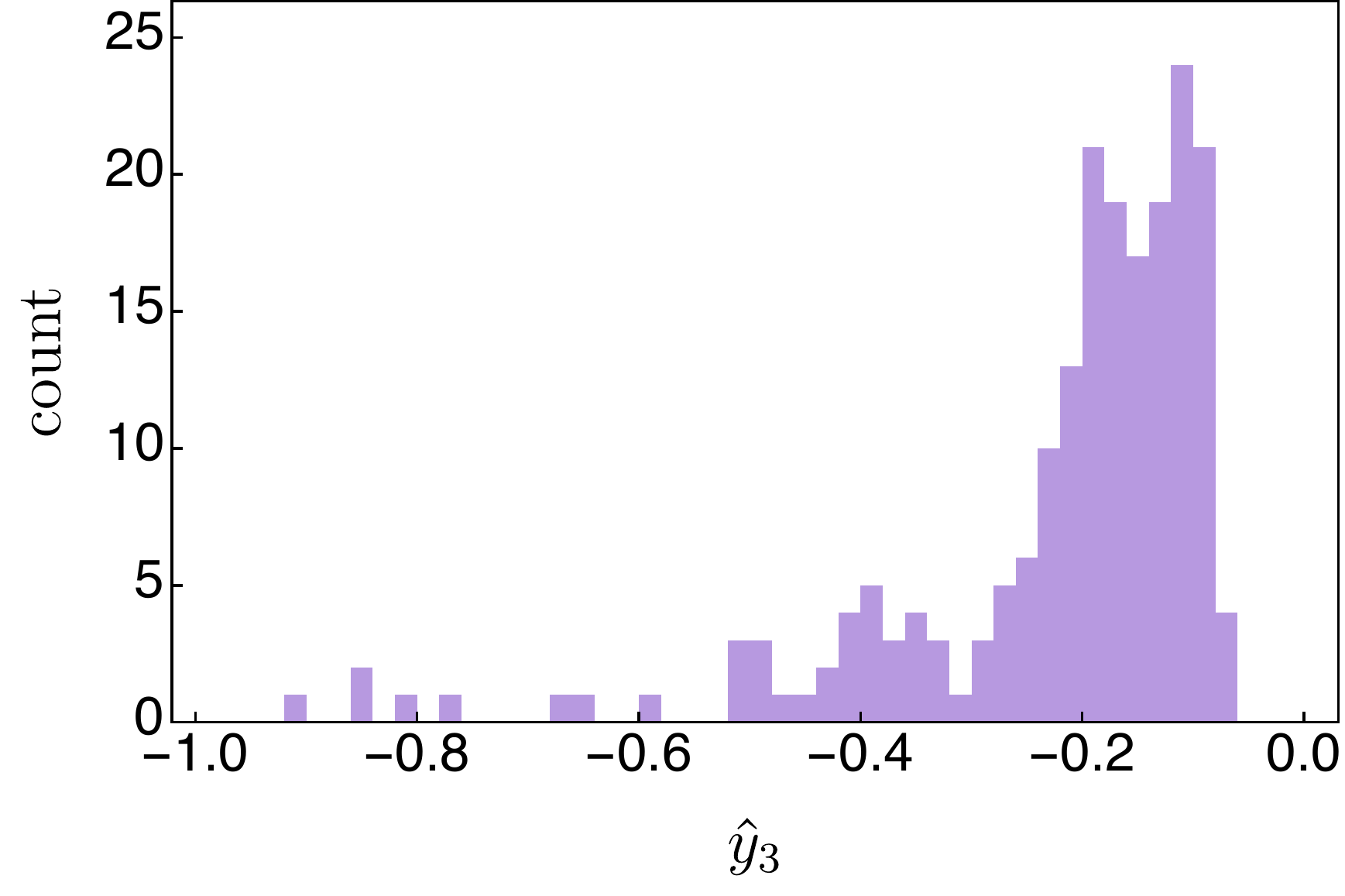}
\includegraphics[width=.49\textwidth]{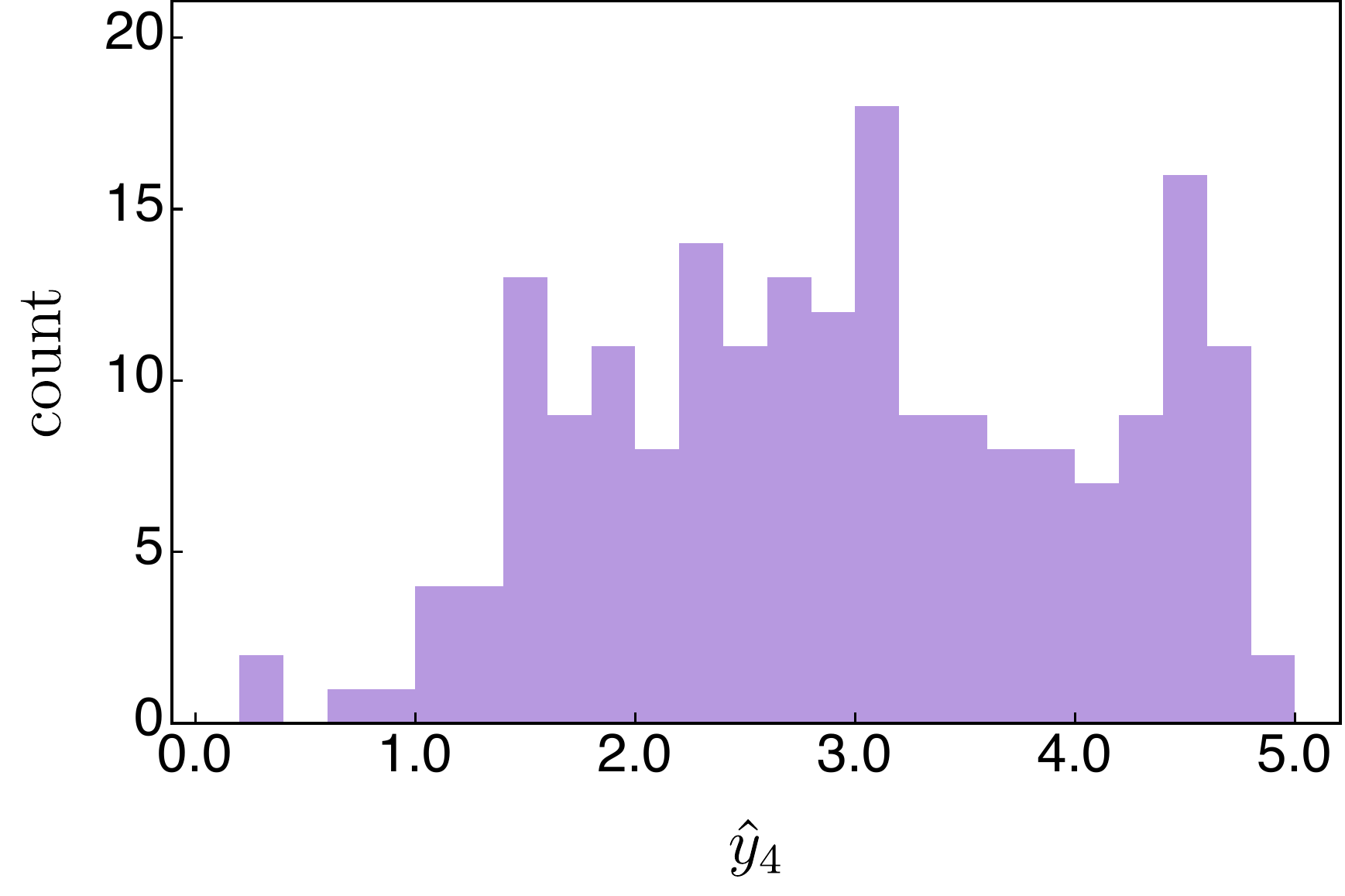}
\caption{%
\label{fig:LEC_NLO_hist}%
Histograms of the distribution of the four  low-energy constants 
appearing in the EFT of Section~\ref{Sec:global_fit} at the 
next-to-the-leading order, as determined from the global fit.
}
\end{center}
\end{figure}

\begin{figure}
\begin{center}
\includegraphics[width=.49\textwidth]{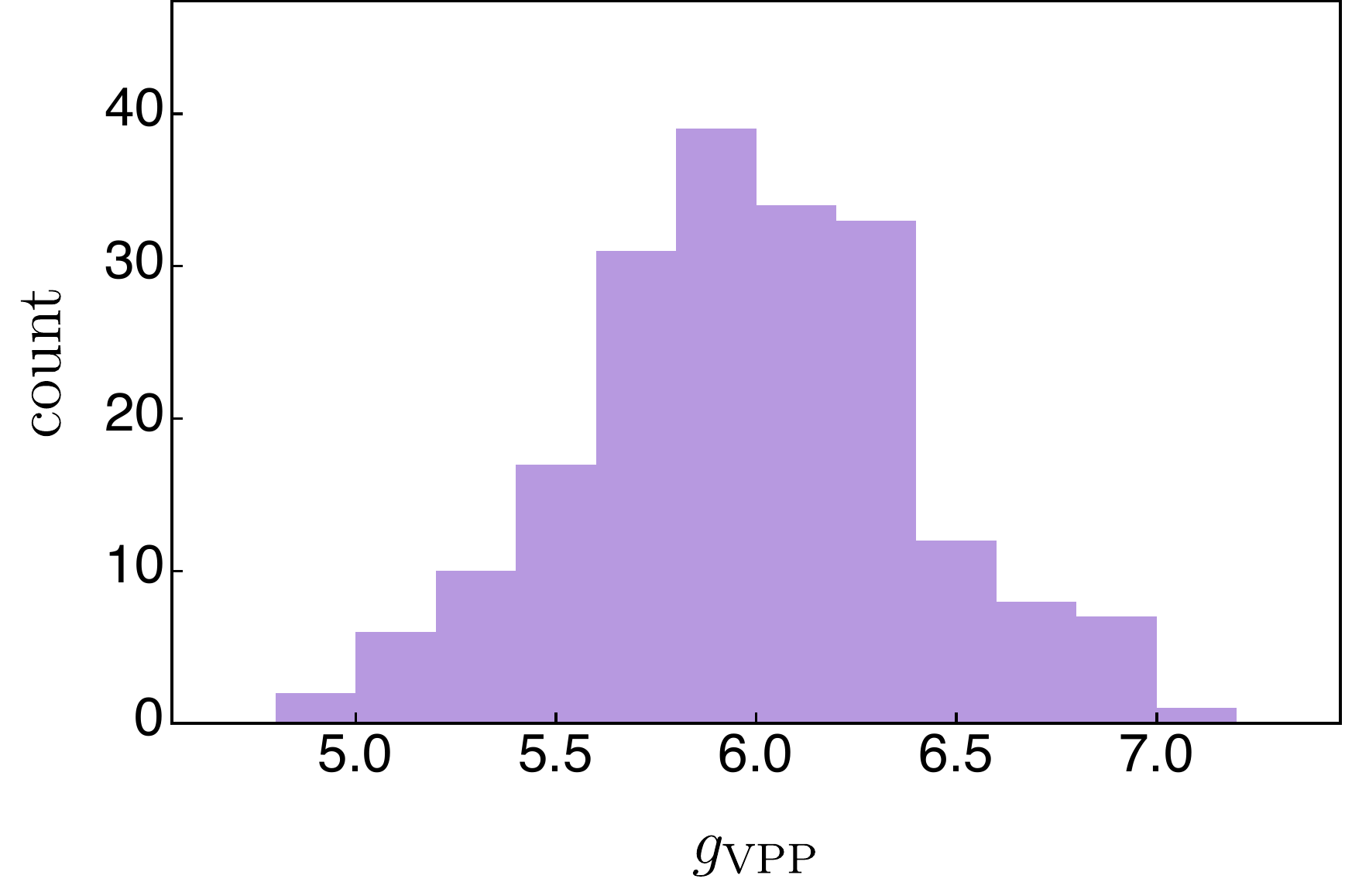}
\caption{%
\label{fig:gpp_hist}%
Histogram of  derived quantity $g_{\rm VPP}$---see Eq.~(\ref{eq:gvpp_massless}) 
in Section~\ref{Sec:global_fit}---based upon the
distributions in Figs.~\ref{fig:LEC_LO_hist} and~\ref{fig:LEC_NLO_hist}.
}
\end{center}
\end{figure}

In this Appendix, we present the numerical results for the LECs 
in Eqs.~(\ref{eq:gfit_func1})-(\ref{eq:gfit_func5}) obtained from the simplified global fit to the data
discussed in Section~\ref{Sec:global_fit}.
As anticipated, we find it instructive to  explicitly
show the histograms associated with the LEC distributions.
Figs. \ref{fig:LEC_LO_hist} and \ref{fig:LEC_NLO_hist} report the 
histograms for the LECs appearing in the EFT  at the leading 
and the next-to-the-leading order, respectively. 
As seen in the figures, some fit parameters do not exhibit  gaussian distributions, 
but rather expose long, flat tails. 
The samples in the tail do not lead to big upwards fluctuations of the value of $\chi^2/N_{\rm{d.o.f}}$,
suggesting that there are some local minima in the parameter space with $\chi^2$
 close to the  global minimum, or possibly a flat direction.
Figure~\ref{fig:gpp_hist} shows the histogram  for the $g_{\rm VPP}$ coupling
as defined and discussed in Eq.~(\ref{eq:gvpp_massless}) 
in Section~\ref{Sec:global_fit}.


\end{document}